\documentclass[twocolumn,twocolappendix]{aastex63}
\usepackage{hyperref}
\usepackage{amsmath,amstext}
\usepackage[T1]{fontenc}
\usepackage{graphicx}
\usepackage[figure,figure*]{hypcap}
\usepackage{multirow}
\usepackage{comment}
\usepackage{nicefrac}

\usepackage{stackengine}

\usepackage[]{algorithm2e} 

\usepackage{lineno}

\def\SPARK{\texttt{SPARK}}
\def\TARDIS{\texttt{TARDIS}}

\def\approxposterior{\texttt{approxposterior}}

\usepackage{soul}

\defcitealias{vieira23}{V23}
\def\V23{\citetalias{vieira23}}

\def\sun{$_{\odot}$}

\def\f28{${f}_{2-8{\rm keV}}$}

\def\sun{$_{\odot}$}

\def\ergscm2{erg s$^{-1}$ cm$^{-2}$}
\def\yr-1{yr$^{-1}$}

\def\eg{{\it e.g.}}

\def\ie{{\it i.e.}}

\def\a{$\&$}
\def\x{$\times$}

\def\w{$\omega$}

\def\asec{\ifmmode^{\prime\prime}\else$^{\prime\prime}$\fi}

\hypersetup{linkcolor=magenta}

\submitjournal{ApJ}\received{(ApJ) August 31, 2023}
\accepted{November 17, 2023}

\shorttitle{Lanthanides in the Inferred Abundance Patterns from the GW170817 Kilonova} 
\shortauthors{Vieira {\it et al.}}

\begin{document}

\title{Spectroscopic $r$-Process Abundance Retrieval for Kilonovae II: Lanthanides in the Inferred Abundance Patterns of Multi-Component Ejecta from the GW170817 Kilonova}

\correspondingauthor{Nicholas~Vieira}
\email{nicholas.vieira@mail.mcgill.ca}

\author[0000-0001-7815-7604]{Nicholas~Vieira}
\affil{Trottier Space Institute at McGill and Department of Physics, McGill University, 3600 rue University, Montreal, Qu{\'e}bec, H3A 2T8, Canada}

\author[0000-0001-8665-5523]{John~J.~Ruan}
\affil{Department of Physics and Astronomy, Bishop's University, 2600 rue College, Sherbrooke, Qu{\'e}bec, J1M 1Z7, Canada}

\author[0000-0001-6803-2138]{Daryl Haggard}
\affil{Trottier Space Institute at McGill and Department of Physics, McGill University, 3600 rue University, Montreal, Qu{\'e}bec, H3A 2T8, Canada}

\author[0000-0001-8921-3624]{Nicole M. Ford}
\affil{Trottier Space Institute at McGill and Department of Physics, McGill University, 3600 rue University, Montreal, Qu{\'e}bec, H3A 2T8, Canada}


\author[0000-0001-7081-0082]{Maria~R.~Drout}
\affil{David A. Dunlap Department of Astronomy and Astrophysics, University of Toronto, 50 St. George St., Toronto, Ontario, M5S 3H4, Canada}

\author[0000-0003-4619-339X]{Rodrigo Fern{\'a}ndez}
\affil{Department of Physics, University of Alberta, Edmonton, Alberta, T6G 2E1, Canada}


\begin{abstract}
In kilonovae, freshly synthesized $r$-process elements imprint features on optical spectra, as observed in AT2017gfo, the counterpart to the GW170817 binary neutron star merger. However, measuring the $r$-process compositions of the merger ejecta is computationally challenging. \cite{vieira23} introduced Spectroscopic $r$-Process Abundance Retrieval for Kilonovae (\SPARK), a software tool to infer elemental abundance patterns of the ejecta and associate spectral features with particular species. Previously, we applied \SPARK~to the 1.4-day spectrum of AT2017gfo and inferred its abundance pattern for the first time, characterized by electron fraction $Y_e=0.31$, a substantial abundance of strontium, and a dearth of lanthanides and heavier elements. This ejecta is consistent with wind from a remnant hypermassive neutron star and/or accretion disk. We now extend our inference to spectra at 2.4 and 3.4 days and test the need for multicomponent ejecta, where we stratify the ejecta in composition. The ejecta at 1.4 and 2.4 days is described by the same single blue component. At 3.4 days, a new redder component with lower $Y_e=0.16$ and a significant abundance of lanthanides emerges. This new redder component is consistent with dynamical ejecta and/or neutron-rich ejecta from a magnetized accretion disk. As expected from photometric modeling, this component emerges as the ejecta expands, the photosphere recedes, and the earlier bluer component dims. At 3.4 days, we find an ensemble of lanthanides, with the presence of cerium most concrete. This presence of lanthanides has important implications for the contribution of kilonovae to the $r$-process abundances observed in the Universe.

\end{abstract}


\section{Introduction}\label{sec:intro}

Approximately half of the elements in the Universe heavier than iron are synthesized by rapid neutron capture nucleosynthesis: the $r$-process (see \citealt{cowan21} for a review). Extreme astrophysical environments---namely, mergers of neutron stars (NS-NS) or an NS and black hole (NS-BH) and other proposed sources like collapsars or magnetorotational supernovae---offer leading candidate sites for this $r$-process nucleosynthesis due to their exceptionally high densities of free neutrons (\citealt{lattimer74, symbalisty82, eichler89, freiburghaus99, goriely11, korobkin12, bauswein13}). However, it is still unclear which of these channels dominates. The NS-NS merger GW170817, first detected in gravitational waves and then across the electromagnetic spectrum (\citealt{abbottLIGO17a, abbottLIGO17b}), has provided some insight. Both photometry (\citealt{andreoni17, arcavi17, coulter17, diaz17, drout17, evans17, hu17, kasliwal17, lipunov17, tanvir17, troja17, utsumi17, valenti17})\footnote{See \cite{villar17} for a compilation of this photometry considering inter-instrument variation.} and spectroscopy (\citealt{chornock17, kasen17, pian17, shappee17, smartt17}) of the optical/near-infrared counterpart, AT2017gfo, matched theoretical expectations for a kilonova: an explosive transient event powered by radioactive decay of freshly synthesized $r$-process elements. However, we do not yet know the precise abundance pattern of the $r$-process elements in the ejecta, nor whether GW170817-like kilonovae could yield the $r$-process abundances seen across the Universe (\citealt{ji19, cowan21}).

The spectra of kilonovae in particular are marked by absorption and emission features from a suite of $r$-process elements and are key for determining the detailed composition of the merger ejecta. Insights gained from spectra are independent of and complementary to light-curve modeling, which has served mostly to infer macroscopic properties of the kilonova such as the heating rates, total ejecta masses, average ejecta velocities, and temperatures of the ejecta (\eg, \citealt{villar17, almualla21, breschi21, ristic22}). By modeling the spectra, we can directly infer the abundances and the conditions of $r$-process nucleosynthesis. These insights further allow us to assess the importance of mergers versus other proposed sites as the source of these elements as well as the physical ejection mechanisms at play during these mergers. 

Spectral modeling has already provided evidence that the GW170817 kilonova ejecta contained $r$-process elements and has enabled associations of certain absorption and/or emission features with individual species. \cite{watson19} analyzed the early-time, optically thick spectra of AT2017gfo and found the imprint of a P Cygni feature arising from \ion{Sr}{2} (strontium, ${}_{38}$Sr) at $\sim$8000~\AA, at 1.4, 2.4, and 3.4 days post-merger. \cite{domoto21, domoto22} similarly ascribe this feature to \ion{Sr}{2}, and find tentative evidence for doubly ionized lanthanides \ion{La}{3} and \ion{Ce}{3} (lanthanum, ${}_{57}$La and cerium, ${}_{58}$Ce) in the near-infrared ($\sim$12,000-14,000 \AA). Similarly, \cite{gillanders22} argue for the presence of \ion{Sr}{2}, but also ions of adjacent first $r$-process peak elements \ion{Y}{2} and \ion{Zr}{2} (yttrium, ${}_{39}$Y and zirconium, ${}_{39}$Zr) at wavelengths $\lesssim$ 6000 \AA. \cite{sneppenwatson23} find that \ion{Y}{2} is present at 4.4 and 5.4 days, producing a P Cygni feature at $\sim$7600~\AA. \cite{gillanders22} also suggest the presence of a modest amount of lanthanide material at these times.  At later times, when the ejecta enters an optically thin regime, the spectrum is dominated by emission features. \cite{gillanders23} find that \ion{Ce}{3} may produce emission at $\sim$15,800~\AA~and $\sim$20,700~\AA~beyond 3.4 days, up to 10.4 days. At $\gtrsim$ 7 days, these may instead arise from intrinsically weak lines, with ions \ion{Te}{3} and \ion{Te}{1} (tellurium, ${}_{52}$Te) and \ion{I}{2} (iodine, ${}_{53}$I) given as the most likely candidates (\citealt{gillanders23, hotokezaka23}). 

While these studies have shed valuable light on some of the species present in the ejecta of AT2017gfo, the abundances of \textit{all} elements in the ejecta are not known. In \cite{vieira23} (hereafter \V23), we fit the spectrum of AT2017gfo at 1.4 days post-merger, using our inference approach and software tool Spectroscopic $r$-Process Abundance Retrieval for Kilonovae (\SPARK). With \SPARK, we inferred the complete elemental abundance pattern of the ejecta. We found that the ejecta was dominated by lighter $r$-process elements, generating a bluer (relatively lower opacity in the UV and optical) kilonova. This ejecta had electron fraction $Y_e = 0.311^{+0.013}_{-0.011}$ and specific entropy per nucleon $s / k_{\mathrm{B}} = 13.6^{+4.1}_{-3.0}$, which yielded an extremely low lanthanide fraction $\log_{10} X_{\mathrm{lan}} = {-6.74}^{+1.15}_{-1.85}$. This lanthanide fraction is inconsistent with the $r$-process abundance pattern seen in the solar system and beyond (\citealt{ji19}). 

We have not yet inferred the abundances at later epochs: 2.4 days, 3.4 days, and beyond. At later epochs, as the kilonova ejecta expands and becomes more optically thin, we expect that the photosphere recedes into the ejecta. This may uncover additional components which power the kilonova at later times, after being physically hidden underneath the photosphere at early times and/or outshined by the early blue emission. The emergence of new components at later times would be consistent with results from light-curve modeling (\eg, \citealt{villar17}), which has indicated the presence of multiple ejecta components, in particular, a redder component emerging at $\sim$3 days post-merger. These different components may originate from different ejection mechanisms during the merger and are characterized by different masses, velocities, and heating rates as a function of time. In an NS-NS merger, we expect a redder component mostly confined to the plane of the initial binary from the tidal ejecta, with little neutrino reprocessing, a bluer squeezed polar component from the collisional interface between the two NSs, and a more isotropic red/blue disk wind from an accretion disk around a merger remnant. Our inferred $Y_e$ and $s$ at 1.4 days are consistent with an outflow produced over timescales longer than the dynamical time that has been substantially reprocessed by neutrinos, \eg, winds from a hypermassive NS remnant or an accretion disk. At later times, the spectrum might be dominated or better described by a different ejecta or a multicomponent ejecta configuration. 

Here we explore the time evolution of the inferred abundance pattern and the need for multicomponent ejecta models to fit the spectra of AT2017gfo, at 1.4, 2.4, and 3.4 days post-merger. We first extend our single-component fitting to 2.4 and 3.4 days to obtain the best single-component models for the ejecta. We then develop a model in the radiative transfer code where the ejecta is stratified and multicomponent. We compare our best multicomponent fits at 1.4, 2.4, and 3.4 days to their single-component equivalents. The abundances of these single- and multicomponent models are then examined to paint a picture of the inferred abundance pattern as a function of time. 

This paper is organized as follows. In Section~\ref{sec:methods}, we briefly review \SPARK~and present the upgrades that enable fitting the later epochs of AT2017gfo with both single- and multicomponent models. In Section~\ref{sec:results}, we present our fits. In Section~\ref{sec:disco}, we explore the time evolution of the inferred abundance pattern, the species present in the ejecta, and the physical origin of different components. We briefly conclude in Section~\ref{sec:conco}.


\section{Methods}\label{sec:methods}

\subsection{Spectroscopic $r$-Process Abundance Retrieval for Kilonovae (\textsc{SPARK})}\label{ssc:spark-summary}

We briefly describe our tool, \SPARK, and refer the reader to \V23 for more detail. \SPARK~(Spectroscopic $r$-Process Abundance Retrieval for Kilonovae) is designed as a modular inference engine for extracting key kilonova parameters from optical spectra, determining the element-by-element abundance pattern of the ejecta, and associating absorption features in the spectra with particular species. 

In \SPARK, we use the 1D \TARDIS~(\citealt{kerzendorf14, kerzendorf23})~radiative transfer code to generate a set of synthetic spectra. In \TARDIS, photon packets are propagated through shell(s) of plasma, where they may undergo either bound-bound processes or electron scattering. To handle bound-bound (matter-radiation) interactions in the ejecta, we require a list of lines for the species in the ejecta. We use a line list of observed lines, obtained through the Vienna Atomic Line Database (VALD; \citealt{ryabchikova15, pakhomov19}). This line list is identical to that used in \V23, and includes only empirical lines, \ie, no lines from theoretical atomic structure calculations. Importantly, this line list is highly incomplete owing to the lack of empirical measurements of transitions and energy levels for many of the heavier elements. In associating any features in the spectrum with elements in our line list, we can only claim that a given element can produce a given feature, but not that it is the only element that could produce this feature. 

Each spectrum is parameterized by a set $\theta_i$ of parameters, including a luminosity, density, inner/outer computational boundary velocities, and three key parameters that set the abundances in the ejecta: the electron fraction $Y_e$, expansion velocity $v_{\mathrm{exp}}$, and specific entropy per nucleon $s / k_{\mathrm{B}}$. The last three parameters describe different abundance patterns output by the nuclear reaction network calculations of \cite{wanajo18}, which use parametric outflow trajectories, allowing us to infer the abundance pattern of the ejecta.  

To perform this inference, we express our likelihood function using the full formalism of \cite{czekala15} for likelihoods involving spectroscopic data. However, because of the considerable computational cost of spectral synthesis with \TARDIS, {the total time for inference with more common methods such as Markov Chain Monte Carlo (MCMC) or nested sampling would be prohibitively long, due to the large numbers of samples required by these methods. Instead, we couple \TARDIS~to the approximate posterior estimation scheme of \approxposterior~(\citealt{fleming18,fleming20}). In this scheme, rather than directly evaluating the likelihood function, we introduce a Gaussian Process (GP) surrogate for the posterior $L_p (\theta)$ and employ Bayesian Active Posterior Estimation (BAPE; \citealt{kandasamy17}) to intelligently sample parameter space. BAPE is a form of active learning in which we maximize an acquisition function with terms including both the mean $\mu(\theta)$ and variance $\sigma^2(\theta)$ of the GP. This acquisition function thus balances exploration (of the parameter space) and exploitation (sampling around the peak(s) of the posterior). The GP is iteratively retrained as new points $(\theta, L_p(\theta))$ are added to a training set, and this GP converges to an approximation of the posterior. 

In all, inference is dramatically accelerated, and we obtain (among the other parameters) the $Y_e$, $v_{\mathrm{exp}}$, and $s / k_{\mathrm{B}}$ that best describe the ejecta, with relatively few forward model evaluations. We begin with a total of $m_0$ baseline samples over all parameter space obtained via Latin Hypercube Sampling (LHS), then follow with $m_{\mathrm{active}}$ active learning samples obtained with BAPE. In \V23, we fit the VLT/X-shooter spectrum of AT2017fgfo (\citealt{pian17, smartt17}) at 1.4 days with $m_0 = 1500$ and $m_{\mathrm{active}} = 1140$ samples. This is a factor of $\sim 10^3$ fewer samples than might be required with a standard MCMC for a similar 6-dimensional fit.

\subsection{Multicomponent, stratified ejecta with \textsc{TARDIS}}\label{ssc:multicomponent-TARDIS}

In \V23, we modeled the kilonova ejecta as a single shell with a uniform abundance pattern. The plasma in this shell is also described by a single temperature and mass/electron density. This configuration is fully described by the luminosity at the outer boundary $L_{\mathrm{outer}}$, (which in fact sets the initial guess for the temperature at the inner boundary), the normalization in the density power law $\rho_0$, the inner and outer boundary velocities $v_{\mathrm{inner}}$ and $v_{\mathrm{outer}}$\footnote{\TARDIS~assumes homologous expansion, in which $v \propto r$, such that velocity can be understood as a spatial coordinate. Given 
a density profile, the velocity range sets the mass of the ejecta.}, and three parameters that set the abundance pattern: electron fraction $Y_e$, expansion velocity $v_{\mathrm{exp}}$, and specific entropy $s / k_{\mathrm{B}}$. A single-component fit is thus 7-dimensional, unless one or more of the parameters are fixed. For example, we fixed $v_{\mathrm{outer}} = 0.35c$ in our fit to the 1.4-day spectrum. This setup can describe a single ejecta component such as dynamical ejecta or some outflow. It may also describe a kilonova in which one component significantly dominates (by mass or by the strength of the absorption/emission features) over the other(s). 
    
Here we implement multicomponent ejecta. \TARDIS~allows for ejecta composed of stratified radial shells, each with a specific temperature, density, plasma conditions, and composition. In this configuration, each shell can have a specific abundance pattern. In single-shell runs, we compute the plasma state only once given $L_{\mathrm{outer}}$. For these multi-shell runs, we must employ multiple \TARDIS~iterations when generating synthetic spectra. Beginning with a guess for the inner boundary temperature based on the user-requested value for $L_{\mathrm{outer}}$, at each iteration the plasma conditions (temperatures and dilution factors in each shell) are perturbed, and electron densities and level populations are recomputed. Photon packets are then propagated through all shells, and an estimate of the emergent spectrum is obtained. With sufficient iterations, the plasma converges to a state where the luminosity emitted at the \textit{outer} boundary matches the user-requested $L_{\mathrm{outer}}$. \TARDIS~employs 20 such shells and 20 such iterations, by default. We use 10 shells to decrease computation times; in our tests, we find no marked difference in the resultant synthetic spectra with this smaller number of shells. We use 30 iterations to guarantee convergence of the plasma to a state described by the input parameters.
    
We begin with a simple two-component ejecta. As with our single-component model, the ejecta has an outer boundary luminosity $L_{\mathrm{outer}}$ and power-law density profile with normalization $\rho_0$. $L_{\mathrm{outer}}$ describes the luminosity at the outer boundary of the overall ejecta, where photon packets are binned into a synthetic spectrum. The ejecta has an inner bound that is the lesser of $v_{\mathrm{inner},1},~v_{\mathrm{inner},2}$ and an outer bound that is the greater of $v_{\mathrm{outer},1},~v_{\mathrm{outer},2}$. Each of two components is thus described by two velocities and three abundance-setting-parameters $Y_e$, $v_{\mathrm{exp}}$, and $s / k_{\mathrm{B}}$, \ie, five parameters each. Two-component fits are thus $2 + 5 + 5 = 12$-dimensional. The two components necessarily overlap in physical space because \TARDIS~cannot simulate a gap between them. The abundance in each shell is then determined by the component(s) that are in a given shell. For shells where there is overlap between two components, the abundance is taken as a sum of the two abundance patterns and renormalized to unity. 

Multiple components allow for additional complexity in the spectral synthesis.\footnote{See \cite{kawaguchi20} for an exploration of the diversity of kilonovae that may be produced when multiple components are present.} In particular, we can produce the effect of reprocessing, where emission from one component is absorbed and reemitted/scattered by another. We can also produce the effect of lanthanide curtaining, in which some outer lower-$Y_e$ ejecta containing the lanthanides masks an inner, bluer, lighter-element ejecta, generating a redder kilonova owing to the considerable opacity of the lanthanides in the near-UV and optical. Some kilonova spectra may be better described by these multicomponent ejecta models. In \V23, we find that the 1.4-day spectrum of AT2017gfo is well-described by a single-component ejecta with {$Y_e = 0.311^{+0.013}_{-0.011}$, where $0.311$ is the median of our posterior distribution for $Y_e$ and the upper/lower bounds are the 97.5th and 2.5th percentiles, respectively. However, at later epochs, as the ejecta expands and becomes more optically thin, the photosphere recedes into the ejecta and we may unmask additional components that were hidden or outshined at early times. Light-curve modeling (\eg, \citealt{villar17}) has shown that the kilonova may indeed be better described by multiple components of different opacities, and some spectral modeling (\eg, \citealt{kasen17}) also invokes multiple components. This motivates our introduction of multicomponent ejecta into \SPARK.

\subsection{Inference setup}\label{ssc:inference-setup}

All \texttt{approxposterior}/BAPE hyperparameters and optimizers used in this work are the same as those used in \V23. We again produce a base set of $m_{0} = 1500$ Latin hypercube sampled points at the beginning of each \SPARK~run. However, the parameter space allowed by our priors differs for the 1.4-, 2.4-, and 3.4-day fits. Table~\ref{tab:priors-single} includes our (all uniform) priors for each of our single-component fits. The bounds on the density $\rho_0$ and abundance-setting parameters $Y_e$, $v_{\mathrm{exp}}$, and $s$ are the same at all epochs. The priors differ in the bounds on the luminosity, as expected given the cooling of the ejecta over time as it expands. {However, all priors on luminosity are quite broad and based on the estimated bolometric luminosity of AT2017gfo (\eg, \citealt{wu19}) and our inference at 1.4 days. We further allow for wider priors on the inner and outer boundary velocities for the fits at later epochs. In \V23, we fixed $v_{\mathrm{outer}} = 0.35c$ during our 1.4-day fit but noted that we observed similar results for $v_{\mathrm{outer}}$ in the range $0.35c - 0.38c$. Here we allow for greater flexibility in $v_{\mathrm{outer}}$ {at later epochs.
    
Our priors for our multicomponent fits are given in Table~\ref{tab:priors-multi}. These are broad, and identical, at 1.4 and 2.4 days. Aside from requiring $v_{\mathrm{inner},i} < v_{\mathrm{outer},i}$, our priors also require some overlap between the two components, \ie, $v_{\mathrm{outer},2} > v_{\mathrm{inner},1}$ or $v_{\mathrm{outer},1} > v_{\mathrm{inner},2}$. To allow for even sampling of parameter space for the $m_0 = 1500$ points in the base training set with these conditional constraints, we use constrained LHS (\citealt{petelet09}). At 3.4 days, we encounter challenges with the convergence of the posterior for these broad, conditional priors in velocities. {These broad, complicated priors result in a highly multimodal posterior, and the GP struggles to capture the relative amplitudes of the different modes, leading to inferred parameters that evolve as more samples are added to the GP. This challenge with GPs has been documented in \cite{elgammal23}. In an effort to obtain a less complex posterior, we use tighter, nonconditional priors on the velocities and find that these lead to a convergent posterior. We thus use tighter priors on the velocities at this epoch, but all other priors are the same as those at 1.4 and 2.4 days.
    
When attempting to fit the full 3.4-day spectrum, we find that the fit converges to a blackbody with little to no absorption to fit the continuum of the spectrum, especially at the shortest wavelengths $\lesssim$6400~\AA. Since we are interested in determining the species involved in the absorption, and especially the prominent feature at $\sim$8000~\AA, we prioritize fitting the $\geqslant$6400~\AA~region of the spectrum by excluding shorter wavelengths in the computation of the likelihood. We perform this exclusion for both single- and multicomponent fits.

Finally, when expressing the likelihood using the \cite{czekala15} formalism, {we introduce a global covariance term in the likelihood that introduces pixel-to-pixel (\ie, wavelength bin-to-wavelength bin) correlations into the likelihood, beyond the standard individual uncorrelated uncertainties that would yield a simple $\chi^2$ likelihood. This global covariance is described by a Mat{\'e}rn-3/2 kernel with some hyperparameters: an amplitude $a_G$ and correlation length scale $\ell$. We use a global covariance term with amplitude $a_G = 10^{-34}~(\mathrm{erg~s^{-1}~cm^{-2}}$~\text{\AA}${}^{-1})^{2}$ and a correlation length scale $\ell = 0.025c$ for all 1.4- and 2.4-day fits. For 3.4-day fits, after some trial and error, we find that a smaller $a_G = 10^{-35}~(\mathrm{erg~s^{-1}~cm^{-2}}$~\text{\AA}${}^{-1})^{2}$ better matches the uncertainties on the observed spectrum at this epoch.

\begin{deluxetable}{c|cc}
\centering
\tablecaption{Uniform priors for the single-component fits at 1.4, 2.4, and 3.4 days. $v_{\mathrm{outer}}$ is fixed to $0.35c$ in the 1.4 day fit (see \V23).}
\tablehead{parameter & 1.4 days & 2.4, 3.4 days}
\startdata\tablewidth{1.0\textwidth}
 \vspace{2pt}
$\log_{10}(L_{\mathrm{outer}} / L_{\odot})$ & $[7.6, 8.0]$ & $[7.2, 7.8]$ \\ 
$\log_{10}(\rho_0 / \mathrm{g~cm^{-3}})$ & $[-16.0, -14.0]$ & same \\
$v_{\mathrm{inner}}/c$& $[0.250, 0.340]$ & $[0.100, 0.275]$ \\
$v_{\mathrm{outer}}/c$& $0.35c$ & $[0.280, 0.400]$\\
$v_{\mathrm{exp}}/c$ & $[0.05, 0.30]$ & same \\
$Y_e$ & $[0.01, 0.40]$ & same \\
$s~[k_{\mathrm{B}}/\mathrm{nucleon}]$ & $[10, 35]$ & same \\
\enddata
\end{deluxetable}\label{tab:priors-single}

\begin{deluxetable}{c|cc}
\centering
\tablecaption{Uniform priors for the multicomponent fits at 1.4, 2.4, and 3.4 days. The priors on the velocities are tighter at 3.4 days.}
\tablehead{parameter & 1.4, 2.4 days & 3.4 days}
\startdata\tablewidth{1.0\textwidth}
 \vspace{2pt}
$\log_{10}(L_{\mathrm{outer}} / L_{\odot})$ & $[7.0, 8.0]$ & same \\ 
$\log_{10}(\rho_0 / \mathrm{g~cm^{-3}})$ & $[-16.0, -14.0]$ & same \\\hline
$v_{\mathrm{inner,1}}/c$& $[0.10, 0.35]$ & $[0.10, 0.29]$ \\
$v_{\mathrm{outer,1}}/c$ &  $[0.25, 0.40]$ & $[0.30, 0.40]$ \\
$v_{\mathrm{exp,1}}/c$ & $[0.05, 0.30]$ & same \\
$Y_{e,1}$ & $[0.01, 0.50]$ & same \\
$s_{1}~[k_{\mathrm{B}}/\mathrm{nucleon}]$ & $[10, 35]$ & same \\\hline
$v_{\mathrm{inner,2}}/c$& $[0.10, 0.35]$ & $[0.20, 0.30]$ \\
$v_{\mathrm{outer,2}}/c$ &  $[0.25, 0.40]$ & $[0.31, 0.40]$ \\
$v_{\mathrm{exp,2}}/c$ & $[0.05, 0.30]$ & same \\
$Y_{e,2}$ & $[0.01, 0.50]$ & same \\
$s_{2}~[k_{\mathrm{B}}/\mathrm{nucleon}]$ & $[10, 35]$ & same \\
\enddata
\end{deluxetable}\label{tab:priors-multi}


\section{Fitting Later Epochs and Multicomponent Ejecta}\label{sec:results}

\begin{figure*}[!ht]
    \centering
    \includegraphics[width=0.98\textwidth]{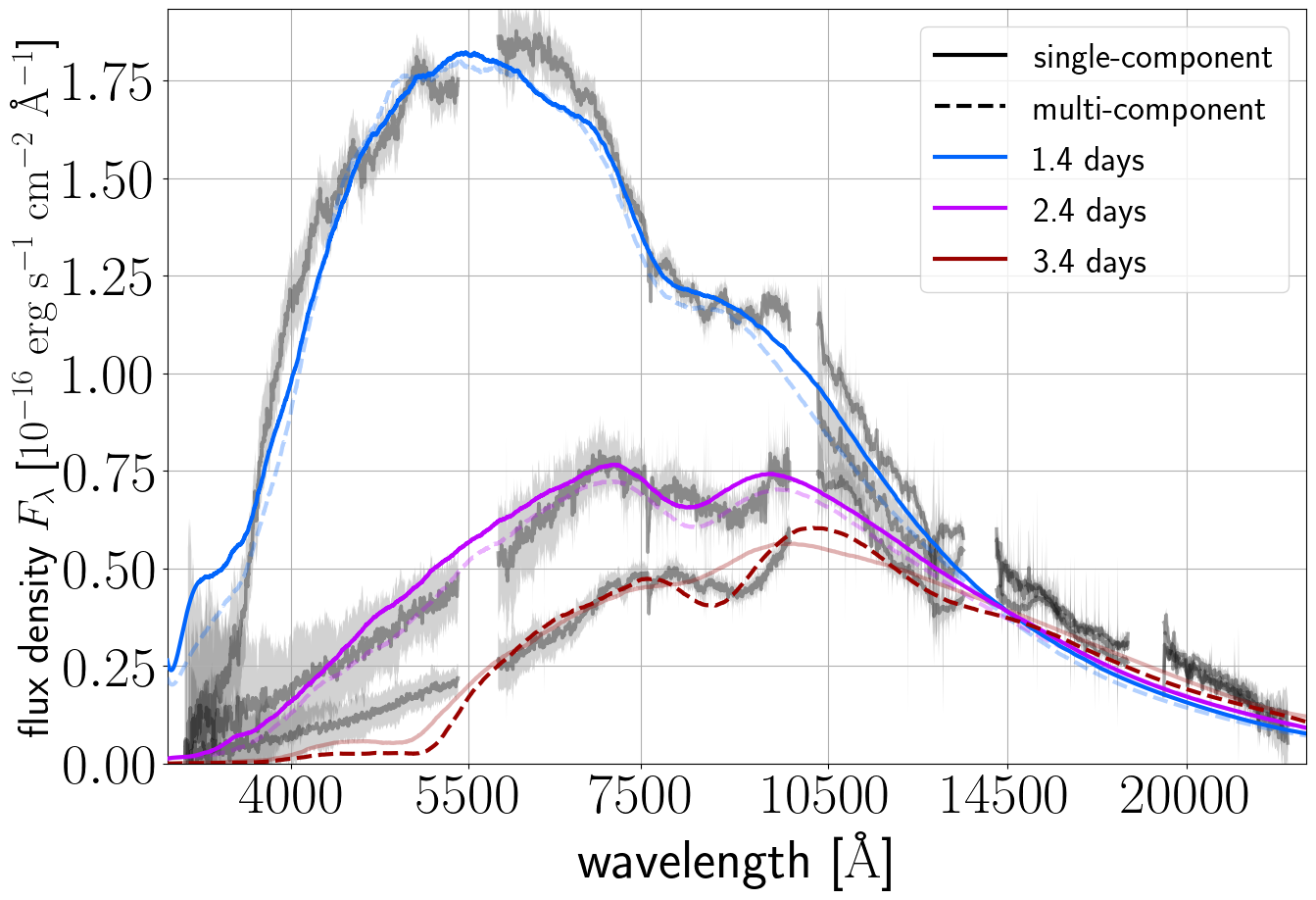}
    \figcaption{\textbf{Compilation of all best-fit models to the spectrum of the GW170817 kilonova from this work: 1.4, 2.4, and 3.4 days post-merger, for both single- and multicomponent models.} The preferred models are highlighted: opaque lines are the preferred fits. We justify their preference in Section~\ref{sec:results} (especially \ref{ssc:favored}). The single-component models are favored at 1.4 and 2.4 days (solid blue and purple lines). At 3.4 days, we require a multicomponent ejecta to adequately fit the spectrum (dashed red line). The abundance patterns corresponding to these preferred models are included in Figure~\ref{fig:abunds_time_evolution}. Zoomed-in versions of these best fits are included in Appendix~\ref{app:allspec_posteriors}. Best fits are obtained as the median of the posterior, or the median of some mode of the posterior; these full posteriors are also included in Appendix~\ref{app:allspec_posteriors}. The preferred fits capture the shape of the continuum and the dominant absorption feature at $\sim$8000~\AA, but struggle to capture the emission bump of this feature in the interpretation that it is a P Cygni feature.}\label{fig:bestfits}
\end{figure*}

\begin{deluxetable}{cccc}
\centering
\tablecaption{Best-fit parameters for single-component fits to the GW170817 kilonova at $1.4$, $2.4$, and $3.4$~days. The 1.4-day fit is the ``purple + warm'' model of \V23. $v_{\mathrm{outer}}$ is fixed to $0.35c$ in the 1.4-day fit. The inner boundary temperature $T_{\mathrm{inner}}$ and lanthanide mass fraction $X_{\mathrm{lan}}$ are derived parameters, \ie, they are not dimensions in $\theta$-space.}
\tablehead{parameter & 1.4 days & 2.4 days & 3.4 days}
\startdata\tablewidth{1.0\textwidth}
 \vspace{2pt}
$\log_{10}(\frac{L_\mathrm{outer}}{L_{\odot}})$ & $7.782^{+0.013}_{-0.014}$ & $7.594^{+0.040}_{-0.061}$ & $7.531^{+0.017}_{-0.016}$ \\ 
$\log_{10}(\frac{\rho_0}{\mathrm{g~cm^{-3}}})$ & $-15.016^{+0.320}_{-0.316}$ & $-15.443^{+0.742}_{-0.463}$ & $-14.586^{+0.313}_{-0.384}$ \\ 
$v_{\mathrm{inner}}/c$ & $0.313^{+0.013}_{-0.014}$ & $0.249^{+0.017}_{-0.032}$ & $0.253^{+0.011}_{-0.021}$ \\
$v_{\mathrm{outer}}/c$ & $0.35$ & $0.342^{+0.047}_{-0.050}$  & $0.309^{+0.032}_{-0.023}$ \\
$v_{\mathrm{exp}}/c$ & $0.240^{+0.055}_{-0.082}$ & $0.172^{+0.107}_{-0.101}$ & $0.132^{+0.094}_{-0.056}$ \\
$Y_e$ & $0.311^{+0.013}_{-0.011}$ & $0.306^{+0.055}_{-0.204}$ & $0.226^{+0.062}_{-0.067}$ \\
$s~[k_{\mathrm{B}}/\mathrm{nuc}]$ & $13.6^{+4.1}_{-3.0}$ & $17.6^{+7.1}_{-6.3}$ & $15.4^{+7.1}_{-6.3}$ \\ \hline
$T_{\mathrm{inner}}~[\mathrm{K}]$ & $3958^{+87}_{-94}$ & $3050^{+126}_{-223}$ & $2455^{+59}_{-104}$ \\ 
$\log_{10} X_{\mathrm{lan}}$ & ${-6.74}^{+1.15}_{-1.85}$ & ${-7.02}^{+3.02}_{-9.81}$ & ${-1.97}^{+1.28}_{-2.84}$ \\
\enddata
\end{deluxetable}\label{tab:bestfit_single}

\begin{deluxetable*}{cccc}
\centering
\tablecaption{Best-fit parameters for multicomponent fits to the GW170817 kilonova at 1.4, 2.4, and 3.4 days. The ejecta mass above the photosphere $M_{\mathrm{phot},1;2}$ (not the total ejecta mass), lanthanide mass fractions $X_{\mathrm{lan,1; 2}}$, and the total $X_{\mathrm{lan,total}}$ are derived parameters.}
\tablehead{parameter & 1.4 days & 2.4 days & 3.4 days}
\startdata\tablewidth{1.0\textwidth}
 \vspace{2pt}
$\log_{10}(\frac{L_\mathrm{outer}}{L_{\odot}})$ & $7.854^{+0.012}_{-0.017}$ & $7.700^{+0.065}_{-0.073}$ & $7.605^{+0.049}_{-0.040}$ \\ 
$\log_{10}(\frac{\rho_0}{\mathrm{g~cm^{-3}}})$ & $-15.095^{+0.127}_{-0.401}$ & $-15.440^{+0.813}_{-0.467}$ & $-14.505^{+0.323}_{-0.372}$ \\ \hline
$v_{\mathrm{inner,1}}/c$ & $0.323^{+0.007}_{-0.020}$ & $0.260^{+0.061}_{-0.039}$ & $0.213^{+0.056}_{-0.035}$ \\
$v_{\mathrm{outer,1}}/c$ & $0.326^{+0.009}_{-0.012}$ & $0.335^{+0.054}_{-0.065}$ & $0.344^{+0.035}_{-0.038}$ \\
$v_{\mathrm{exp,1}}/c$ & $0.118^{+0.029}_{-0.025}$ & $0.170^{+0.108}_{-0.100}$ & $0.198^{+0.092}_{-0.102}$ \\
$Y_{e,1}$ & $0.139^{+0.028}_{-0.114}$ & $0.288^{+0.129}_{-0.187}$ & $0.228^{+0.073}_{-0.088}$ \\
$s_{1}~[k_{\mathrm{B}}/\mathrm{nuc}]$ & $24.6^{+3.4}_{-5.1}$ & $21.1^{+10.5}_{-9.5}$ & $14.9^{+8.0}_{-4.5}$ \\ \hline
$v_{\mathrm{inner,2}}/c$ & $0.281^{+0.011}_{-0.013}$ & $0.268^{+0.049}_{-0.045}$ & $0.232^{+0.038}_{-0.027}$ \\
$v_{\mathrm{outer,2}}/c$ & $0.355^{+0.015}_{-0.015}$ & $0.335^{+0.045}_{-0.070}$ & $0.334^{+0.037}_{-0.022}$ \\
$v_{\mathrm{exp,2}}/c$ & $0.195^{+0.034}_{-0.036}$ & $0.183^{+0.089}_{-0.110}$ & $0.134^{+0.091}_{-0.069}$ \\
$Y_{e,2}$ & $0.340^{+0.022}_{-0.021}$ & $0.261^{+0.163}_{-0.124}$ & $0.161^{+0.149}_{-0.104}$ \\
$s_{2}~[k_{\mathrm{B}}/\mathrm{nuc}]$ & $20.1^{+3.8}_{-3.6}$ & $22.0^{+10.0}_{-10.1}$ & $21.5^{+8.3}_{-9.5}$ \\ \hline\hline
$M_{\mathrm{phot},1}~[M_{\odot}]$ & $6.9^{+22.1}_{-6.0} \times 10^{-7}$ & $2.3^{+12.5}_{-0.7}\times 10^{-5}$ & $5.3^{+1.8}_{-1.3}~\times 10^{-4}$ \\
$\log_{10} X_{\mathrm{lan,1}}$ & ${-1.24}^{+0.68}_{-0.20}$ & ${-4.10}^{+3.68}_{-13.17}$ & ${-1.78}^{+1.11}_{-3.69}$ \\ \hline
$M_{\mathrm{phot},2}~[M_{\odot}]$ & $7.0^{+0.8}_{-1.0} \times 10^{-5}$ & $4.0^{+4.8}_{-0.9}\times 10^{-5}$ & $3.8^{+1.1}_{-0.9}~\times 10^{-4}$ \\
$\log_{10} X_{\mathrm{lan,2}}$ & ${-13.05}^{+4.74}_{-4.82}$ & ${-2.61}^{+2.11}_{-8.86}$ & ${-0.53}^{+0.08}_{-2.86}$
 \\ \hline
$\log_{10} X_{\mathrm{lan,total}}$ & ${-13.05}^{+9.88}_{-3.32}$ & ${-3.55}^{+3.00}_{-5.75}$ & ${-1.58}^{+0.86}_{-1.46}$ \\
\enddata
\end{deluxetable*}\label{tab:bestfit_multi}

\begin{figure*}[!ht]
    \centering
    \includegraphics[width=\textwidth]{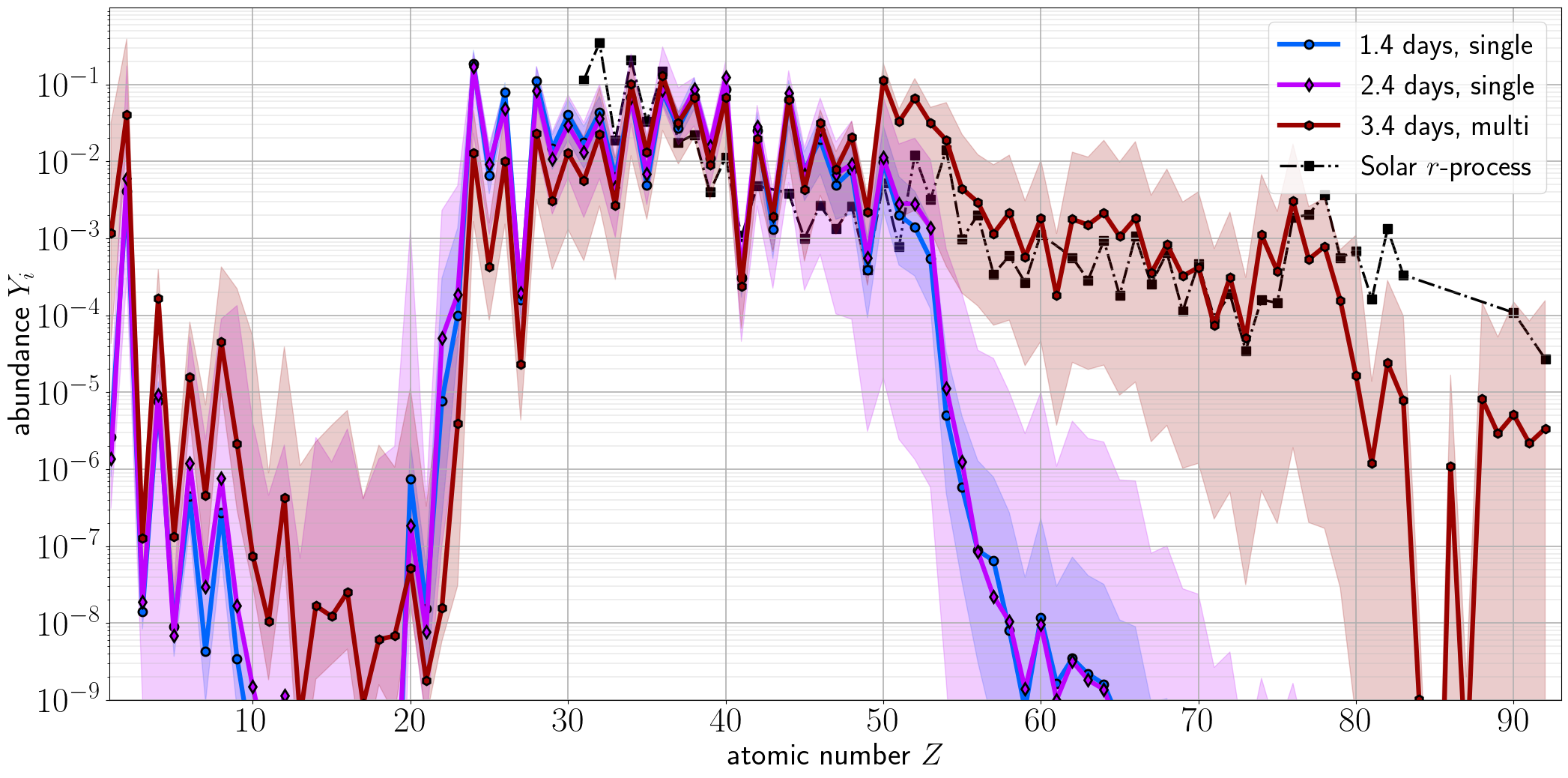}
    \figcaption{\textbf{Best-fit abundance patterns at 1.4, 2.4, and 3.4 days.} The abundances at 1.4 and 2.4 days are taken from the preferred single-component fits, while the abundance at 3.4 days is taken from the preferred multicomponent fit (see Figure~\ref{fig:bestfits}). The overall abundance of the multicomponent ejecta at 3.4 days is the mass-weighted sum over all 10 shells in the stratified ejecta. A new, redder kilonova component emerges at 3.4 days post-merger. Uncertainty bands are obtained by taking additional samples from the posterior, effectively propagating the uncertainty on $Y_e$, $v_{\mathrm{exp}}$, and $s$ into the abundances. We also show the solar $r$-process pattern, computed using data from \cite{lodders09} subtracted by the $s$-process residuals of \cite{bisterzo14}. The best-fit abundances are evidently nonsolar at the first two epochs but are closer to solar at the third.}\label{fig:abunds_time_evolution}
\end{figure*}

\begin{figure}[!ht]
    \centering
    \includegraphics[width=0.47\textwidth]{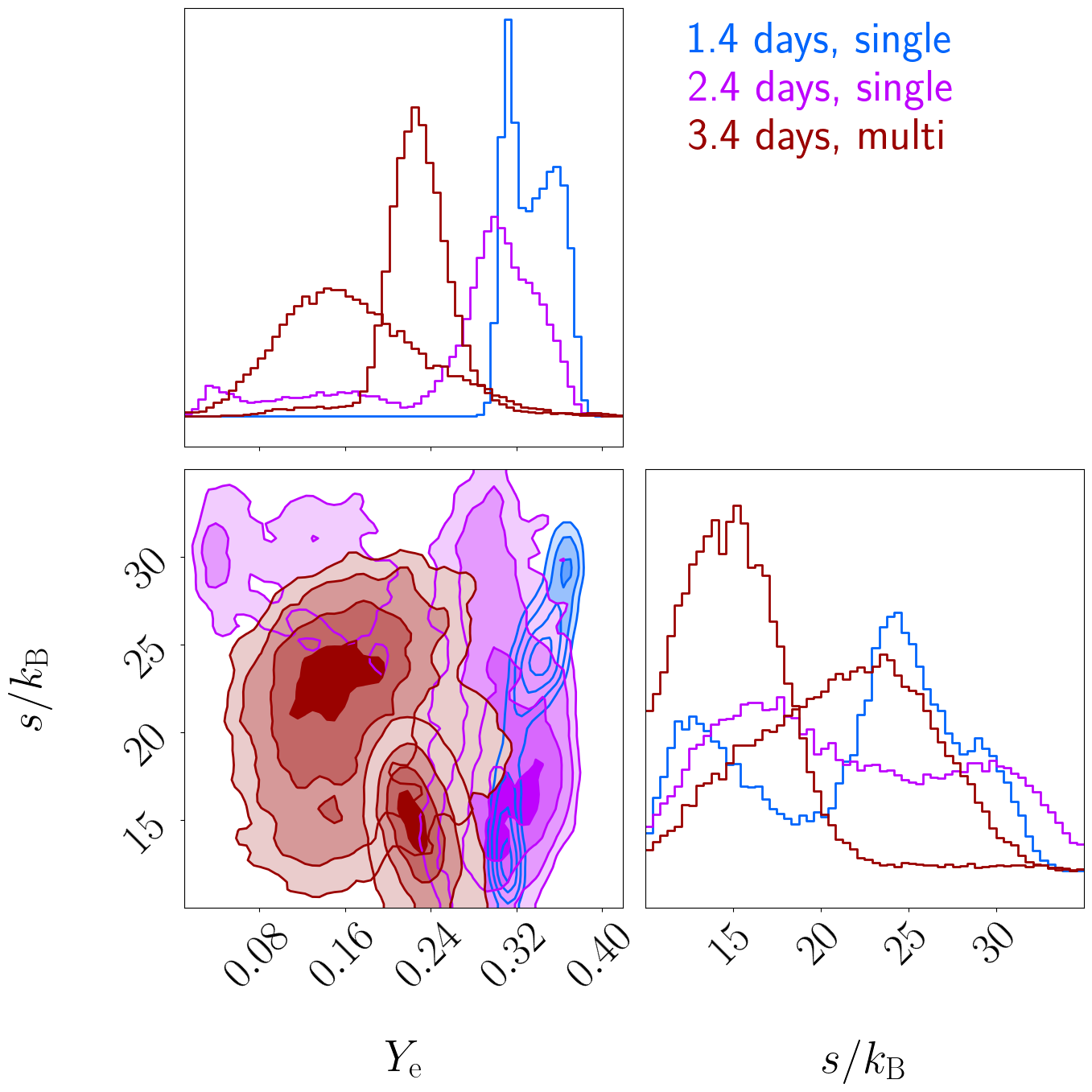}
    \figcaption{\textbf{Comparison of the posterior distributions of $(Y_e,~s)$ for the favored models: single component at 1.4 and 2.4 days, and multicomponent at 3.4 days.} The lower-$Y_e$ mode at 1.4 days is the preferred ``purple + warm'' model of \V23. The two contours at 3.4 days describe the two different components, one of which has a higher $Y_e$. The posteriors broaden at later epochs, but nonetheless, one of the components at 3.4 days has a substantially lower $Y_e$ than at any earlier epoch. Posteriors for all other ejecta parameters are included in Appendix~\ref{app:allspec_posteriors}.}\label{fig:posterior-Ye-s}
\end{figure}

\begin{figure*}[!ht]
    \centering {\ } \\ 
    \centering {\ } \\ 
    \centering 1.4 days, multicomponent \\
    \includegraphics[width=\textwidth]{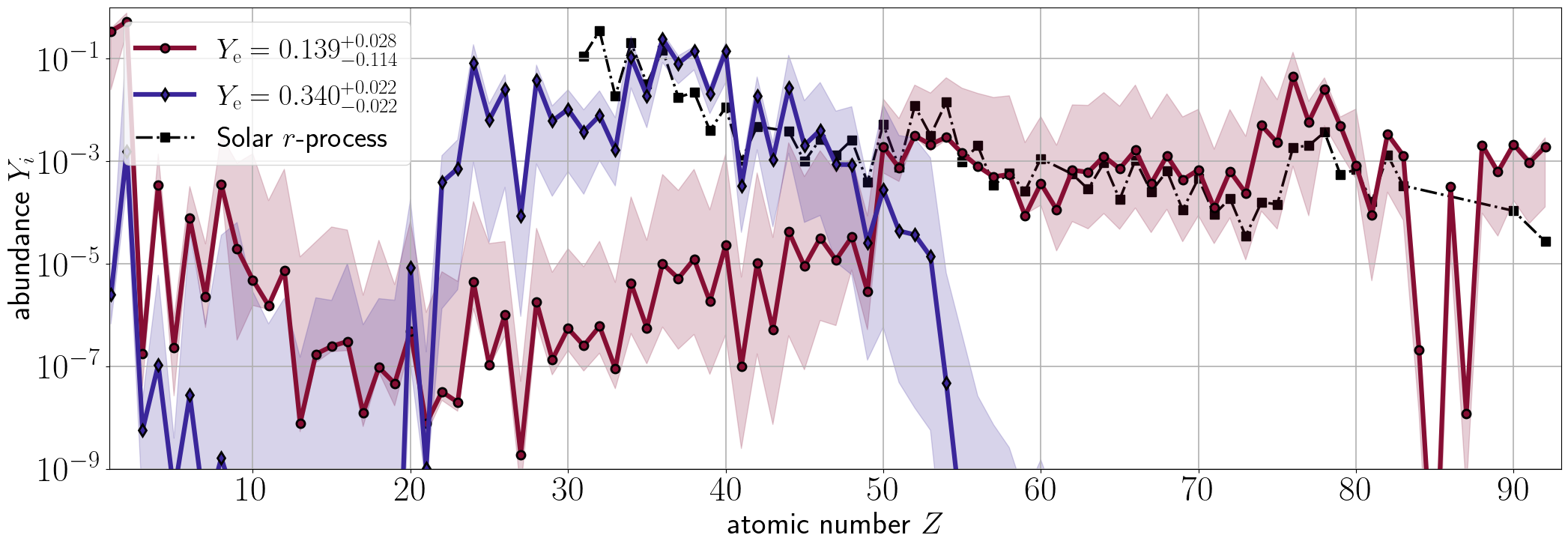} \\
    \centering 2.4 days, multicomponent \\
    \includegraphics[width=\textwidth]{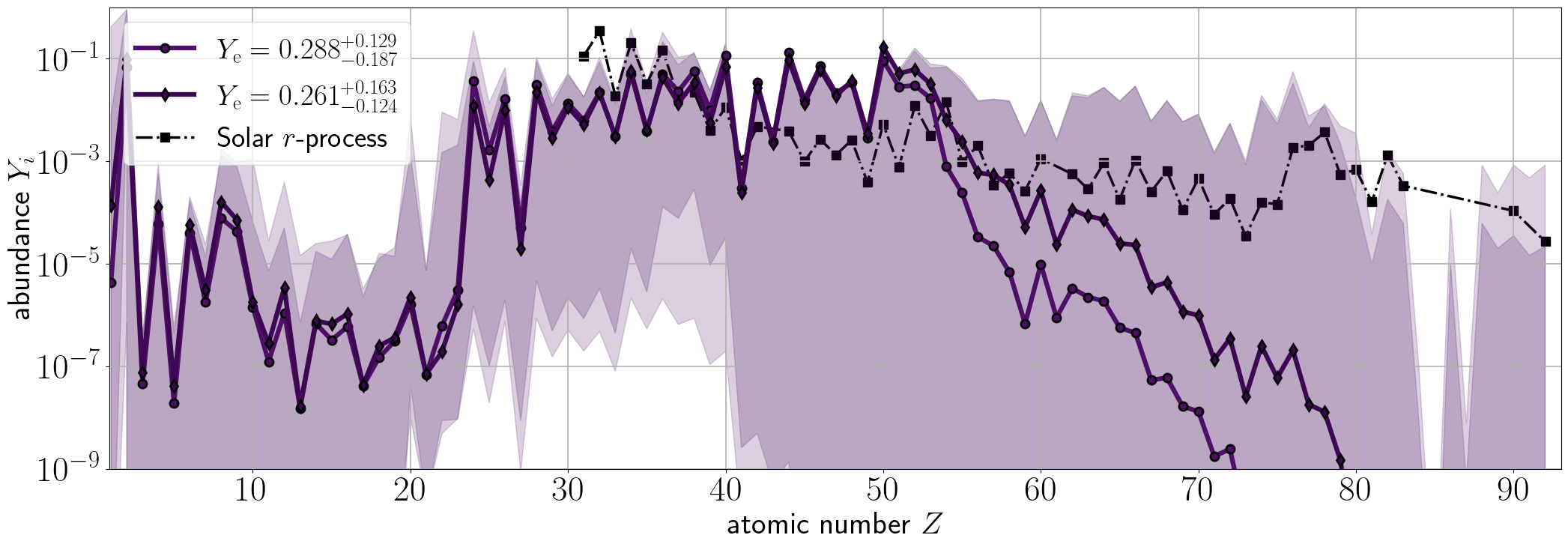} \\
    \centering 3.4 days, multicomponent \\
    \includegraphics[width=\textwidth]{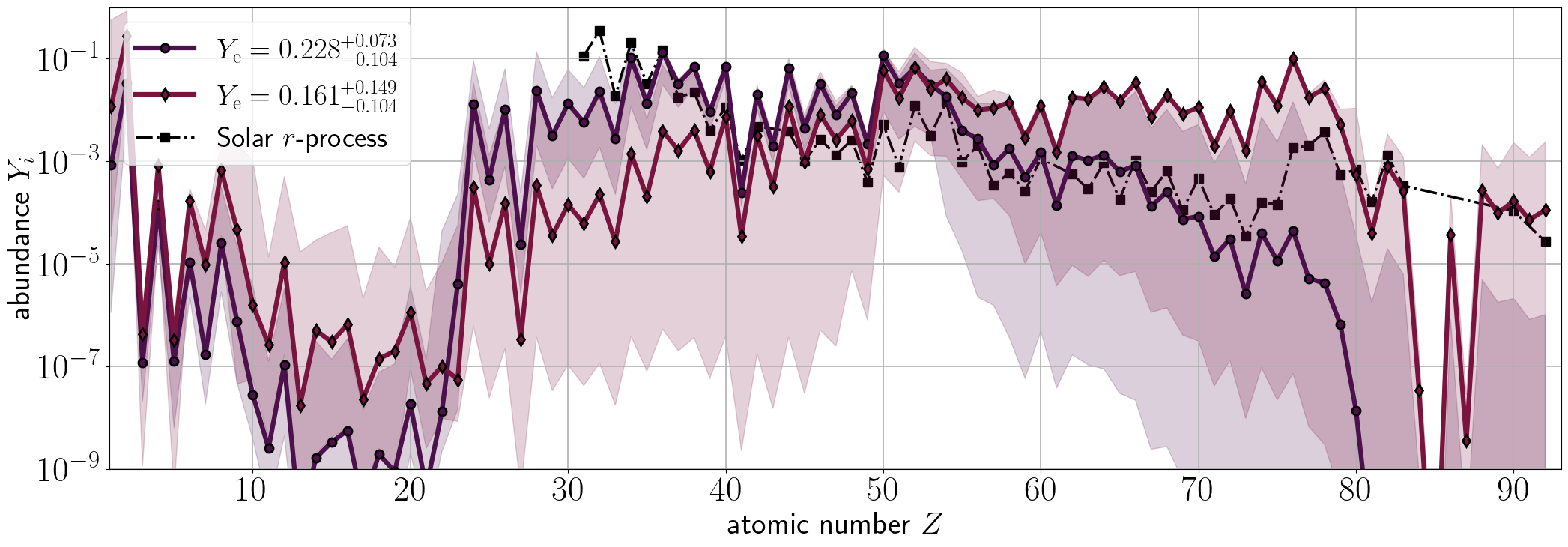}\\
    \figcaption{\textbf{Comparison of the abundances of different components for the multicomponent models.} \textit{Top to bottom}: 1.4, 2.4, and 3.4 days. Importantly, these plots show abundances, and not mass fractions. The red component at 1.4 days has $\sim$100$\times$~less mass than the blue one and its contribution to the emergent spectrum is negligible, despite having a substantially different abundance pattern. In contrast, the abundance patterns for the two components are similar at 2.4 days. At 3.4 days, there is a lower-$Y_e$ component with more lanthanides and a relative dearth of Sr, while the higher-$Y_e$ component contains $\sim$10$\times$~as much Sr.  The components are approximately equipartitioned in mass at 2.4 and 3.4 days, such that both contribute to the emergent spectrum. Single-component models are favored over the multicomponent models shown here at 1.4 and 2.4 days, while the multicomponent model is favored at 3.4 days.}\label{fig:abunds_components_compare}
\end{figure*}

\begin{figure}[!t]
    \centering 1.4 days \\
    \includegraphics[width=0.325\textwidth]{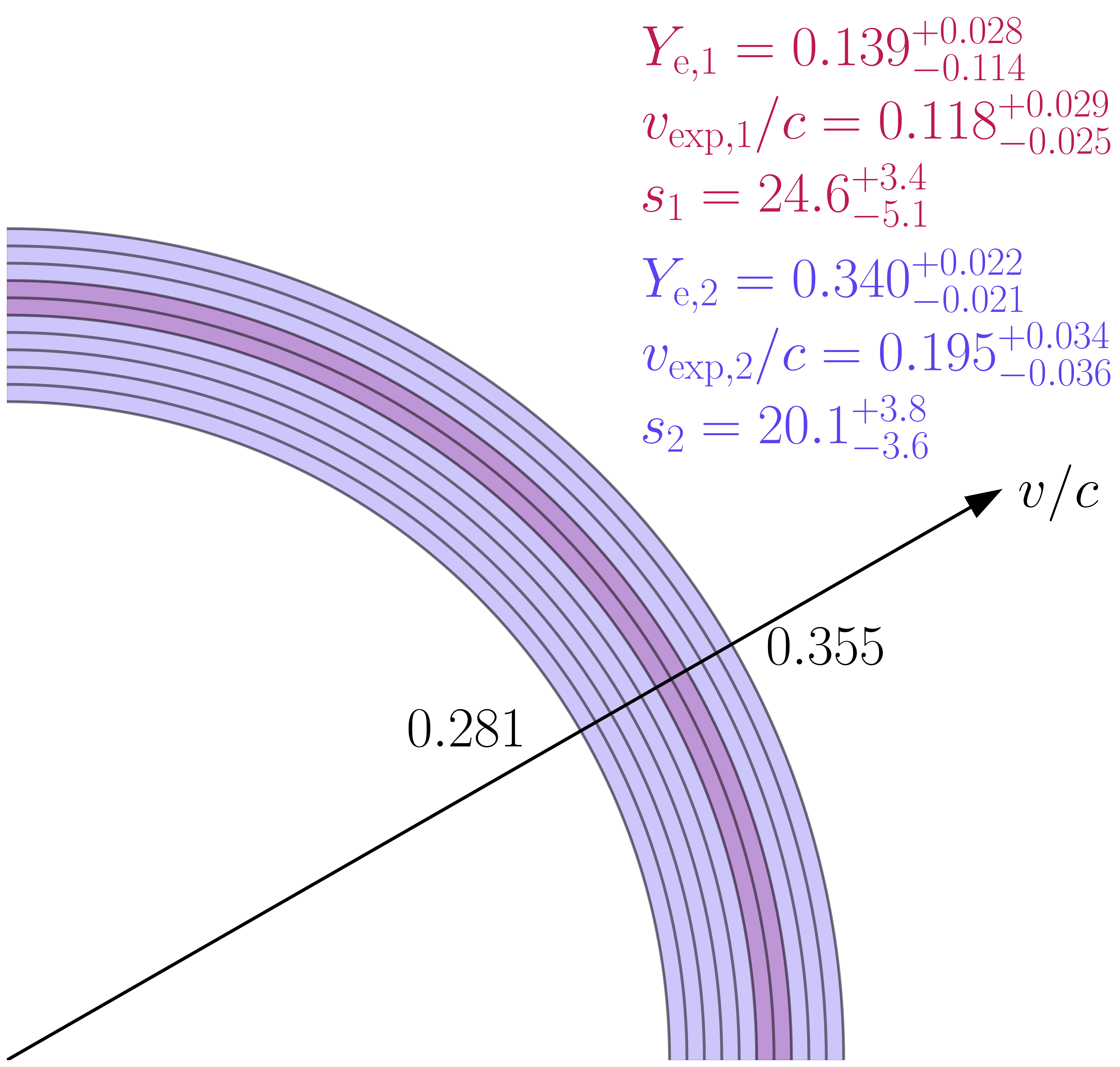} \\
    2.4 days \\
    \includegraphics[width=0.325\textwidth]{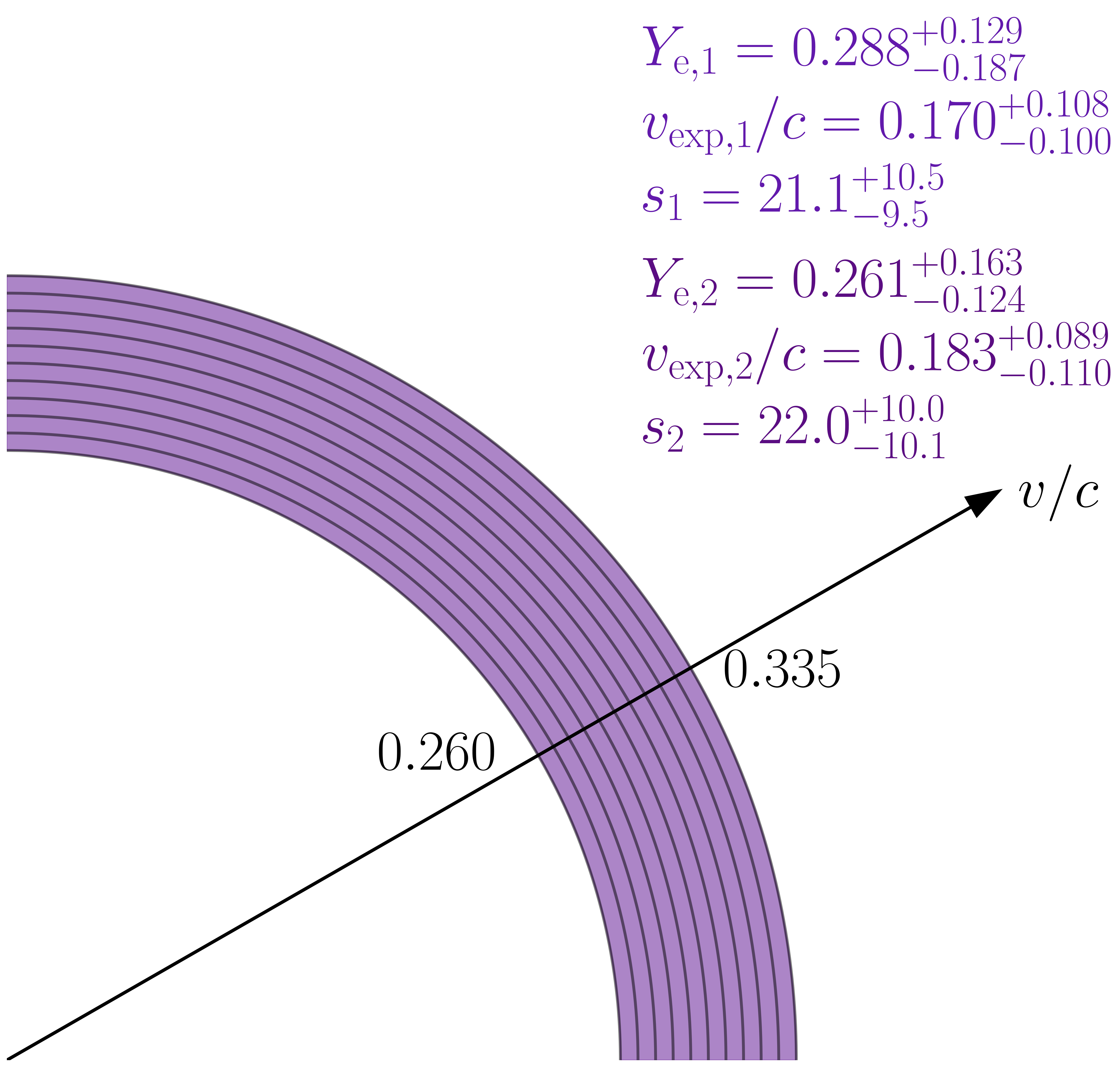} \\ 
    3.4 days \\
    \includegraphics[width=0.325\textwidth]{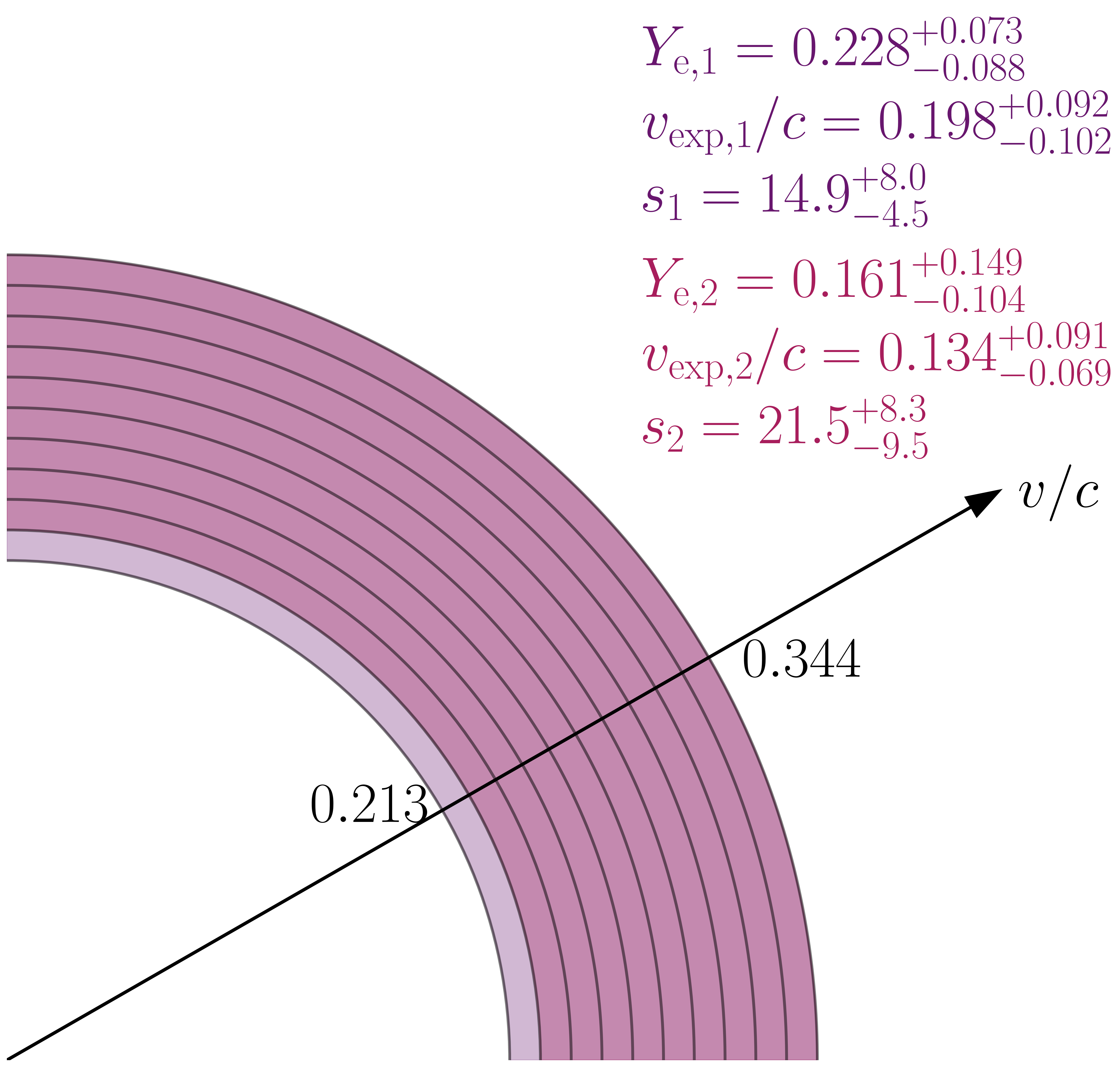}
    \figcaption{\textbf{Inferred physical ejecta configuration and the composition-setting parameters for the multicomponent fits.} \textit{Top to bottom:} 1.4, 2.4, and 3.4 days. At 1.4 days, a thin red component containing $\sim1$\% of the total mass is inferred. This component has a negligible impact on the emergent spectrum, indicating that this model is effectively a single-component one. At 2.4 days, two components with highly similar compositions completely overlap in space, also effectively indicating a single-component ejecta. At 3.4 days, two components with different compositions are present; the ejecta is multicomponent. The 1.4- and 2.4-day fits are disfavored compared to the single-component equivalents, while the multicomponent fit shown here is favored for 3.4 days.}\label{fig:infer_multicomp_wedges}
\end{figure}

We model the 1.4-, 2.4-, and 3.4-day spectra of AT2017gfo with both single- and multicomponent (two-component) models, and we compare the results at each epoch. In Figure~\ref{fig:bestfits}, we compile our best-fit spectra, and highlight our favored models. The parameters for our best-fit single- and multicomponent models are included in Tables~\ref{tab:bestfit_single}~and~\ref{tab:bestfit_multi}, respectively. In all cases, we use the median of the posterior as our best-fit parameter and the 2.5th / 97.5th quantiles of the posterior as the lower/upper bounds, respectively. For multimodal posteriors where we have selected some mode as our best fit, we use median and percentiles of that mode. We find that a single-component blue ejecta model best describes the 1.4- and 2.4-day spectra. At 3.4 days, we require an additional red, lanthanide-rich component in a multicomponent ejecta configuration. The element-by-element abundances with uncertainties for the best-fit models are included in Figure~\ref{fig:abunds_time_evolution}, and a detailed list of the mass fractions of each element is included in Appendix~\ref{app:massfracs_detailed}. In the following sections, we present and examine each of the best fits in turn, at each epoch. 

In all \SPARK~runs, we obtain a sufficient number of $m_0$ baseline samples and $m_{\mathrm{active}}$ BAPE active learning samples such that the posterior converges, according to the $z$ convergence diagnostic given in \cite{fleming20}. Appendix~\ref{app:allspec_posteriors} contains compilations of all $m_0 + m_{\mathrm{active}}$ spectra (generated by LHS and BAPE active learning sampling, respectively) in a given \SPARK~run, for the 1.4-, 2.4-, and 3.4-day single- and multicomponent runs. Posteriors generated by the single- and multicomponent runs are also included in Appendix~\ref{app:allspec_posteriors}. 

Among the parameters that set the abundances, we find that electron fraction $Y_e$ and entropy $s$ are the most tightly constrained, while expansion velocity $v_{\mathrm{exp}}$ is poorly constrained at all epochs. Given the importance of $Y_e$ and $s$ in setting the abundances, we highlight the posteriors for $(Y_e, s)$ at all epochs for the favored models in Figure~\ref{fig:posterior-Ye-s}. We discuss the inferred $Y_e$, $s$ in greater detail in the following sections. We emphasize that the posterior distributions of the ejecta parameters should not be understood as the distribution of the ejecta parameters themselves, but rather a probabilistic distribution of the unique best-fit parameters for the ejecta. Our model still assumes a single $Y_e$ (or two, in the case of two components); the posterior for $Y_e$ is not the distribution of $Y_e$ in the ejecta itself, but rather quantifies the confidence of our best-fit parameters.

\subsection{Single-and multicomponent modeling at 1.4 days}\label{ssc:1.4}

In \V23, we fit the 1.4-day spectrum of AT2017gfo with a single-component ejecta model. This fit required $m_0 + m_{\mathrm{active}} = 1500 + 1140$ points to obtain a good fit and converged posterior. We describe the best-fit ``purple + warm'' model here and refer the reader to \V23 for details. The purple + warm model is characterized by a photospheric velocity of $v_{\mathrm{inner}}/c = 0.313^{+0.013}_{-0.014}$; while the outer boundary velocity was fixed to $v_{\mathrm{outer}} = 0.35c$. For abundance-setting parameters, we found an electron fraction $Y_e = 0.311^{+0.013}_{-0.011}$, expansion velocity of $v_{\mathrm{exp}}/c = 0.240^{+0.055}_{-0.082}$, and entropy of $s / k_{\mathrm{B}} = 13.6^{+4.1}_{-3.0}$. In \V23, we referred to this model as ``purple + warm'', but we emphasize that this is nonetheless substantially blue ejecta within the broader context of kilonova components. These parameters correspond to an abundance pattern that is poor in lanthanides but rich in strontium (${}_{38}$Sr), allowing for a good fit to the prominent $\sim$8000~\AA~feature. The fit struggled at shorter wavelengths: over $3500-4500$~\AA, we overestimated the absorption from yttrium (${}_{39}$Y) and zirconium (${}_{40}$Zr), while at $\lesssim$3500~\AA, we underestimated. Nonetheless, the single-component model remains our preferred model at 1.4 days, as we elaborate on below.

In our 1.4-day multicomponent \SPARK~run, we require $m_0 + m_{\mathrm{active}} = 1500 + 1440$ points to obtain a good fit and converged posterior. More active learning samples are required than for the single-component model owing to the increased dimensionality of the multicomponent fits. The resultant fit generally captures the shape of the spectrum and the broad absorption at $\sim$8000~\AA, like the single-component equivalent. However, the fit does not capture the continuum at wavelengths $\gtrsim$10,500~\AA~nor $\sim$ 3500 - 5000~\AA. The fit is marginally closer to the observed spectrum at the shortest wavelengths ($\lesssim$ 3500 \AA) but this region lies at the edge of the X-shooter spectrograph sensitivity, and overall the single-component model is favored by visual inspection. 

We infer an outer boundary luminosity of $\log_{10} (L_{\mathrm{outer}}/L_{\odot}) = 7.854^{+0.012}_{-0.017}$. This is brighter than is inferred in the single-component equivalent. This is as expected: due to the use of 30 \TARDIS~iterations (rather than 1), $L_{\mathrm{outer}}$ is updated at each iteration. Hence, in our multicomponent models, the meaning of $L_{\mathrm{outer}}$ is different and cannot be mapped directly to $T_{\mathrm{inner}}$. ($L_{\mathrm{outer}}$ is also higher in the 2.4- and 3.4-day multicomponent models, compared to the single-component equivalents).

This multicomponent run yields two components with substantially different abundance patterns. A comparison of the abundance patterns is given in Figure~\ref{fig:abunds_components_compare}. The first component has a lower electron fraction $Y_{e,1} = 0.139^{+0.028}_{-0.114}$ (and specific entropy $s_1 / k_{\mathrm{B}} = 24.6^{+3.4}_{-5.1}$) and could be described as red. The second has higher $Y_{e,2} = 0.340^{+0.022}_{-0.021}$ (and $s_2 / k_{\mathrm{B}} = 20.1^{+3.8}_{-3.6}$), distinctly bluer. Indeed, the $Y_{e,2}$ of this second, bluer component is consistent\footnote{In this section, we describe two inferred parameters as consistent if their uncertainties overlap, where the lower and upper bounds on a parameter are given by the 2.5th and 97.5th percentiles of a given parameter's posterior, respectively.} with the single-component fit at 1.4 days. As in the single-component fit, expansion velocities ($v_{\mathrm{exp},1}$ and $v_{\mathrm{exp},2}$) are poorly constrained.  Interestingly, the better-constrained entropies ($s_1$ and $s_2$) are both higher than and inconsistent with the single-component equivalent.

However, despite the substantial differences between the two components, one component dominates over the other in producing the emergent spectrum. A physical picture for the components is included in Figure~\ref{fig:infer_multicomp_wedges}. We see that the red component is confined to only two shells, while the blue component extends over a greater radius. The mass contained in these two components is $M_{\mathrm{phot},1}/M_{\odot} = 6.9^{+22.1}_{-6.0} \times 10^{-7}$ and $M_{\mathrm{phot},2}/M_{\odot} = 7.0^{+0.8}_{-1.0} \times 10^{-5}$, respectively. We emphasize that this is the mass above the photosphere, and not the total ejecta mass. Nonetheless, the bluer component contains $\sim$100$\times$~as much mass as the redder component and dominates the absorption/emission in the spectrum. Indeed, while the redder component has a lanthanide mass fraction of $\log_{10} X_{\mathrm{lan},1} = -1.24^{+0.68}_{-0.20}$, the total lanthanide mass fraction of both components, computed as the mass-weighted sum over all shells, is $\log_{10} X_{\mathrm{lan,total}} = -13.05^{+9.88}_{-3.32}$ due to the much larger mass of the bluer component. This small presence of lanthanides in the redder component might mark the photometric evolution of the kilonova, but it does not shape the spectrum. In this sense, despite the different compositions of the two components, this multicomponent \SPARK~run has effectively converged to a single-component model.

\subsection{Single-and multicomponent modeling at 2.4 days}\label{ssc:2.4}

In our 2.4-day single-component \SPARK~run, we require $m_0 + m_{\mathrm{active}} = 1500 + 600$ points to obtain a good fit and converged posterior. The resultant posterior shows some bimodality, most evident in the $s/k_{\mathrm{B}}$ dimension. If we select the higher-entropy mode by selecting all samples with $s/k_{\mathrm{B}} \geqslant 25.0$, the resultant posterior has median $Y_e = 0.25$ and $v_{\mathrm{outer}}/c = 0.33$. The synthetic spectrum generated at the median of this higher-entropy mode resembles a blackbody with some absorption at $\sim$5000~\AA, which is not seen in the observed spectrum, and is a poor fit. Instead selecting samples from the posterior with $s/k_{\mathrm{B}} \leqslant 25.0$ (selecting the lower-entropy mode), we obtain a superior fit, and we use this as our preferred model. There is some arbitrariness in the cutoff $s/k_{\mathrm{B}}$; we obtain similar results for cutoff $s/k_{\mathrm{B}}$ in the range $22-28$. This preferred fit captures most of the continuum of the observed spectrum and most of the $\sim$8000 \AA~absorption. If this absorption belongs to a P Cygni feature (as has been argued in \citealt{watson19, sneppen23}), we partially miss the red wing of this feature. Furthermore, we overestimate the absorption, or underestimate the continuum, at wavelengths $\lesssim$4500 \AA. 

We infer n $\log_{10} (L_{\mathrm{outer}}/L_{\odot}) = 7.594^{+0.040}_{-0.061}$. In the case of single-shell, single-iteration \TARDIS~runs, we can use these values to determine an inner boundary temperature of $T_{\mathrm{inner}} = 3050^{+126}_{-223}~\mathrm{K}$. As expected, the kilonova has dimmed and cooled since the previous epoch. The inferred velocities also match expectations: we obtain inner and outer boundary velocities $v_{\mathrm{inner}}/c = 0.249^{+0.017}_{-0.032}$ and $v_{\mathrm{outer}} = 0.342^{+0.047}_{-0.050}$. This $v_{\mathrm{outer}}$ is consistent with the fixed $v_{\mathrm{outer}} = 0.35c$ from 1.4 days, while $v_{\mathrm{inner}}$ (effectively the photospheric velocity) has receded into the ejecta, compared to $0.31c$ at 1.4 days. This is evidence for the ejecta expanding, cooling, and becoming optically thinner, as expected.

Finally, we obtain $Y_e = 0.306^{+0.055}_{-0.204}$ and $s/k_{\mathrm{B}} = 17.6^{+7.1}_{-6.3}$. Similar to our results at 1.4 days, $v_{\mathrm{exp}}$ is poorly constrained, while $s$ is better constrained. This $s$ is also consistent with the previous epoch. The posterior distribution in all dimensions is wider at 2.4 days; in particular, $Y_e$ exhibits a tail extending to smaller electron fractions. The posterior is nonetheless peaked at $Y_e = 0.306^{+0.055}_{-0.204}$, which is slightly lower than at 1.4 days, but consistent to within the uncertainties. 

In our 2.4-day, multicomponent \SPARK~run, we require $m_0 + m_{\mathrm{active}} = 1500 + 1020$ points to obtain a good fit and converged posterior. The resultant fit once again captures the broad shape and $\sim$8000~\AA~absorption, but the depth of the $\sim$8000~\AA~absorption feature is overestimated. The multicomponent fit does, however, achieve a better fit to the continuum at wavelengths $\lesssim$7000~\AA. 

In contrast with the 1.4-day, multicomponent run, we see that the two inferred components are remarkably similar. Figure~\ref{fig:abunds_components_compare} compares the abundance patterns of both individual components. The first component is described by $Y_{e,1} = 0.288^{+0.129}_{-0.187}$ and $s_1 / k_{\mathrm{B}} = 21.1^{+10.5}_{-9.5}$, while the latter has $Y_{e,2} = 0.261^{+0.163}_{-0.124}$ and $s_2 / k_{\mathrm{B}} = 22.0^{+10.0}_{-10.1}$. Again, expansion velocities are poorly constrained, while the better-constrained entropies are higher than and inconsistent with the single-component equivalent. Both entropies are consistent with the entropies inferred in the 1.4-day, multicomponent model. Finally, $Y_{e,1}$ and $Y_{e,2}$ are both lower than, but still consistent with, the dominant blue component in the 1.4-day, multicomponent model and the 1.4-day, single-component (purple + warm) model.

Our inferred geometry for the two components is shown in Figure~\ref{fig:infer_multicomp_wedges}. As expected, the photosphere recedes into the ejecta. The two components almost completely overlap in physical space; indeed, the mass contained in these two components is roughly equally split into $M_{\mathrm{phot},1}/M_{\odot} = 2.3^{+12.5}_{-0.7} \times 10^{-5}$ and $M_{\mathrm{phot},2}/M_{\odot} = 4.0^{+4.8}_{-0.9} \times 10^{-5}$, respectively. Given the two components' similar compositions and equal contributions to the emergent spectrum, this multicomponent~\SPARK~run has also converged to an effectively single-component model.

\subsection{Single-and multicomponent modeling at 3.4 days}\label{ssc:3.4}

Our best single-component model for 3.4 days is in general a poor fit to AT2017gfo. Nonetheless, we review it here for completeness. As at 2.4 days, we require $m_0 + m_{\mathrm{active}} = 1500 + 600$ points to obtain a converged posterior. The resultant posterior shows a high degree of multimodality, most evident in the $\rho_0$, $Y_e$, and $s / k_{\mathrm{B}}$ dimensions. Of the multiple modes, we obtain our best (but still poor) fit with a higher-$\rho_0$, mid-$Y_e$, lower-$s$ mode. This fit produces some, but not all, of the dominant absorption feature at $\sim$8000 \AA. This fit also overestimates the absorption, or underestimates the continuum, for wavelengths $\lesssim$5500 \AA. 

We find $\log_{10} (\rho_0 / \mathrm{g~cm^{-3}}) = -14.586^{+0.313}_{-0.384}$. This density is larger than that inferred at earlier epochs. We also infer $v_{\mathrm{outer}} = 0.309^{+0.032}_{-0.023}$. This $v_{\mathrm{outer}}$ is much smaller than the fixed value at 1.4 days or the inferred value at 2.4 days. This may explain the insufficient absorption at $\sim$8000 \AA, if the photon packets are not interacting with enough matter to be absorbed in large numbers before escaping the ejecta. Finally, we find $Y_e = 0.226^{+0.062}_{-0.067}$ and $s/k_{\mathrm{B}} = 15.4^{+7.2}_{-3.1}$ at this epoch. $Y_e$ is lower and inconsistent with both previous epochs. Interestingly, despite the lower $Y_e$, the inferred abundance of Sr is within $1$\% of that of our single-component model at 1.4 days, and in fact $\sim$25\% larger than that of our single-component model at 2.4 days. Nonetheless, Sr does not yield the expected prominent absorption feature at $\sim$8000 \AA. This suggests that the shortcomings of this model may be due not to the inferred abundance pattern, but rather to the density, velocity structure, and/or some inherent shortcoming of the single-component model.

For our preferred, multicomponent model, we require $m_0 + m_{\mathrm{active}} = 1500 + 1140$ points to obtain a good fit and converged posterior at 3.4 days. The resultant fit is better than the single-component equivalent. This fit better captures the $\sim$8000 \AA~absorption, and the red wing of the nominal P Cygni feature. However, this fit also overestimates the absorption, or underestimates the continuum, for wavelengths $\lesssim$5500 \AA. This over/underestimation is more severe than in the single-component model.

Interestingly, we see evolution in the inferred densities. Indeed, we find $\log_{10}(\rho_0 / \mathrm{g~cm^{-3}})$ of $-15.016^{+0.320}_{-0.316}$, $-15.443^{+0.742}_{-0.463}$ in our preferred models at 1.4 and 2.4 days, respectively, and $-14.505^{+0.323}_{-0.372}$ in this new multicomponent fit at 3.4 days. While this density at 3.4 days is higher than that inferred at earlier epochs, which may hint at the emergence of a new mass component, these densities are consistent with each other, within uncertainties. Note that the inferred total mass above the photosphere increases with time. However, this is due not to a lack of mass preservation, but rather to the recession of the photosphere into the ejecta. 

Most significantly, we infer two components with substantially different properties. The first component is described by $Y_{e,1} = 0.228^{+0.073}_{-0.088}$ and $s_1 / k_{\mathrm{B}} = 14.9^{+8.0}_{-4.5}$, while the latter has $Y_{e,2} = 0.161^{+0.149}_{-0.104}$ and $s_2 / k_{\mathrm{B}} = 21.5^{+8.3}_{-9.5}$. Again, expansion velocities are poorly constrained. Both entropies are better constrained and consistent with the entropies inferred in the single-component (and multicomponent) models at 1.4 and 2.4 days. The $Y_{e,1}$ and $Y_{e,2}$ are both substantially lower than those of the single- and multicomponent fits at 1.4 and 2.4 days. This leads to an ejecta that is substantially more lanthanide-rich: the two components have $\log_{10} X_{\mathrm{lan},1} = -1.78^{+1.11}_{-3.69}$ and $\log_{10} X_{\mathrm{lan},2} = -0.53^{+0.08}_{-2.86}$, respectively. This yields a total lanthanide fraction of $\log_{10} X_{\mathrm{lan,total}} = -1.58^{+0.86}_{-1.46}$. Figure~\ref{fig:abunds_components_compare} shows the abundances of the two components, with uncertainties. Both components contain some substantial abundance of lanthanides, but the higher $Y_e$ of the two has $\sim$10$\times$~more of the crucial element Sr.

In Figure~\ref{fig:infer_multicomp_wedges}, we show a physical picture of the inferred ejecta. The photosphere further recedes compared to 1.4 and 2.4 days. Both components mostly overlap in space, and the mass is roughly equally split: they have  masses $M_{\mathrm{phot},1}/M_{\odot} = 5.3^{+1.8}_{-1.3} \times 10^{-4}$ and $M_{\mathrm{phot},2}/M_{\odot} = 3.8^{+1.1}_{-0.9} \times 10^{-4}$, respectively. Given the roughly equal contributions from a component rich in lanthanides and another with a similar abundance of lanthanides but $\sim$10$\times$~as much Sr, as well as the fact that the multicomponent model is a substantially better fit than the single-component model at 3.4 days, we interpret the ejecta as being genuinely composed of two unique components at this epoch. As we will see, the equal contribution from two components with distinct properties shapes the emergent spectrum.

\subsection{Favored models}\label{ssc:favored}

\begin{figure}[!ht]
    \includegraphics[width=0.47\textwidth]{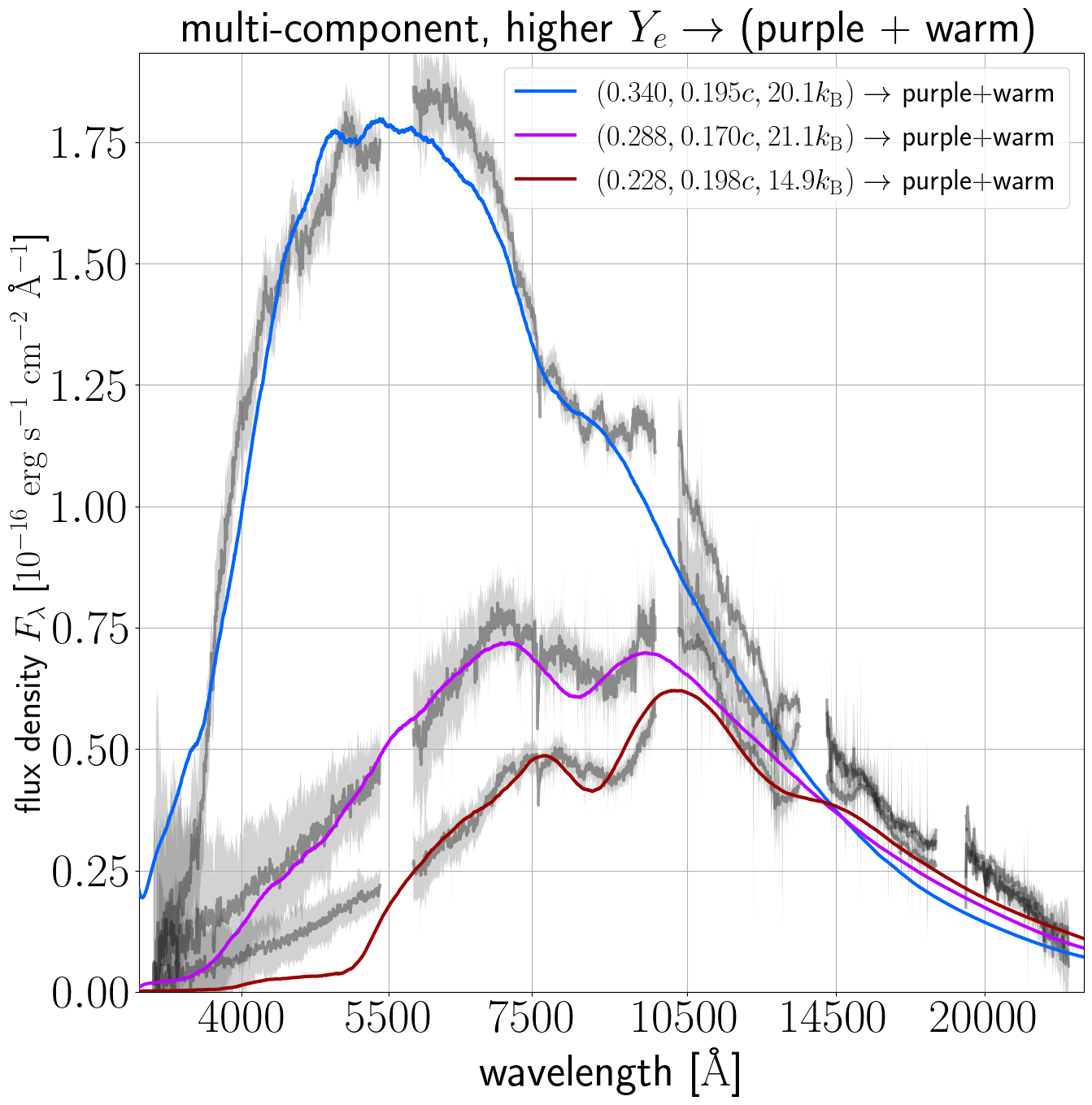}
    \figcaption{\textbf{Hypothetical models at 1.4, 2.4, and 3.4 days, where the higher-$Y_e$ component of each multicomponent fit has been replaced by that of the single-component, 1.4-day (purple + warm) model of \V23}. Here we take the best-fit parameters from our multicomponent runs (Table~\ref{tab:bestfit_multi}) and, after fitting, swap out the higher-$Y_e$ component with the purple + warm component $(Y_e = 0.311,~v_{\mathrm{exp}}/c=0.240,~s / k_{\mathrm{B}}=13.6)$. The legend indicates the parameters that have been changed to those of the purple + warm model. $L_{\mathrm{outer}}$, $\rho_0$, and the velocities are left unchanged, and thus the mass in each component is also unchanged. Comparing to the original multicomponent models (dashed lines, Figure~\ref{fig:bestfits}), the differences are negligible at all epochs.}\label{fig:swap-purplewarm}
\end{figure}

Thus far, we have selected our favored models by visual inspection, seeing which produces a better fit to the observed spectrum at a given epoch. Here we explore the physical consistency of our favored models. 

We have found that the 1.4- and 2.4-day spectra are well fit by a single bluer component. Even when allowed to yield multiple components, the multicomponent \SPARK~runs yield effectively single-component best fits at 1.4 and 2.4 days. At 1.4 days, the bluer component contains $\sim$100$\times$~the mass of the redder one, and the presence of the red component is thus inconsequential for the spectrum. At 2.4 days, the two components are roughly equal in mass and characterized by similar abundance patterns; this fit is thus also effectively a single-component fit.

In contrast, at 3.4 days, the two components in the multicomponent fit have different abundance patterns. In particular, an additional lower-$Y_e$ component rich in lanthanides is required to fit the observed spectrum. The lower-$Y_e$ component contains as much as $\sim$10$\times$~more lanthanides than the higher-$Y_e$ component, given the broadness of the posterior. The remaining higher-$Y_e$ component, however, contains $\sim$10$\times$~as much Sr. Given our atomic data, line list, and (most importantly) abundance patterns from the reaction network calculations of \cite{wanajo18}, no single component is able to simultaneously yield the abundance of Sr and lanthanides needed to reproduce the observed 3.4-day spectrum of AT2017gfo. As we will see in the following sections, the role of the higher-$Y_e$ component is to provide enough Sr to yield the $\sim$8000 \AA~absorption, while the lower-$Y_e$ component yields most of the lanthanides needed to produce the short-wavelength $\lesssim$7500 \AA~absorption.  

Aside from asking whether a single- or multicomponent ejecta is individually preferred at each epoch, we also ask, is the higher-$Y_e$ component of the multicomponent models interchangeable with the purple + warm ejecta model obtained at 1.4 days? To test this, we replace the higher-$Y_e$ components of the multicomponent models at 1.4, 2.4, and 3.4 days with this purple + warm component, while leaving the lower-$Y_e$ component intact. (These higher-$Y_e$ components have $Y_{e} = $ $ 0.340^{+0.022}_{-0.021}$, $0.288^{+0.129}_{-0.187}$, and $0.228^{+0.073}_{-0.088}$ at 1.4, 2.4, and 3.4 days, respectively). Figure~\ref{fig:swap-purplewarm} shows the spectra that result from replacing the higher-$Y_e$ components of each multicomponent fit with the parameters of the purple + warm model $(Y_e = 0.311,~v_{\mathrm{exp}}/c=0.240,~s / k_{\mathrm{B}}=13.6)$ at each epoch. All other parameters are unchanged, \ie, we only change the abundance pattern of the higher-$Y_e$ component. This amounts to testing whether we produce a good fit with some hypothetical two-component model that includes the purple + warm component plus the lower-$Y_e$ component of each respective fit. At all epochs, there is minimal impact on the resultant spectrum and we retain a good fit.  This is not surprising at 1.4 and 2.4 days, where the inferred higher $Y_e$ is consistent with the purple + warm model. It is, however, surprising at 3.4 days. Crucially, replacing the higher-$Y_e$ component at 3.4 days changes the overall abundance pattern, but the overall abundance of Sr remains relatively unchanged owing to the presence of substantial Sr in the purple + warm ejecta. This further suggests that the role of the higher-$Y_e$ component at 3.4 days is primarily to provide enough Sr to produce the $\sim$8000 \AA~absorption, while the lower-$Y_e$ component provides sufficient lanthanides. 

Overall, we infer that a new, redder, lanthanide-rich component emerges in addition to the previously inferred bluer ejecta at 3.4 days. This new red ejecta component has photospheric velocity $0.21c$ and extends out to $0.35c$, suggesting that this component was in fact present at earlier epochs but was partially buried underneath the photosphere and/or outshined by the brighter blue component.


\section{Discussion}\label{sec:disco}

\subsection{Abundances of the favored models}\label{ssc:disco-abundances}

Given our favored inferred models and their best fit ($Y_e$, $v_{\mathrm{exp}}$, $s$), we study the inferred abundance pattern of the ejecta as a function of time. We favor the single-component models at 1.4 and 2.4 days and the multicomponent model at 3.4 days. We show the abundance patterns of these favored models in Figure~\ref{fig:abunds_time_evolution}. The overall multicomponent abundance pattern at 3.4 days is computed as the mass-weighted sum over all shells, including both components. 

The 1.4- and 2.4-day abundance patterns are remarkably similar, but the 2.4-day model has greater uncertainties owing to the broader overall posterior. These 2.4-day abundances show evidence for $\sim$$10^{-9}$ to $10^{-6}$~mass fractions for some of the lighter lanthanides. However, as we see in the following section, lanthanides do not leave any clear imprint on the spectrum at 2.4 days. Overall, the 1.4- and 2.4-day spectra are well described by a single component dominated by light $r$-process elements around the first $r$-process peak.

In contrast, the 3.4-day abundance pattern indicates the clear presence of heavier elements and especially lanthanides in the ejecta. The total lanthanide fraction is $\log_{10} X_{\mathrm{lan,total}} = -1.58^{+0.86}_{-1.46}$, several orders of magnitude larger than inferred at 1.4 and 2.4 days. The higher-$Y_e$ component of this multicomponent ejecta then provides Sr, the role of which is crucial, as we will see in the following section. Indeed, we infer abundances of $Y_{\mathrm{Sr},1.4} = 0.069^{+0.053}_{-0.022}$ and $Y_{\mathrm{Sr},2.4} = 0.087^{+0.035}_{-0.059}$ at 1.4 and 2.4 days and $Y_{\mathrm{Sr},3.4} = 0.067^{+0.017}_{-0.046}$ (mass-weighted sum over all shells, including both components) at 3.4 days. These are remarkably consistent with each other.

The inferred abundance patterns at 1.4 and 2.4 days are markedly nonsolar in Figure~\ref{fig:abunds_time_evolution}, where the solar $r$-process abundance pattern is taken using the data of \cite{lodders09} with the $s$-process residuals subtracted according to \cite{bisterzo14}. If all NS-NS/BH mergers produced such blue ejecta, this would point to the inability of these mergers to yield the $r$-process abundances seen in several astrophysical settings. The 3.4-day abundance pattern, in contrast, is much closer to solar up to $Z \sim 80$. In particular, the presence of lanthanides improves this agreement. 

Our abundance results are in broad agreement with inference on the light curves and spectral fitting of AT2017gfo (see \citealt{ji19}, and references therein). While light-curve modeling typically uses gray (wavelength-independent) opacities, this modeling generally recovers the emergence of multiple components, including a higher-opacity redder component at $\gtrsim$3 days, as we find here. Our results for lanthanide fractions deviate somewhat from previous works. In particular, previous works find lanthanide fractions $\log_{10} X_{\mathrm{lan}} \sim -6$ to $\sim$$-4$ for a blue component and $\sim$$-2$ for a red component (\eg, \citealt{chornock17, kasen17}). We find a more lanthanide-poor bluer component ($\log_{10} X_{\mathrm{lan}} \sim -7$ as per the inference at 1.4 and 2.4 days) and a more lanthanide-rich ($\log_{10} X_{\mathrm{lan}} \sim -1.6$, as per the inference at 3.4 days) redder component. Notably, \cite{ji19} compile the inferred lanthanide fractions of AT2017gfo from several studies and find that the $\log_{10} X_{\mathrm{lan}} \sim -2$ inferred in previous works falls on the low end of the lanthanide fractions measured in metal-poor stars. If our higher inferred lanthanide fraction is correct, we relieve some of the tension between the lanthanide fractions of the AT2017gfo ejecta and metal-poor stars. However, we caution that we do not probe all of the ejecta, and in particular are not sensitive to the ejecta below the photosphere. This ejecta could be more or less lanthanide-rich, complicating these comparisons.

\subsection{Elements and ions present in the favored models}\label{ssc:disco-elements}

\begin{figure}[!ht]
    \centering
    \includegraphics[width=0.36\textwidth]{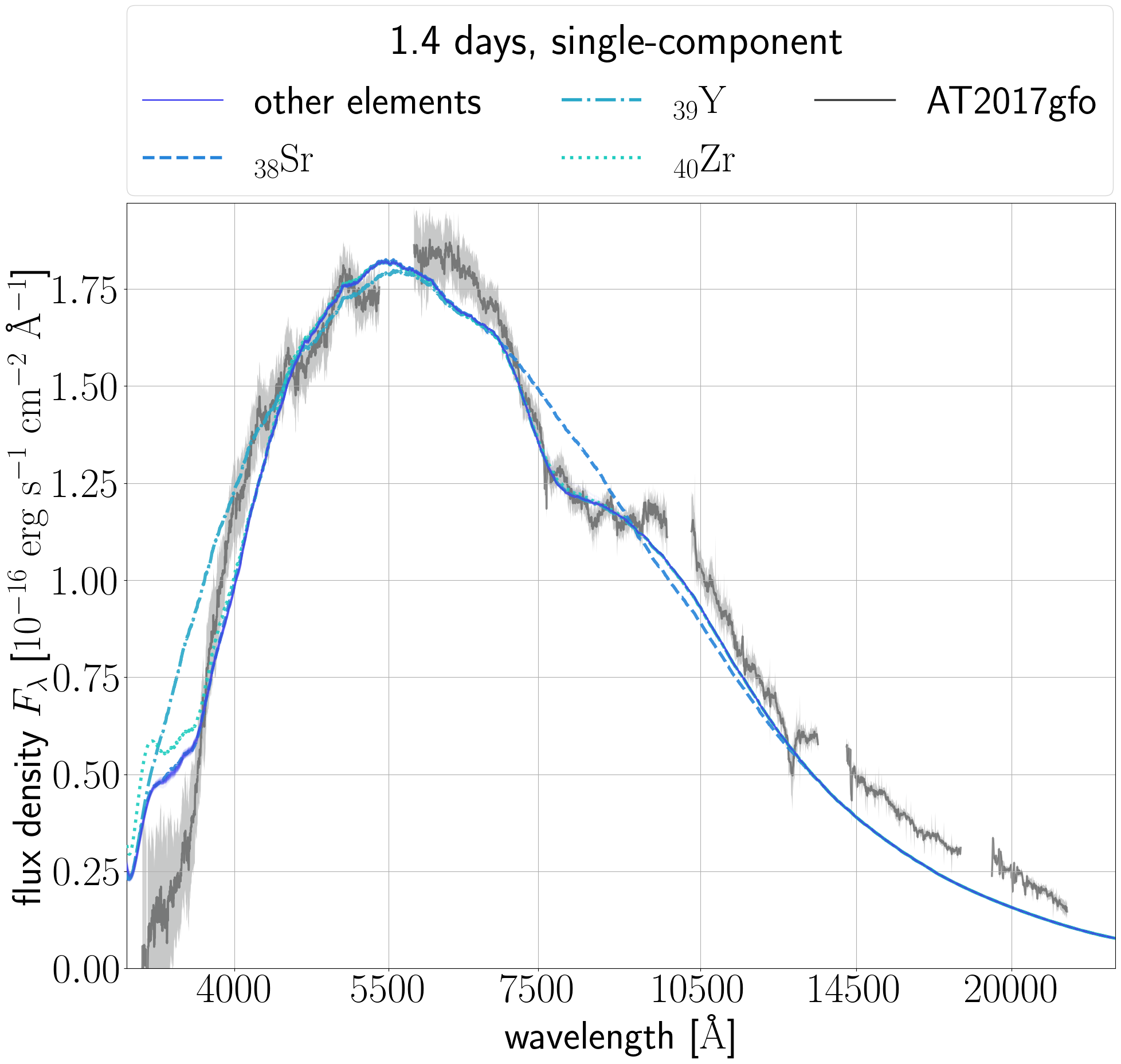}
    \includegraphics[width=0.36\textwidth]{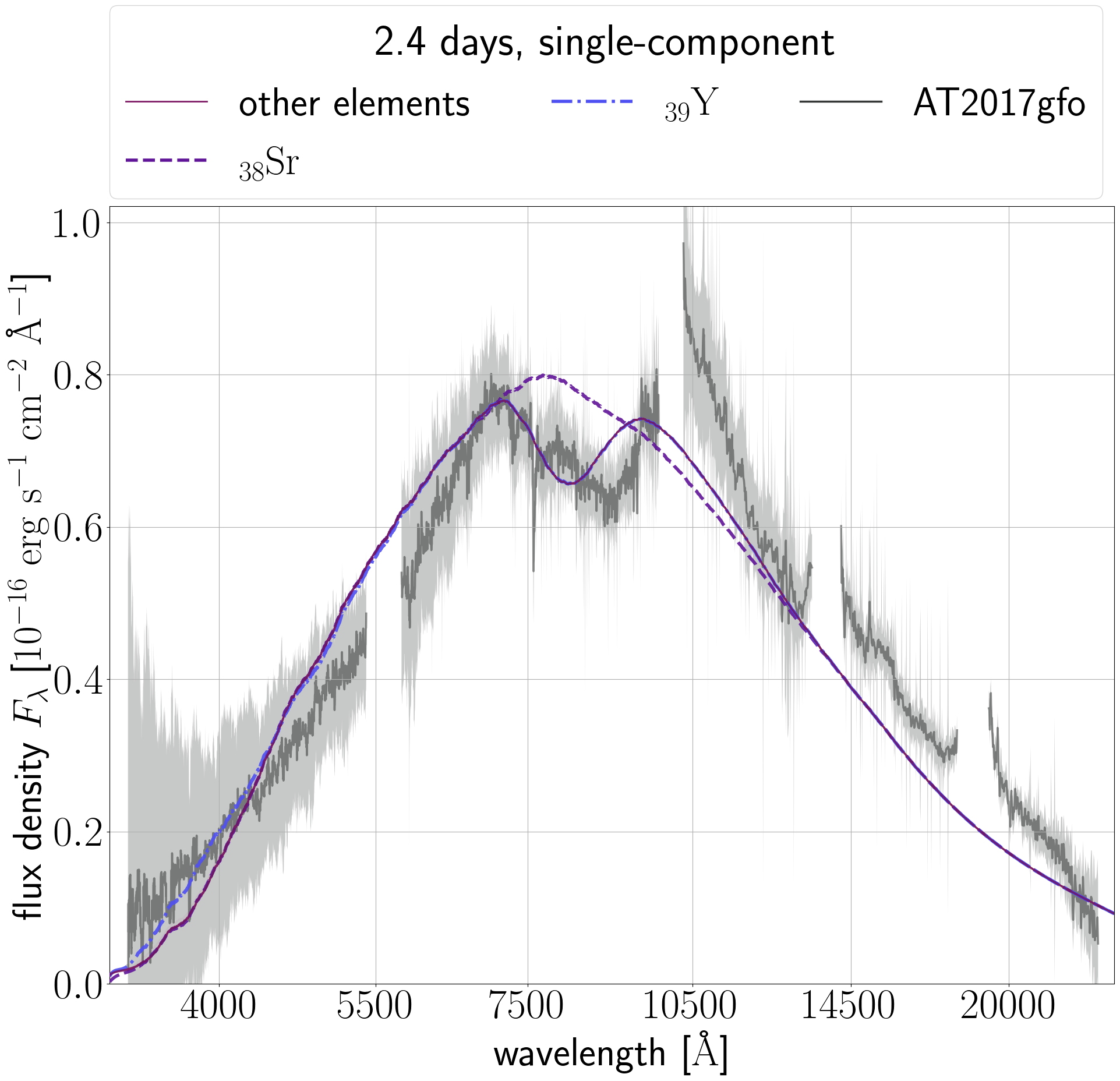}
    \includegraphics[width=0.36\textwidth]{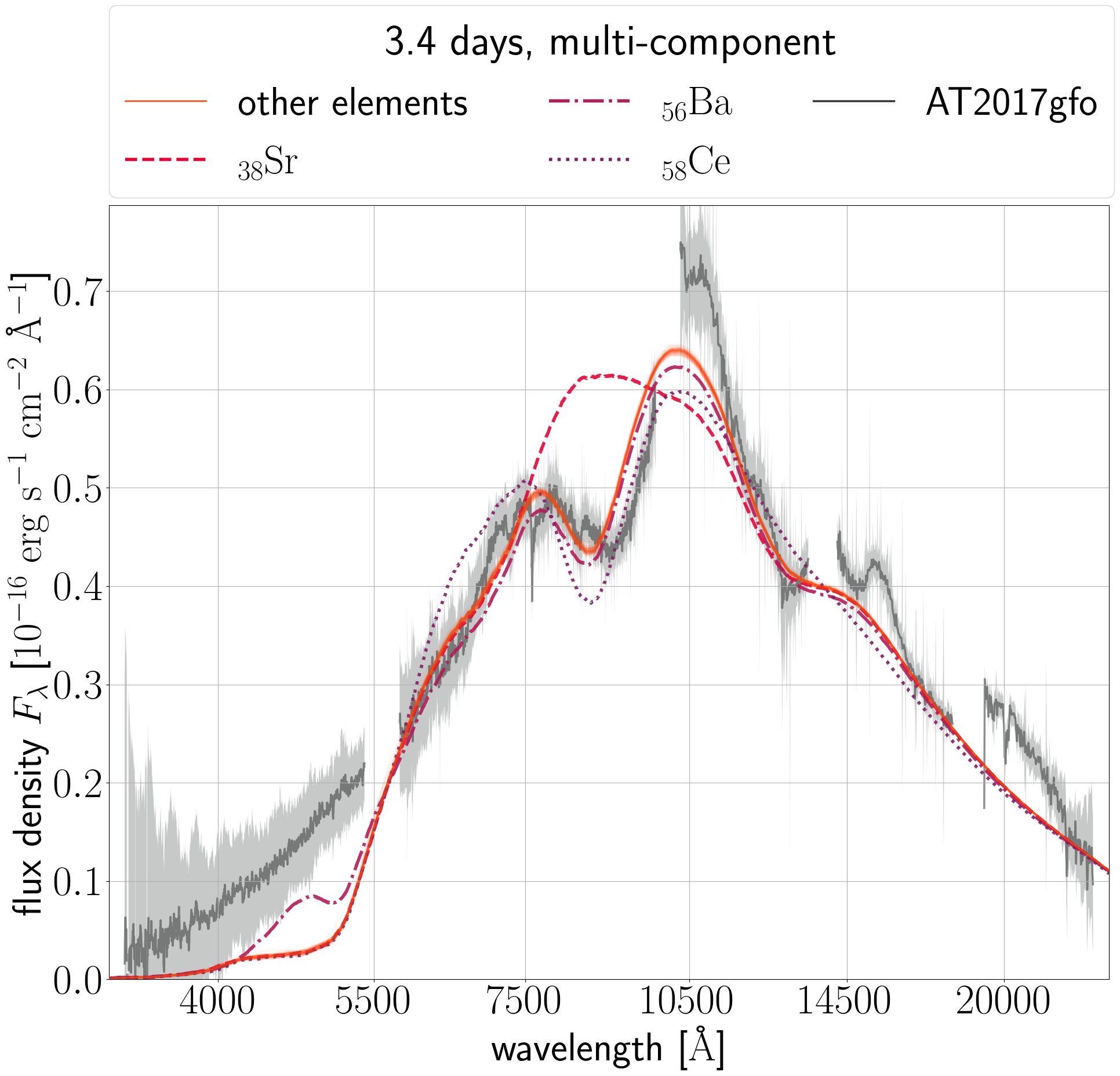}
    \figcaption{\textbf{Leave-one-out spectra for the favored models: single component for 1.4 (top) and 2.4 (center) days, and multicomponent for 3.4 days (bottom).} All models show clear absorption from strontium (${}_{38}$Sr) at $\sim$8000 \AA. At 3.4 days, when the ejecta is richer in heavier elements, we also see absorption (at $\sim$7000 and $\sim$12,000 \AA) and emission (at $\sim$8000 \AA) from the lanthanide cerium (${}_{58}$Ce). However, we also see overabsorption from barium (${}_{56}$Ba) at $\sim$4500\AA. SDEC plots in Figure~\ref{fig:SDEC} provide a complementary view of the dominant species.}\label{fig:leaveoneout_pref}
\end{figure}


\newlength{\SDECheight} 
\setlength{\SDECheight}{+5.6cm} 

\begin{figure*}[!ht]
    \centering
    \begin{tabular}{ccc}
    single-component & multicomponent & \\
    & & \\
    \includegraphics[width=0.47\textwidth]{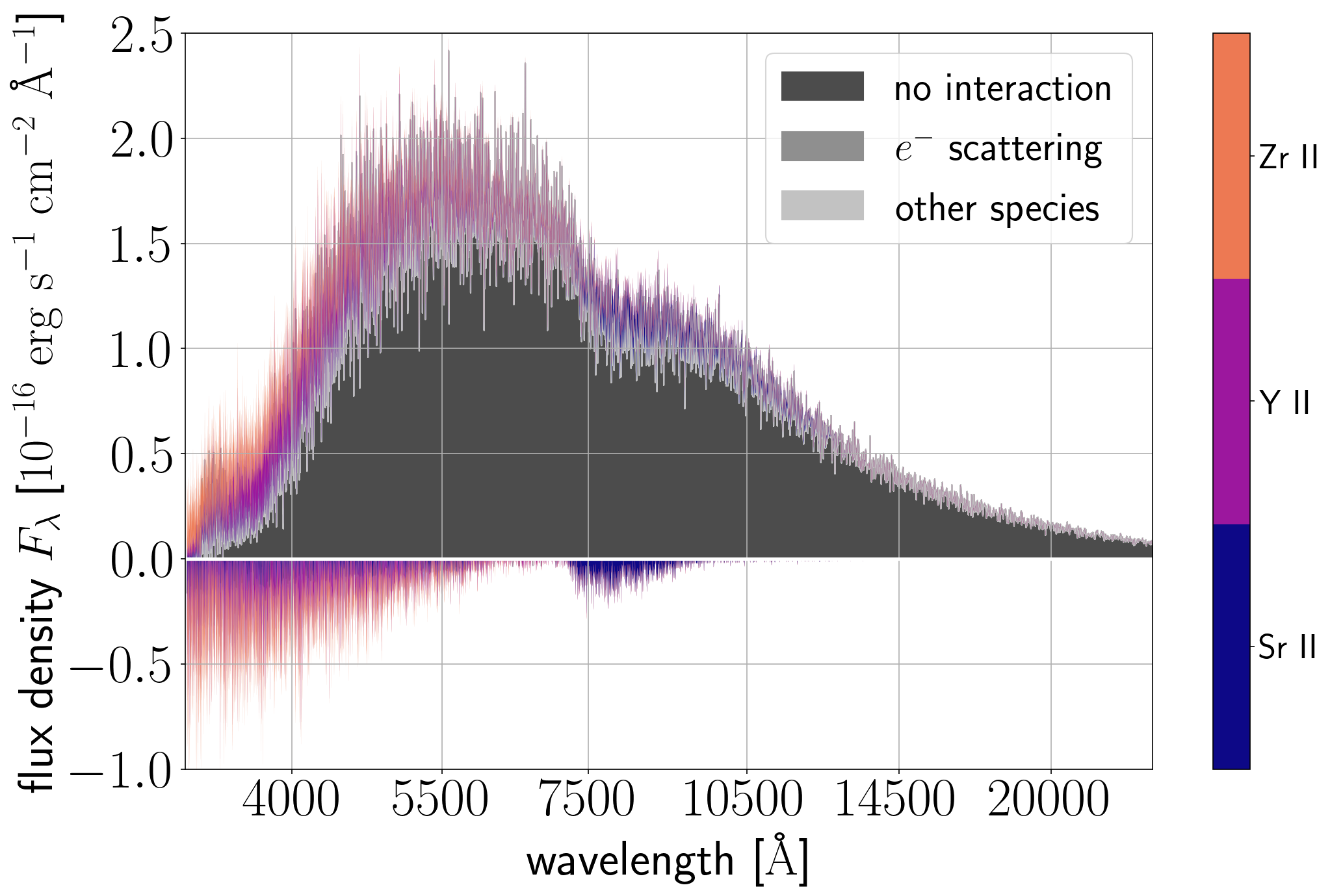} & \includegraphics[width=0.47\textwidth] {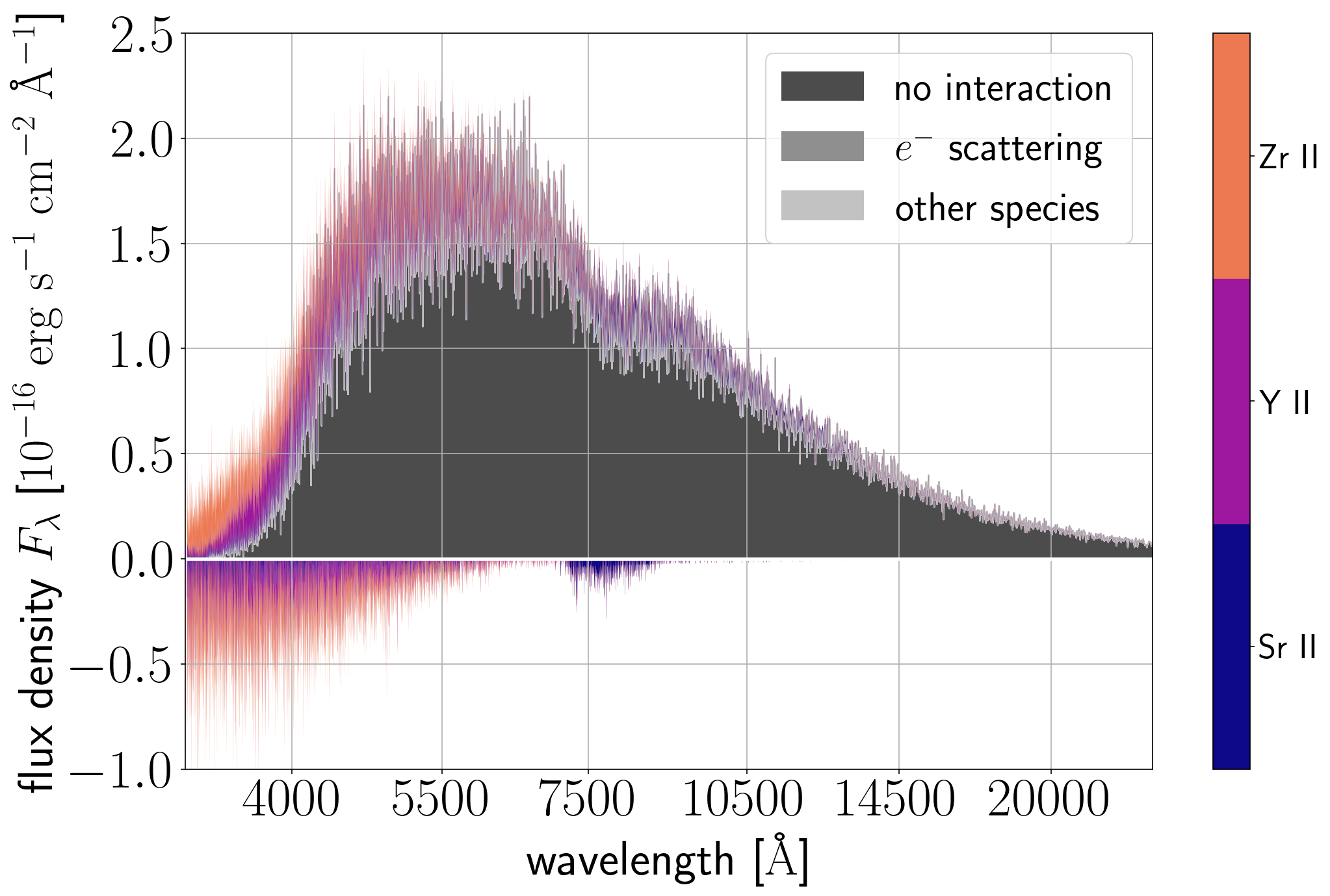} & \addstackgap{\raisebox{0.5\SDECheight}{1.4 days}} \\
    \includegraphics[width=0.47\textwidth]{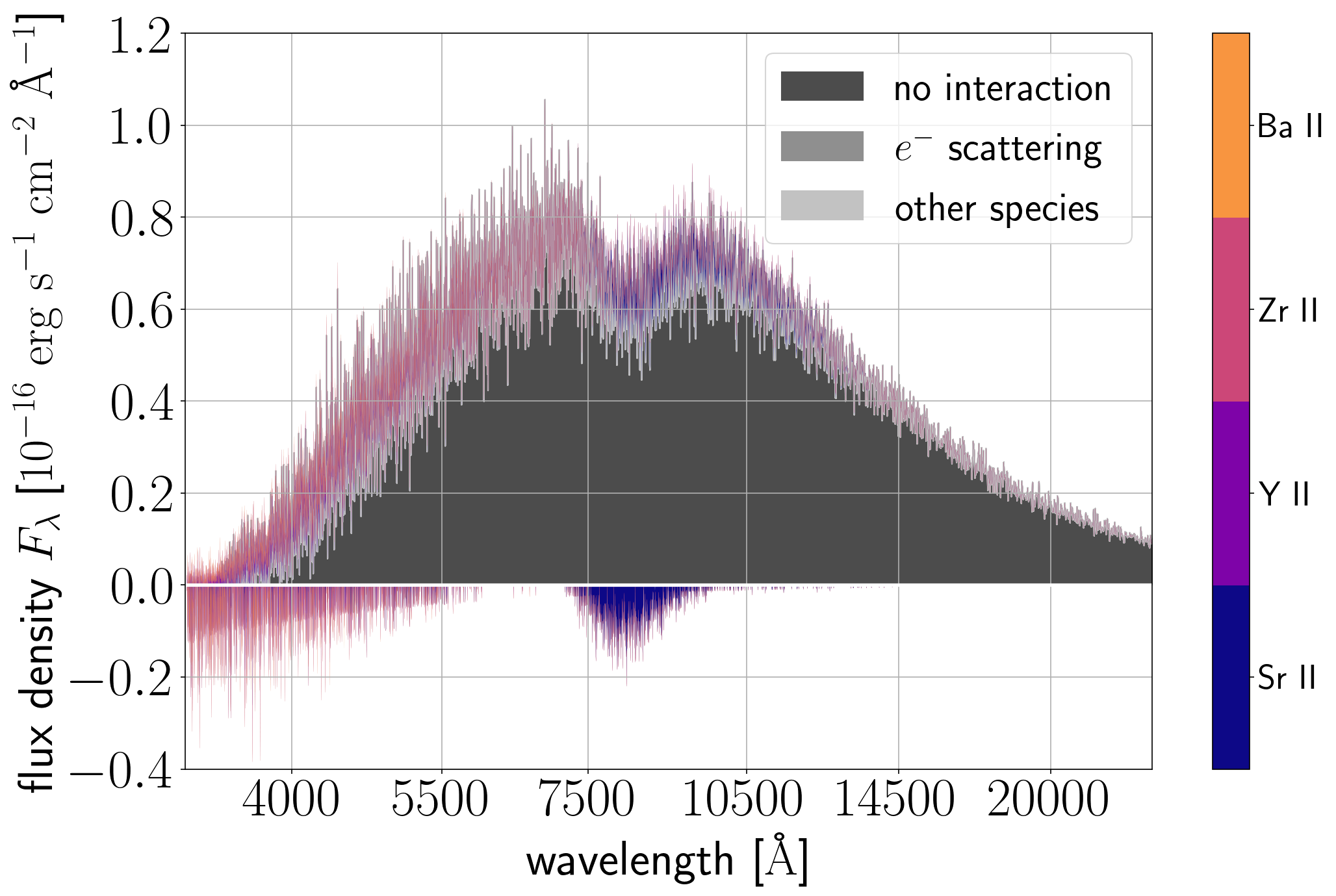} & \includegraphics[width=0.47\textwidth] {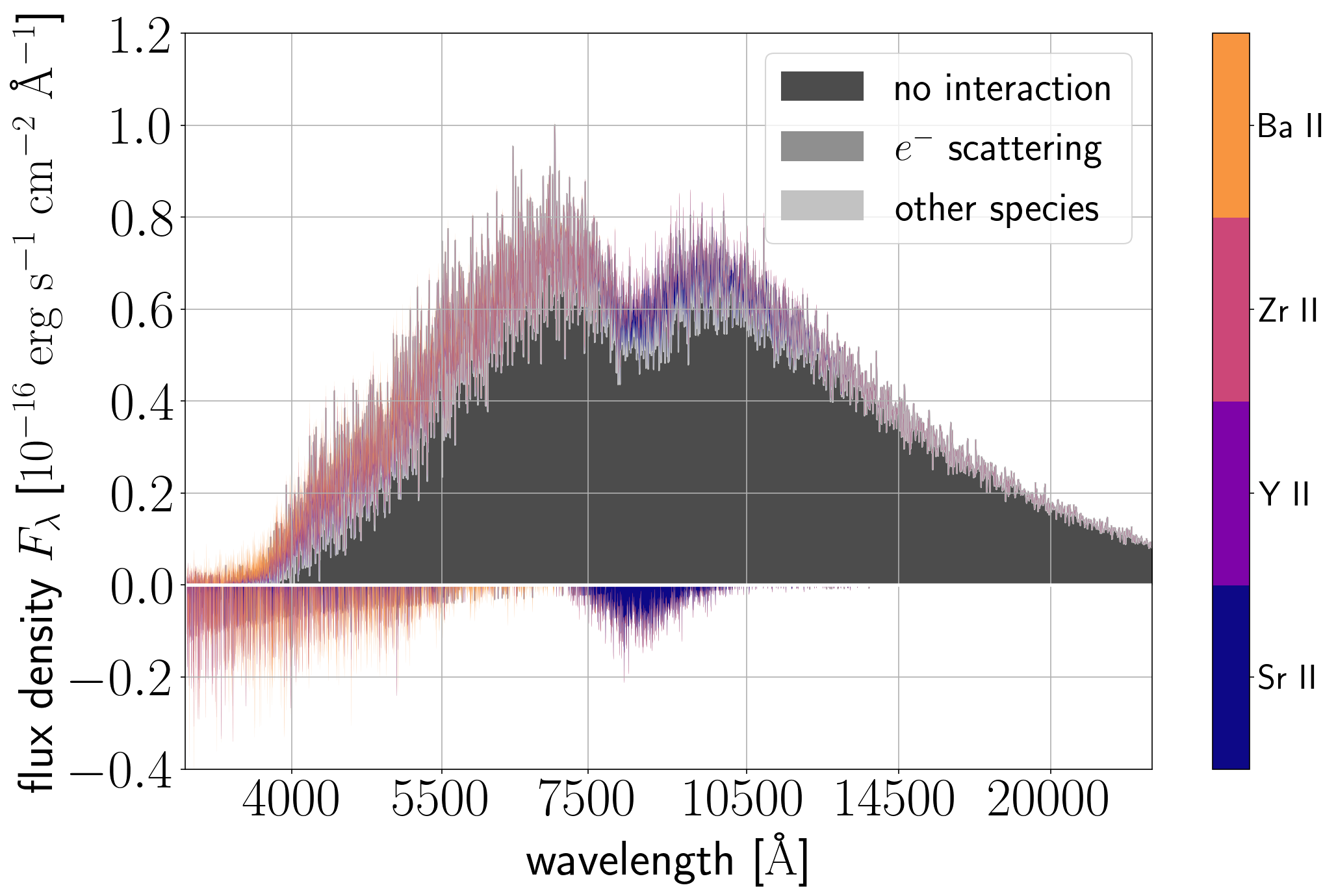} & \addstackgap{\raisebox{0.5\SDECheight}{2.4 days}} \\ 
    \includegraphics[width=0.47\textwidth]{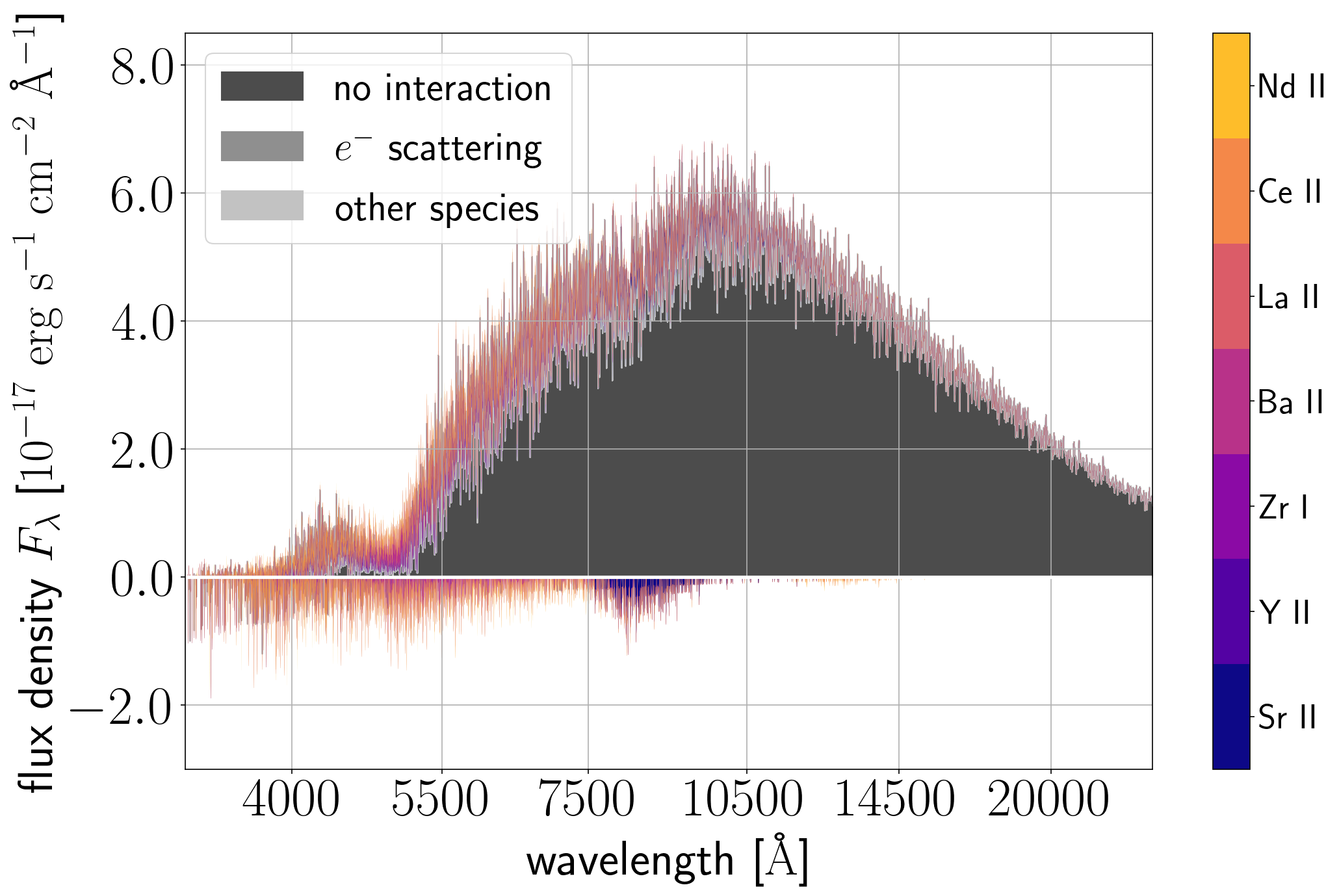} & \includegraphics[width=0.47\textwidth] {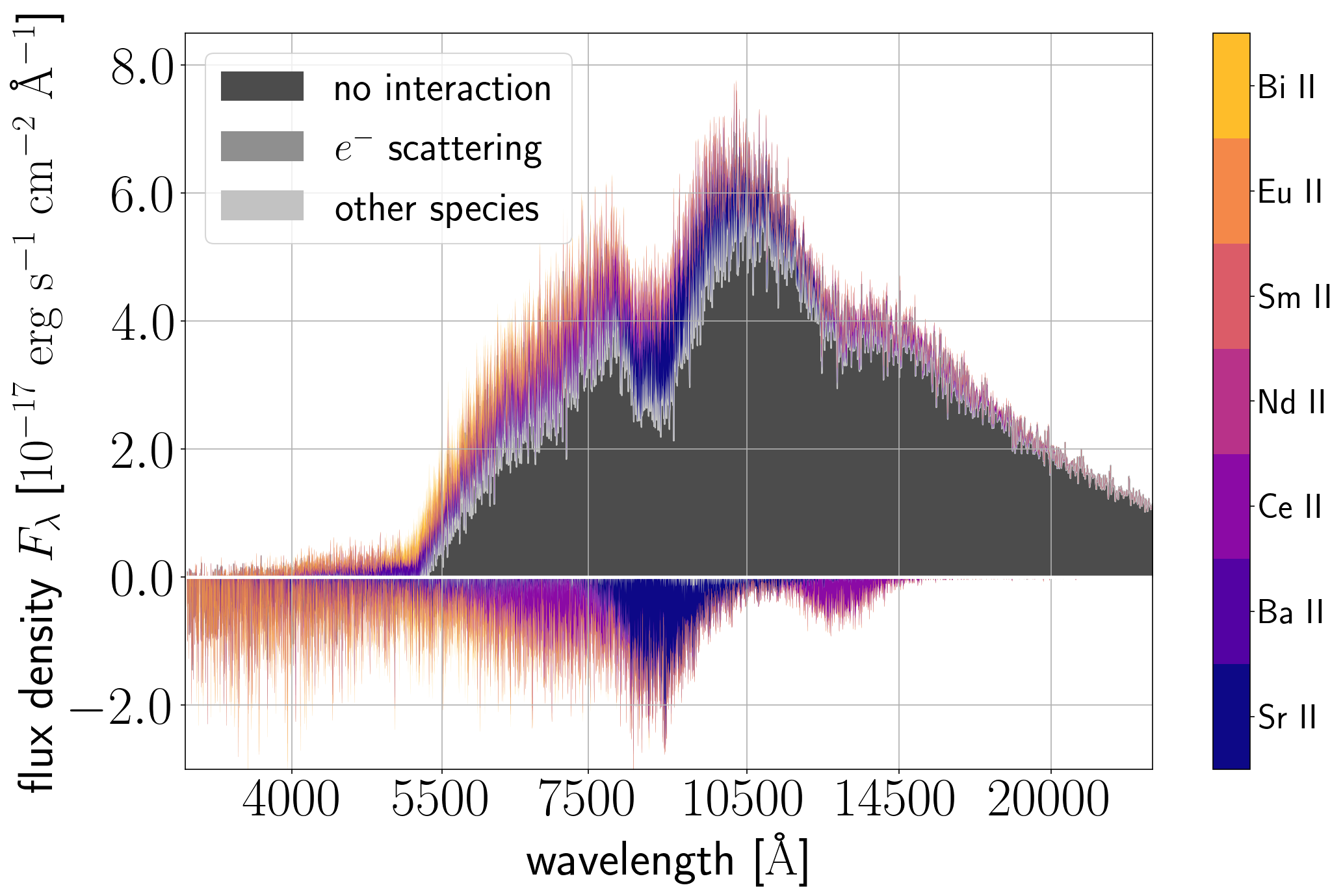} & \addstackgap{\raisebox{0.5\SDECheight}{3.4 days}} \\
    \end{tabular}
    \figcaption{\textbf{SDEC plots for all best fits, at 1.4, 2.4, and 3.4 days, for single- and multicomponent models}. The left column contains single-component models for 1.4, 2.4, and 3.4 days; the right contains multicomponent equivalents. The single-component models are favored at 1.4 and 2.4 days; the multicomponent is favored at 3.4 days. Above the horizontal axis, the height of the histogram indicates the total emitted luminosity in some wavelength bin, across different species, including electron scattering, and including photon packets that escape without interaction. Below the horizontal axis, the depth of the histogram indicates the total absorbed luminosity. These plots also show how emission may be reprocessed, \ie, absorbed in some wavelength range and reemitted in another. The ions \ion{Sr}{2}, \ion{Y}{2}, and \ion{Zr}{2} produce significant absorption in all 1.4-day and 2.4-day models. The 2.4-day models also show absorption from \ion{Ba}{2}. The 3.4-day single-component model (which is a poor fit to the observed spectrum, but nonetheless the best fit possible with a single component) exhibits absorption from all four of these ions, except \ion{Zr}{1} in the place of \ion{Zr}{2}, due to the cooler ejecta. The spectrum also shows moderate absorption from ions of three lanthanides: \ion{La}{2}, \ion{Ce}{2}, and \ion{Nd}{2}. The 3.4-day multicomponent model, which is favored over the single-component model, shows absorption from two open $s$-shell elements in the same periodic table group (\ion{Sr}{2} and \ion{Ba}{2}), four lanthanide ions (\ion{Ce}{2}, \ion{Nd}{2}, \ion{Sm}{2}, and \ion{Eu}{2}), and one ion of an especially heavy element, \ion{Bi}{2}. Furthermore, unlike at earlier epochs, ions of Y and Zr do not dominate absorption in the 3.4-day multicomponent model. Electron scattering is negligible compared to bound-bound interactions, in all models.}\label{fig:SDEC}
\end{figure*}

To assess the impact of different elements on the best fits, we generate leave-one-out spectra. In each, we iteratively leave out a single element by setting its abundance to 0 and transferring that original abundance to a filler element. We use helium (He) as our filler element, since it should not have a marked impact on the emergent spectrum when the ejecta remains optically thick and local thermodynamic equilibrium (LTE) is a valid approximation (\citealt{perego22, tarumi23}).

In Figure~\ref{fig:leaveoneout_pref}, we show leave-one-out spectra for the new favored models at 2.4 and 3.4 days and the 1.4-day model from \V23. We see the clear imprint of Sr at $\sim$8000 \AA~at all epochs. This is consistent with \cite{watson19}, \cite{domoto21}, \cite{domoto22}, \cite{gillanders22}, \cite{sneppenwatson23}, and \cite{sneppen23}, which all argue for the importance of Sr in the spectra. In our new model for 2.4 days, we also see some light evidence for absorption from an adjacent first $r$-process peak element, yttrium (${}_{39}$Y), at short wavelengths $\lesssim$4500~\AA. Considering our previous identification of Y at 1.4 days in \V23, this identification at 2.4 days strengthens our previous claim of the importance of Y and is in agreement with \cite{gillanders22}. Interestingly, Y does not have the same pronounced impact at 3.4 days. This is in agreement with \cite{sneppenwatson23}, which finds that absorption from Y is less prominent at 3.4 days, before a P Cygni feature emerges at $\sim$7600~\AA~at 4.4 and 5.4 days. In our favored multicomponent model, the absorption from the lanthanides is much stronger at these (and shorter) wavelengths. 

At 3.4 days, we identify two new, heavier elements: barium (${}_{56}$Ba), an open $s$-shell element in the same periodic table group as Sr, and cerium (${}_{58}$Ce), a lanthanide. Ba is actually responsible for some of the overestimated absorption at $\lesssim$5000 \AA. However, recall that we neglect wavelengths $\leqslant$6400 \AA~in our computation of the likelihood at 3.4 days to obtain a fit that accurately captures the Sr absorption. Nonetheless, the omission of Ba improves the fit, suggesting that we overestimate the abundance of Ba in our best-fit abundance pattern. The other new element, Ce, produces absorption at $\sim$7000 \AA~that is reprocessed and emitted at $\sim$8000 \AA, and its omission worsens the fit. Ce also introduces some absorption at $\sim$12,000 \AA. Interestingly, this absorption may originate from astrophysically measured \ion{Ce}{2} lines with rest wavelengths in the range of $\sim$$15,200$ $-$ $16,700$~\AA~(\citealt{cunha17, majewski17}), blueshifted owing to the expansion of the ejecta. \cite{domoto21} similarly noted that these \ion{Ce}{2} lines may be prominent in kilonova spectra. The observation of Ce is also consistent with the kilonova parameter space clustering analysis of \cite{ford23}, which finds that \ion{Ce}{2} may broadly be an important ion in kilonova spectra from 1.4 to 3.4 days. \ion{Ce}{3} may also be important at later epochs (\citealt{gillanders23}). 

We complement these leave-one-out analyses using the Spectral DEComposition (SDEC) tool of \TARDIS. SDEC allows us to measure which elements or ions absorb and/or emit the greatest luminosity during a given \TARDIS~run. All SDEC plots, for favored and disfavored models, are compiled in Figure~\ref{fig:SDEC}. In both models at 1.4 days, we see that the absorption of photon packets is dominated by just three singly ionized species: \ion{Sr}{2}, \ion{Y}{2}, and \ion{Zr}{2}. Indeed, 98.9\% of all absorbed luminosity is absorbed by just these three species in the favored single-component model. These ions remain important at 2.4 days, though the impact of \ion{Zr}{2} is less clear. We also see some minor absorption from \ion{Ba}{2} at 2.4 days, at the shortest wavelengths.

At 3.4 days, while \ion{Sr}{2} remains important, the SDEC plots reveal the presence of several new, heavier ions. In the favored multicomponent model, we see clearer evidence for \ion{Ba}{2}, with the caveat that the abundance of Ba is likely overestimated. More interestingly, we see absorption from four singly ionized lanthanides: \ion{Ce}{2}, \ion{Nd}{2}, \ion{Sm}{2}, and \ion{Eu}{2} (cerium, ${}_{58}$Ce; neodymium, ${}_{60}$Nd; samarium, ${}_{62}$Sm; europium, ${}_{63}$Eu). The absorption from Ce is strongest. Altogether, \ion{Ce}{2} is responsible for 27.5\% of the absorbed luminosity at this epoch, compared to 20.4\% from \ion{Nd}{2} + \ion{Sm}{2} + \ion{Eu}{2}, and 30.3\% from \ion{Sr}{2}. This is a clear indication of the presence of lanthanides in the ejecta. Interestingly, Eu is a ``pure'' $r$-process element: at the epoch of solar system formation, Eu was likely produced in negligible quantities by the $s$-process (\textit{e.g.}, \citealt{bisterzo14}). The presence of Eu is thus further proof for the operation of the $r$-process in NS-NS merger ejecta. Finally, at 3.4 days, we also see some light (5.2\% of the total) absorption from an ion of bismuth (${}_{83}$Bi), \ion{Bi}{2}. This is the heaviest element detected in any of our models, but we caution that its detection is marginal (we measure {$Y_{\mathrm{Bi},3.4} \sim 7.8 \times 10^{-6}$), and it does not leave a clear imprint on the leave-one-out spectra. Recall that all lines in our line list are empirical. 

Considering both the leave-one-out and SDEC analyses, we confidently identify Sr at all epochs. We solidify our identification of Y at 1.4 and 2.4 days. Ba may also be present in the ejecta, but its abundance is likely overestimated at 3.4 days, complicating this claim. Finally, at 3.4 days, we infer the presence of singly ionized species of the lanthanides Ce, Nd, Sm, and Eu, with the detection of Ce being most concrete. This ensemble of lanthanides, which was not present at 1.4 and 2.4 days, emerges at 3.4 days to significantly shape the emergent spectrum.

\subsection{Physical origin of the ejecta components}\label{ssc:disco-origins}

Inferring the parameters of the ejecta component(s) also allows us to map these components to different ejection mechanisms in the NS-NS merger that generated the kilonova. Given the inferred $Y_e$ and $s$ see Figure~\ref{fig:posterior-Ye-s}; we neglect $v_{\mathrm{exp}}$ here since it is poorly constrained), the ejecta at 1.4 and 2.4 days can be attributed to matter that has undergone strong neutrino reprocessing. This ejecta can arise from the neutrino-reprocessed wind of a hypermassive NS remnant plus disk, either magnetized (\eg, \citealt{combi23,curtis23b,fahlman23,kiuchi23}) or unmagnetized (\eg, \citealt{fujibayashi23,just23}). Alternatively, accretion disks around BHs---evolved in MHD---also yield ejecta with a broad distribution of electron fractions (\eg, \citealt{siegel18, fernandez19, christie19, miller19, just22, fahlman22, hayashi23, curtis23a}). An outflow from such an accretion disk could in principle provide both the ejecta at 1.4 and 2.4 days and the neutron-rich component that emerges at 3.4 days. 
 
The lanthanide-bearing component at 3.4 days ($Y_{e,2} = 0.161^{+0.149}_{-0.104}$, $s_2 / k_{\mathrm{B}} = 21.5^{+8.3}_{-9.5}$) can also be accounted for by dynamical ejecta. Numerical relativity simulations that include neutrino absorption typically find $Y_e \sim 0.15 - 0.25$ and $s / k_{\mathrm{B}} \sim 20$ (\eg, \citealt{zappa23}), consistent with our inferred  parameters. The slightly higher $Y_e$ component at 3.4 days ($Y_{e,1} = 0.228^{+0.073}_{-0.088}$, $s_1 / k_{\mathrm{B}} = 14.9^{+8.0}_{-4.5}$) may be consistent with a wind from an accretion disk around a BH or longer-lived NS, but it is also consistent with dynamical ejecta. However, the dynamical ejecta alone is unlikely to account for most of the mass generating the kilonova, as numerical relativity simulations {that employ parameters consistent with GW170817 produce as little as $\lesssim 10^{-2} M_{\odot}$ in this dynamical component and more than this amount in post-merger ejecta (\eg, \citealt{shibata17, most19, nedora21}). Our inferred masses $M_{\mathrm{phot},1} \sim M_{\mathrm{phot},2} \sim 10^{-4} M_{\odot}$ are consistent with this bound, but recall that we are only sensitive to the mass above the photosphere. Moreover, some two-component light-curve modeling has found that more mass is contained in the red than the blue component (\eg, \citealt{chornock17, cowperthwaite17, villar17}), which is challenging to accomplish if our red component is indeed dynamical ejecta. Modeling the later spectra of AT2017gfo, when we are sensitive to more of the ejecta mass, will be important for understanding whether this new, redder, lanthanide-rich component is more consistent with dynamical ejecta or an outflow from an accretion disk.

\subsection{Impact of nuclear and atomic physics}\label{ssc:nuclearatomic}

All of our results rely on (1) the use of a particular list of elements' energy levels and lines, \ie, a particular atomic physics, and (2) the results of the \cite{wanajo18} nuclear network calculations, \ie, a particular nuclear physics. Changing this nuclear or atomic physics might alter the inferred parameters. As highlighted in \cite{barnes21}, the photometric evolution of kilonovae is sensitive to the employed nuclear physics. Whether the nuclear physics inputs also impact spectral evolution remains to be seen, but is likely. In particular, the assumed nuclear mass models, decay rates, prescription for fission, and treatment of neutrino transport (if included) all enter into the nuclear reaction network calculations. These inputs in turn affect the mapping between elemental abundances and $Y_e$, $v_{\mathrm{exp}}$, and $s$. 

With respect to the atomic physics, the incompleteness of our line list, which uses only empirical lines as obtained from VALD, may bias the inference. Injecting theoretical lines into our list might worsen this bias, as it has been observed that different theoretical atomic structure calculations generate a large spread in line transition wavelengths and energies given the same initial configurations (\eg, \citealt{tanaka20, flors23}). In practice, the use of a larger line list also significantly increases the runtime for spectral syntheses. Nevertheless, given our use of an incomplete but empirical line list, we can claim that a particular element can produce a particular feature in the observed spectra. We cannot claim that a particular element is the only element that can produce said feature.


\section{Conclusions}\label{sec:conco}

We fit single- and multicomponent ejecta models to the spectra of the GW170817 kilonova, AT2017gfo, during the early, optically thick phase at 1.4, 2.4, and 3.4 days post-merger. With these fits, we infer the element-by-element abundance patterns at each of these epochs. We find that a single-component model is favored at 1.4 and 2.4 days, while a multicomponent model is favored at 3.4 days. 

This single component at 1.4 and 2.4 days is characterized by a high electron fraction $Y_e \sim 0.3$ and moderate specific entropy $s / k_{\mathrm{B}} \sim 13 - 18$, yielding an ejecta dominated by lighter $r$-process elements and a blue kilonova. This ejecta is consistent with material that has undergone substantial neutrino reprocessing, \eg, winds from a remnant hypermassive NS and/or an accretion disk. The multicomponent ejecta at 3.4 days contains a higher $Y_e \sim 0.23$ component and lower $Y_e \sim 0.16$ component, with entropies in the range $s / k_{\mathrm{B}}\sim 15 - 22$. These new components contain heavier elements, especially the lanthanides, yielding a redder kilonova. The most substantial contribution of lanthanides comes from the lower-$Y_e$ component, which is consistent with either dynamical ejecta or a neutron-rich outflow from a remnant accretion disk. 

The emergence of a new red component at 3.4 days is broadly in agreement with modeling of the light curves of AT2017gfo. Physically, we infer that the photosphere recedes into the ejecta over time: from $0.31c$ at 1.4 days, to $0.25c$ at 2.4 days, to $0.21c$ at 3.4 days. This recession of the photosphere, as well as the dimming of the earlier, blue component, reveals this new red component that was not inferred at earlier epochs. 

Using both leave-one-out and SDEC analyses, we assess the contributions of individual elements to the emergent spectra. We find that strontium Sr produces the $\sim$8000~\AA~absorption at each epoch. We also strengthen our identification of yttrium Y at short wavelengths $\lesssim$4500~\AA~at 1.4 and 2.4 days. At 3.4 days, this absorption from Y is less clear, as absorption from an ensemble of lanthanides (cerium Ce, neodymium Nd, samarium Sm, and europium Eu) dominates the absorption at these short wavelengths. The identification of Ce is most concrete, producing absorption at $\sim$7000~\AA~and $\sim$12,000~\AA.

The abundance patterns at 1.4 and 2.4 days show a dearth of lanthanides and heavier elements, which makes these inconsistent with the solar $r$-process abundance pattern. However, at 3.4 days, the emergence of ejecta with lanthanide fraction $\log_{\mathrm{10}} X_{\mathrm{lan}} \sim -1.6$~substantially improves the agreement between the inferred abundance pattern and the solar $r$-process, as well as the distribution of lanthanide fractions measured in metal-poor stars. The better agreement between the solar $r$-process/metal-poor stars and inferred abundance pattern at 3.4 days, as well as the possible presence of multiple components (remnant/disk wind and dynamical ejecta), lends more credence to the ability of NS-NS/BH mergers to dominate the $r$-process in the Universe. 

All of our inference relies on the assumption of a particular atomic physics (line list of transitions and energy levels for ions) and nuclear physics (reaction network calculations). \SPARK~has been designed in a modular way such that either the atomic physics or nuclear physics, could be swapped out for another to quantify how the different physics impact the inference. A further simplification in our model is the use of single ejecta parameters (especially single $Y_e$) in generating synthetic spectra. In reality, the ejecta is described by a distribution of such parameters. In future models, it could be possible to introduce a toy distribution in the ejecta parameters (based on, \eg, the results of GRMHD simulations) and fit for the parameters of this distribution.

While we have fit spectra of AT2017gfo at 1.4, 2.4, and 3.4 days, the observed spectra of AT2017gfo extend to 10.4 days. At later times, the ejecta is expected to leave the photospheric phase and enter the optically thin nebular phase, when non-LTE effects become nonnegligible. Given the modular nature of \SPARK, we could swap out \TARDIS~for a code specifically suited to non-LTE radiative transfer, or use the non-LTE functionality already available in \TARDIS. Our Bayesian inference framework would also enable quantitative model comparison between models with and without non-LTE effects at epochs where their relative importance is not yet understood (\eg, 4.4 and 5.4 days). Fitting later epochs will also be crucial for inferring the abundances and masses of all of the ejecta components, including those hidden below the photosphere at 3.4 days. Beyond the later epochs of AT2017gfo, \SPARK~can perform inference on the growing zoo of kilonovae yielded by NS-NS/BH binaries with different parameters, allowing us to understand the connections between binary parameters, ejection mechanisms, elemental compositions, and fundamental $r$-process conditions in the ejecta.

\acknowledgments

N.V. works in Tiohti{\'a}:ke/Mooniyang, also known as Montr{\'e}al, which lies on the unceded land of the Haudenosaunee and Anishinaabeg nations. This work made use of high-performance computing resources in Tiohti{\'a}:ke/Mooniyang and in Burnaby, British Columbia, the unceded land of the Coast Salish peoples, including the Tsleil-Waututh, Kwikwetlem, Squamish, and Musqueam nations. We acknowledge the ongoing struggles of Indigenous peoples on this land, and elsewhere on Turtle Island, and hope for a future marked by true reconciliation.

We thank the referee, whose extensive comments have strengthened this study.

We thank the attendees of a 2023 April workshop at University of California--Santa Cruz, for fruitful discussions on kilonovae and the $r$-process. We are also grateful to Shinya Wanajo for kindly sharing their reaction network calculations. Finally, we thank Jessica Birky and David Fleming for useful discussions on approximate Bayesian inference and the use of \href{https://dflemin3.github.io/approxposterior/index.html}{\approxposterior}.

This work made extensive use of the \href{https://docs.alliancecan.ca/wiki/Narval/en}{\texttt{Narval}} and \href{https://docs.alliancecan.ca/wiki/Cedar}{\texttt{Cedar}} clusters of the \href{https://alliancecan.ca/en}{Digital Research Alliance of Canada} at the {\'E}cole de technologie sup{\'e}rieure and Simon Fraser University (with regional partner \href{https://www.westgrid.ca/}{WestGrid}), respectively. We thank the support staff of Calcul Qu{\'e}bec in particular for their assistance at various steps in this project. 

This work made use of the \href{http://vald.astro.uu.se/~vald/php/vald.php}{Vienna Atomic Line Database (VALD)}, operated at Uppsala University, the Institute of Astronomy RAS in Moscow, and the University of Vienna. We thank Nikolai Piskunov and Eric Stempels for help in obtaining the VALD data.

This research also made use of \href{https://tardis-sn.github.io/tardis/index.html}{\TARDIS}, a community-developed software package for spectral synthesis in supernovae (\citealt{kerzendorf14}). The development of \TARDIS~received support from the Google Summer of Code initiative and from the European Space Agency's (ESA) Summer of Code in Space program. \TARDIS~is a fiscally
sponsored project of NumFOCUS. \TARDIS~makes extensive use of \href{https://docs.astropy.org/en/stable/}{\texttt{astropy}} and \href{https://pyne.io/}{\texttt{PyNE}}. We thank Andrew Fullard, Wolfgang Kerzendorf, and the entire \TARDIS~development team for their assistance and their commitment to the development and maintenance of the code. 

N.V. acknowledges funding from the Natural Sciences and Engineering Research Council of Canada (NSERC) Canada Graduate Scholarship - Doctoral (CGS-D), the Murata Family Fellowship, and the Bob Wares Science Innovation Prospectors Fund. J.J.R.\ and D.H.\ acknowledge support from the Canada Research Chairs (CRC) program, the NSERC Discovery Grant program, the FRQNT Nouveaux Chercheurs Grant program, and the Canadian Institute for Advanced Research (CIFAR). J.J.R.\ acknowledges funding from the Canada Foundation for Innovation (CFI) and the Qu\'{e}bec Ministère de l’\'{E}conomie et de l’Innovation. N.M.F. acknowledges funding from the Fondes de Recherche Nature et Technologies (FRQNT) Doctoral research scholarship No. 328732. M.R.D. acknowledges support from the NSERC through grant RGPIN-2019-06186, the Canada Research Chairs Program, and the Dunlap Institute at the University of Toronto.
R.F. acknowledges support from NSERC of Canada through Discovery Grant RGPIN-2022-03463.
\newline

\software{
\href{https://dflemin3.github.io/approxposterior/index.html}{\approxposterior}: \cite{fleming18};
\href{https://docs.astropy.org/en/stable/}{\texttt{astropy}}: \cite{astropy18};
\href{https://cmasher.readthedocs.io/}{\texttt{cmasher}}: \cite{velden20};
\href{https://corner.readthedocs.io/en/latest/index.html}{\texttt{corner}}: \cite{foreman-mackey16};
\href{https://dynesty.readthedocs.io/en/latest/index.html}{\texttt{dynesty}}: 
\cite{speagle20};
\href{https://george.readthedocs.io/en/latest/}{\texttt{george}}: \cite{ambikasaran15};
\href{https://tardis-sn.github.io/tardis/index.html}{\TARDIS}: \cite{kerzendorf14, kerzendorf23};
\href{https://johannesbuchner.github.io/UltraNest/}{\texttt{UltraNest}}: \cite{buchner21}
} 

\clearpage

\bibliographystyle{apj}

\appendix{}

\section{All Training Spectra and Posteriors}\label{app:allspec_posteriors}

Figure~\ref{fig:all_spec_single} shows all of the spectra used to construct the training sets in the single-component \SPARK~runs at 1.4, 2.4, and 3.4 days post-merger. Figure~\ref{fig:all_spec_multi} shows the same for the multicomponent equivalents. These include the $m_0 = 1500$ Latin hypercube samples, $m_{\mathrm{active}}$ active learning samples, and the best-fit spectrum for each \SPARK~run.

Posterior probabilities are computed for each of these spectra, and these are used by the GP to construct a surrogate posterior. All posteriors are obtained through dynamic nested sampling of the surrogate GP using \texttt{dynesty} (\citealt{speagle20}). Figures \ref{fig:corner_single_2.4}~and \ref{fig:corner_single_3.4} show the full posteriors for the new single-component fits to the 2.4- and 3.4-day spectra, respectively. Both posteriors are multimodal. We highlight the modes that correspond to the best fits in Table~\ref{tab:bestfit_single}. Figures~\ref{fig:corner_multi_1.4}, \ref{fig:corner_multi_2.4}, and~\ref{fig:corner_multi_3.4} show the full posteriors for the new multicomponent fits to the 1.4-, 2.4-, and 3.4-day spectra, respectively. The medians of these posteriors are given as the best-fit parameters in Table~\ref{tab:bestfit_multi}.

\section{ELEMENTAL MASS FRACTIONS}\label{app:massfracs_detailed}

In Table~\ref{tab:massfracs}, we compile the mass fractions of all elements and their uncertainties, for the favored models: single component at 1.4 and 2.4 days, and multicomponent at 3.4 days. We separately show the abundance of the individual components at 3.4 days and the mass-weighted sum of the two components. We omit elements' mass fractions when these are below $10^{-9}$. For elements for which the relative uncertainty on the mass fraction is $>100\%$ the best-fit parameter, we use a tilde ($\sim$) to indicate the large uncertainty. For elements for which the uncertainty is small ($<1\%$), we report just the best-fit value (\ie, posterior median).

We have used boldface for certain elements of interest, namely, Sr, Y, Zr, Ba, and Ce. We caution that we overestimate the absorption of Ba in our 3.4-day, multicomponent fit. The mass fraction of Ba should thus be taken conservatively as an upper limit.

\startlongtable
\begin{deluxetable*}{cc|cccccc}
\centering
\tablecaption{Mass fractions $X_i$ for the best-fit single-component models at 1.4 and 2.4 days, the two individual components at 3.4 days, and the mass-weighted sum of both components at 3.4 days.}\label{tab:massfracs}
\tablehead{$Z$ & element & \textbf{1.4 days} & \textbf{2.4 days} & 3.4 days, higher $Y_e$ & 3.4 days, lower $Y_e$ & \textbf{3.4 days, total}}
\startdata\tablewidth{1.0\textwidth}
1 & H & ${3.55}^{+4.30}_{-2.45} \times 10^{-8}$ & ${1.79}^{+8.38}_{-1.38} \times 10^{-8}$ & $\sim {9.09} \times 10^{-6}$ & ${1.09}^{+52.19}_{-1.09} \times 10^{-4}$ & ${1.25}^{+32.83}_{-1.21} \times 10^{-5}$ \\
2 & He & ${2.24}^{+1.43}_{-0.92} \times 10^{-4}$ & ${3.10}^{+86.53}_{-2.94} \times 10^{-4}$ & ${1.41}^{+3.47}_{-1.06} \times 10^{-3}$ & ${1.04}^{+2.09}_{-0.87} \times 10^{-2}$ & ${1.72}^{+14.78}_{-0.88} \times 10^{-3}$ \\
3 & Li & ${1.34}^{+0.95}_{-0.55} \times 10^{-9}$ & ${1.69}^{+1.81}_{-1.60} \times 10^{-9}$ & ${8.66}^{+12.04}_{-7.10} \times 10^{-9}$ & ${2.78}^{+1.73}_{-2.78} \times 10^{-8}$ & ${9.32}^{+9.69}_{-7.65} \times 10^{-9}$ \\
4 & Be & ${9.32}^{+7.33}_{-4.86} \times 10^{-7}$ & ${1.07}^{+4.47}_{-1.07} \times 10^{-6}$ & ${1.39}^{+3.34}_{-1.17} \times 10^{-5}$ & ${6.96}^{+2.77}_{-6.64} \times 10^{-5}$ & ${1.58}^{+2.17}_{-1.37} \times 10^{-5}$ \\
5 & B & ${1.32}^{+1.42}_{-0.78} \times 10^{-9}$ & - & ${1.45}^{+4.94}_{-1.38} \times 10^{-8}$ & ${3.23}^{+5.92}_{-3.22} \times 10^{-8}$ & ${1.51}^{+3.54}_{-1.30} \times 10^{-8}$ \\
6 & C & ${7.21}^{+11.91}_{-4.37} \times 10^{-8}$ & ${1.87} \times 10^{-7}$ & ${1.40}^{+9.76}_{-1.11} \times 10^{-6}$ & ${1.90}^{+1.38}_{-1.74} \times 10^{-5}$ & ${2.00}^{+8.40}_{-1.41} \times 10^{-6}$ \\
7 & N & - & $\sim {5.34} \times 10^{-9}$ & ${2.59}^{+90.74}_{-2.31} \times 10^{-8}$ & ${1.24}^{+4.65}_{-1.22} \times 10^{-6}$ & ${6.75}^{+95.28}_{-5.98} \times 10^{-8}$ \\
8 & O & ${5.90}^{+0.60}_{-0.06} \times 10^{-8}$ & ${1.56}^{+4.37}_{-0.05} \times 10^{-7}$ & ${4.34}^{+49.22}_{-3.86} \times 10^{-6}$ & ${1.00}^{+2.91}_{-0.95} \times 10^{-4}$ & ${7.63}^{+64.56}_{-5.87} \times 10^{-6}$ \\
9 & F & - & $\sim {4.13} \times 10^{-9}$ & ${1.54}^{+43.31}_{-1.44} \times 10^{-7}$ & $\sim {8.22} \times 10^{-6}$ & $\sim {4.31} \times 10^{-7}$ \\
10 & Ne & - & - & $\sim {6.04} \times 10^{-9}$ & $\sim {2.95} \times 10^{-7}$ & $\sim {1.59} \times 10^{-8}$ \\
11 & Na & - & - & - & $\sim {5.70} \times 10^{-8}$ & $\sim {2.55} \times 10^{-9}$ \\
12 & Mg & - & - & ${2.82}^{+3.03}_{-0.06} \times 10^{-8}$ & $\sim {2.38} \times 10^{-6}$ & ${1.09}^{+99.15}_{-0.90} \times 10^{-7}$ \\
13 & Al & - & - & - & $\sim {4.35} \times 10^{-9}$ & - \\
14 & Si & - & - & - & ${1.32}^{+74.57}_{-1.31} \times 10^{-7}$ & $\sim {4.99} \times 10^{-9}$ \\
15 & P & - & - & $\sim {1.11} \times 10^{-9}$ & $\sim {8.61} \times 10^{-8}$ & $\sim {4.02} \times 10^{-9}$ \\
16 & S & - & - & ${1.93}^{+3.91}_{-0.02} \times 10^{-9}$ & $\sim {1.95} \times 10^{-7}$ & $\sim {8.56} \times 10^{-9}$ \\
17 & Cl & - & - & - & $\sim {7.59} \times 10^{-9}$ & - \\
18 & Ar & - & - & - & $\sim {5.20} \times 10^{-8}$ & $\sim {2.59} \times 10^{-9}$ \\
19 & K & - & - & - & $\sim {7.21} \times 10^{-8}$ & $\sim {2.83} \times 10^{-9}$ \\
20 & Ca & ${4.03}^{+1.35}_{-0.43} \times 10^{-7}$ & $\sim {9.67} \times 10^{-8}$ & ${7.97}^{+92.26}_{-0.44} \times 10^{-9}$ & $\sim {4.20} \times 10^{-7}$ & ${2.21}^{+86.42}_{-0.19} \times 10^{-8}$ \\
21 & Sc & ${9.41}^{+4.46}_{-0.70} \times 10^{-9}$ & $\sim {4.48} \times 10^{-9}$ & - & $\sim {1.93} \times 10^{-8}$ & - \\
22 & Ti & ${5.00} \times 10^{-6}$ & ${3.13} \times 10^{-5}$ & ${6.79}^{+44.38}_{-1.02} \times 10^{-9}$ & $\sim {4.44} \times 10^{-8}$ & $\sim {8.08} \times 10^{-9}$ \\
23 & V & ${6.88} \times 10^{-5}$ & ${1.23} \times 10^{-4}$ & ${2.18} \times 10^{-6}$ & $\sim {2.62} \times 10^{-8}$ & ${2.11}^{+4.77}_{-0.01} \times 10^{-6}$ \\
24 & Cr & ${1.30} \times 10^{-1}$ & ${1.13} \times 10^{-1}$ & ${7.30} \times 10^{-3}$ & $\sim {1.51} \times 10^{-4}$ & ${7.05} \times 10^{-3}$ \\
25 & Mn & ${4.91} \times 10^{-3}$ & ${6.53} \times 10^{-3}$ & ${2.56} \times 10^{-4}$ & $\sim {5.14} \times 10^{-6}$ & ${2.47} \times 10^{-4}$ \\
26 & Fe & ${5.99} \times 10^{-2}$ & ${3.49} \times 10^{-2}$ & ${6.12} \times 10^{-3}$ & $\sim {7.81} \times 10^{-5}$ & ${5.92} \times 10^{-3}$ \\
27 & Co & ${1.27} \times 10^{-4}$ & ${1.47} \times 10^{-4}$ & ${1.50} \times 10^{-5}$ & $\sim {1.83} \times 10^{-7}$ & $\sim {1.45} \times 10^{-5}$ \\
28 & Ni & ${8.84} \times 10^{-2}$ & ${6.22} \times 10^{-2}$ & ${1.50} \times 10^{-2}$ & $\sim {1.85} \times 10^{-4}$ & ${1.45}^{+0.07}_{-0.02} \times 10^{-2}$ \\
29 & Cu & ${1.26} \times 10^{-2}$ & ${8.90} \times 10^{-3}$ & ${2.13} \times 10^{-3}$ & $\sim {2.09} \times 10^{-5}$ & ${2.06}^{+20.80}_{-5.01} \times 10^{-3}$ \\
30 & Zn & ${3.58} \times 10^{-2}$ & ${2.49} \times 10^{-2}$ & ${9.25} \times 10^{-3}$ & $\sim {8.71} \times 10^{-5}$ & ${8.93}^{+0.07}_{-0.01} \times 10^{-3}$ \\
31 & Ga & ${1.65} \times 10^{-2}$ & ${1.19} \times 10^{-2}$ & ${4.28}^{+43.96}_{-6.39} \times 10^{-3}$ & $\sim {3.96} \times 10^{-5}$ & ${4.14}^{+44.25}_{-12.53} \times 10^{-3}$ \\
32 & Ge & ${4.22} \times 10^{-2}$ & ${3.37} \times 10^{-2}$ & ${1.79}^{+0.23}_{-0.02} \times 10^{-2}$ & $\sim {1.52} \times 10^{-4}$ & ${1.73}^{+0.59}_{-0.18} \times 10^{-2}$ \\
33 & As & ${6.34} \times 10^{-3}$ & ${4.43}^{+1.40}_{-0.03} \times 10^{-3}$ & ${2.21}^{+31.57}_{-5.65} \times 10^{-3}$ & $\sim {1.89} \times 10^{-5}$ & ${2.13}^{+23.06}_{-8.03} \times 10^{-3}$ \\
34 & Se & ${7.23} \times 10^{-2}$ & ${6.91}^{+0.31}_{-0.01} \times 10^{-2}$ & ${8.79} \times 10^{-2}$ & ${1.04}^{+51.30}_{-1.04} \times 10^{-3}$ & ${8.49}^{+1.12}_{-0.38} \times 10^{-2}$ \\
35 & Br & ${5.31}^{+0.87}_{-0.04} \times 10^{-3}$ & ${7.11}^{+60.81}_{-97.41} \times 10^{-3}$ & ${1.14}^{+7.01}_{-1.37} \times 10^{-2}$ & ${1.51}^{+48.07}_{-1.51} \times 10^{-4}$ & ${1.11}^{+5.30}_{-1.53} \times 10^{-2}$ \\
36 & Kr & ${8.97}^{+6.60}_{-5.26} \times 10^{-2}$ & ${9.11}^{+0.51}_{-0.57} \times 10^{-2}$ & ${1.19}^{+0.10}_{-0.02} \times 10^{-1}$ & ${3.04}^{+55.05}_{-3.04} \times 10^{-3}$ & ${1.15}^{+0.07}_{-0.02} \times 10^{-1}$ \\
37 & Rb & ${3.12}^{+0.59}_{-0.21} \times 10^{-2}$ & ${3.51}^{+3.17}_{-2.79} \times 10^{-2}$ & ${2.98}^{+3.34}_{-0.84} \times 10^{-2}$ & ${1.29}^{+19.14}_{-1.29} \times 10^{-3}$ & ${2.88}^{+7.63}_{-7.50} \times 10^{-2}$ \\
\textbf{38} & \textbf{Sr} & $\mathbf{{8.26}^{+1.94}_{-2.78} \times 10^{-2}}$ & $\mathbf{{9.90}^{+0.01}_{-0.01} \times 10^{-2}}$ & $\mathbf{{6.44}^{+1.59}_{-0.39} \times 10^{-2}}$ & $\mathbf{{3.27}^{+41.61}_{-3.27} \times 10^{-3}}$ & $\mathbf{{6.23}^{+0.61}_{-0.96} \times 10^{-2}}$ \\
\textbf{39} & \textbf{Y} & $\mathbf{{1.34} \times 10^{-2}}$ & $\mathbf{{1.81}^{+6.77}_{-5.00} \times 10^{-2}}$ & $\mathbf{{8.71}^{+67.05}_{-16.45} \times 10^{-3}}$ & $\mathbf{{5.22}^{+67.61}_{-5.22} \times 10^{-4}}$ & $\mathbf{{8.43}^{+22.95}_{-93.12} \times 10^{-3}}$ \\
\textbf{40} & \textbf{Zr} & $\mathbf{{1.07}^{+0.45}_{-0.44} \times 10^{-1}}$ & $\mathbf{{1.45}^{+0.11}_{-0.07} \times 10^{-1}}$ & $\mathbf{{6.77}^{+0.83}_{-0.20} \times 10^{-2}}$ & $\mathbf{{6.29}^{+73.58}_{-6.29} \times 10^{-3}}$ & $\mathbf{{6.56}^{+1.04}_{-2.04} \times 10^{-2}}$ \\
41 & Nb & ${3.83}^{+65.81}_{-57.89} \times 10^{-4}$ & $\sim {3.88} \times 10^{-4}$ & $\sim {2.42} \times 10^{-4}$ & ${2.97}^{+24.47}_{-2.97} \times 10^{-5}$ & $\sim {2.35} \times 10^{-4}$ \\
42 & Mo & ${3.25}^{+2.11}_{-1.92} \times 10^{-2}$ & ${3.49}^{+1.68}_{-1.02} \times 10^{-2}$ & ${2.09}^{+0.49}_{-1.00} \times 10^{-2}$ & ${2.78}^{+24.31}_{-2.78} \times 10^{-3}$ & ${2.03}^{+0.30}_{-0.54} \times 10^{-2}$ \\
43 & Tc & ${1.73}^{+10.35}_{-8.81} \times 10^{-3}$ & ${2.48}^{+60.54}_{-28.03} \times 10^{-3}$ & ${2.05}^{+29.97}_{-91.72} \times 10^{-3}$ & ${2.90}^{+23.67}_{-2.90} \times 10^{-4}$ & ${1.99}^{+21.16}_{-50.65} \times 10^{-3}$ \\
44 & Ru & ${8.70}^{+2.66}_{-2.27} \times 10^{-2}$ & ${1.01}^{+0.13}_{-0.03} \times 10^{-1}$ & ${6.96}^{+0.61}_{-1.94} \times 10^{-2}$ & ${1.09}^{+8.57}_{-1.09} \times 10^{-2}$ & ${6.76}^{+0.01}_{-0.02} \times 10^{-2}$ \\
45 & Rh & ${1.06}^{+0.40}_{-0.35} \times 10^{-2}$ & $\sim {8.97} \times 10^{-3}$ & ${4.87}^{+6.41}_{-39.47} \times 10^{-3}$ & ${9.33}^{+66.89}_{-9.33} \times 10^{-4}$ & ${4.74}^{+10.11}_{-15.07} \times 10^{-3}$ \\
46 & Pd & ${2.76}^{+2.13}_{-1.30} \times 10^{-2}$ & ${3.53}^{+1.61}_{-0.43} \times 10^{-2}$ & ${3.67}^{+0.15}_{-0.49} \times 10^{-2}$ & ${8.04}^{+46.64}_{-8.04} \times 10^{-3}$ & ${3.57}^{+0.14}_{-0.13} \times 10^{-2}$ \\
47 & Ag & ${7.19}^{+2.12}_{-0.53} \times 10^{-3}$ & $\sim {9.44} \times 10^{-3}$ & ${9.21}^{+15.36}_{-29.26} \times 10^{-3}$ & ${2.62}^{+12.37}_{-2.62} \times 10^{-3}$ & ${8.98}^{+25.89}_{-56.37} \times 10^{-3}$ \\
48 & Cd & ${1.17}^{+3.78}_{-2.58} \times 10^{-2}$ & ${1.34}^{+6.47}_{-1.71} \times 10^{-2}$ & ${2.55}^{+0.01}_{-0.01} \times 10^{-2}$ & ${6.39}^{+24.93}_{-6.39} \times 10^{-3}$ & ${2.48}^{+0.18}_{-0.42} \times 10^{-2}$ \\
49 & In & $\sim {6.03} \times 10^{-4}$ & $\sim {8.34} \times 10^{-4}$ & ${2.71}^{+7.09}_{-10.32} \times 10^{-3}$ & ${7.56}^{+22.71}_{-7.55} \times 10^{-4}$ & ${2.64}^{+17.11}_{-31.65} \times 10^{-3}$ \\
50 & Sn & ${1.58}^{+6.23}_{-2.59} \times 10^{-2}$ & ${1.72}^{+0.68}_{-1.27} \times 10^{-2}$ & ${1.45}^{+0.01}_{-0.01} \times 10^{-1}$ & ${6.60}^{+9.43}_{-6.54} \times 10^{-2}$ & ${1.42}^{+0.06}_{-0.08} \times 10^{-1}$ \\
51 & Sb & ${3.30}^{+11.90}_{-3.98} \times 10^{-3}$ & $\sim {4.45} \times 10^{-3}$ & ${4.31}^{+1.83}_{-3.72} \times 10^{-2}$ & ${1.92}^{+4.36}_{-1.90} \times 10^{-2}$ & ${4.23}^{+1.46}_{-2.14} \times 10^{-2}$ \\
52 & Te & ${2.42}^{+60.55}_{-38.32} \times 10^{-3}$ & ${4.63}^{+0.42}_{-0.33} \times 10^{-3}$ & ${9.01}^{+0.24}_{-0.31} \times 10^{-2}$ & ${7.98}^{+11.56}_{-7.68} \times 10^{-2}$ & ${8.97}^{+0.18}_{-0.23} \times 10^{-2}$ \\
53 & I & ${9.30}^{+2.46}_{-1.67} \times 10^{-4}$ & ${2.27}^{+31.28}_{-31.61} \times 10^{-3}$ & ${4.25}^{+1.36}_{-2.49} \times 10^{-2}$ & ${2.99}^{+7.29}_{-2.84} \times 10^{-2}$ & ${4.21}^{+9.58}_{-13.32} \times 10^{-2}$ \\
54 & Xe & $\sim {8.86} \times 10^{-6}$ & $\sim {1.93} \times 10^{-5}$ & ${2.56}^{+0.46}_{-0.62} \times 10^{-2}$ & ${4.89}^{+4.97}_{-4.73} \times 10^{-2}$ & ${2.64}^{+1.92}_{-3.99} \times 10^{-2}$ \\
55 & Cs & $\sim {1.05} \times 10^{-6}$ & $\sim {2.13} \times 10^{-6}$ & ${5.71}^{+8.89}_{-19.15} \times 10^{-3}$ & ${2.14}^{+4.35}_{-2.08} \times 10^{-2}$ & ${6.25}^{+71.52}_{-86.60} \times 10^{-3}$ \\
\textbf{56} & \textbf{Ba} & $\mathbf{\sim {1.63} \times 10^{-7}}$ & $\mathbf{\sim {1.52} \times 10^{-7}}$ & $\mathbf{{4.00}^{+1.10}_{-2.16} \times 10^{-3}}$ & $\mathbf{{1.27}^{+1.06}_{-1.25} \times 10^{-2}}$ & $\mathbf{{4.30}^{+25.77}_{-40.38} \times 10^{-3}}$ \\
57 & La & $\sim {1.21} \times 10^{-7}$ & $\sim {4.00} \times 10^{-8}$ & $\sim {1.26} \times 10^{-3}$ & ${1.40}^{+1.75}_{-1.38} \times 10^{-2}$ & ${1.70}^{+54.54}_{-25.76} \times 10^{-3}$ \\
\textbf{58} & \textbf{Ce} & $\mathbf{\sim {1.54} \times 10^{-8}}$ & $\mathbf{\sim {1.89} \times 10^{-8}}$ & $\mathbf{{2.65}^{+14.38}_{-39.58} \times 10^{-3}}$ & $\mathbf{{1.82}^{+0.76}_{-1.81} \times 10^{-2}}$ & $\mathbf{{3.18}^{+25.37}_{-5.97} \times 10^{-3}}$ \\
59 & Pr & $\sim {1.43} \times 10^{-9}$ & $\sim {2.52} \times 10^{-9}$ & $\sim {7.55} \times 10^{-4}$ & ${3.75}^{+0.94}_{-3.73} \times 10^{-3}$ & $\sim {8.57} \times 10^{-4}$ \\
60 & Nd & $\sim {2.32} \times 10^{-8}$ & $\sim {1.81} \times 10^{-8}$ & ${2.31}^{+38.33}_{-38.15} \times 10^{-3}$ & ${1.64}^{+0.57}_{-1.64} \times 10^{-2}$ & ${2.79}^{+11.61}_{-1.59} \times 10^{-3}$ \\
61 & Pm & $\sim {3.23} \times 10^{-9}$ & $\sim {1.89} \times 10^{-9}$ & $\sim {2.15} \times 10^{-4}$ & ${2.01}^{+0.66}_{-2.00} \times 10^{-3}$ & $\sim {2.77} \times 10^{-4}$ \\
62 & Sm & $\sim {7.05} \times 10^{-9}$ & $\sim {6.21} \times 10^{-9}$ & ${2.06}^{+38.86}_{-5.68} \times 10^{-3}$ & ${2.45}^{+0.30}_{-2.45} \times 10^{-2}$ & ${2.83}^{+3.73}_{-0.82} \times 10^{-3}$ \\
63 & Eu & $\sim {4.47} \times 10^{-9}$ & $\sim {3.57} \times 10^{-9}$ & ${1.69}^{+16.03}_{-4.00} \times 10^{-3}$ & ${2.28}^{+0.26}_{-2.28} \times 10^{-2}$ & ${2.41}^{+13.09}_{-2.72} \times 10^{-3}$ \\
64 & Gd & $\sim {3.44} \times 10^{-9}$ & $\sim {2.79} \times 10^{-9}$ & ${2.24}^{+18.11}_{-1.26} \times 10^{-3}$ & ${4.07}^{+0.31}_{-4.07} \times 10^{-2}$ & ${3.56}^{+1.50}_{-0.27} \times 10^{-3}$ \\
65 & Tb & $\sim {1.13} \times 10^{-9}$ & - & ${1.07}^{+16.83}_{-2.65} \times 10^{-3}$ & ${2.20}^{+0.21}_{-2.20} \times 10^{-2}$ & ${1.78}^{+18.19}_{-2.79} \times 10^{-3}$ \\
66 & Dy & - & - & ${1.44}^{+4.36}_{-0.75} \times 10^{-3}$ & ${5.12}^{+0.56}_{-5.12} \times 10^{-2}$ & ${3.15}^{+15.93}_{-2.38} \times 10^{-3}$ \\
67 & Ho & - & - & $\sim {2.34} \times 10^{-4}$ & ${1.14}^{+0.22}_{-1.14} \times 10^{-2}$ & $\sim {6.16} \times 10^{-4}$ \\
68 & Er & - & - & ${4.51}^{+19.41}_{-2.15} \times 10^{-4}$ & ${3.07}^{+0.06}_{-3.07} \times 10^{-2}$ & ${1.49}^{+14.73}_{-1.77} \times 10^{-3}$ \\
69 & Tm & - & - & $\sim {1.33} \times 10^{-4}$ & ${1.32}^{+0.02}_{-1.32} \times 10^{-2}$ & $\sim {5.80} \times 10^{-4}$ \\
70 & Yb & - & - & $\sim {1.57} \times 10^{-4}$ & ${1.80}^{+0.21}_{-1.80} \times 10^{-2}$ & ${7.67}^{+57.35}_{-6.13} \times 10^{-4}$ \\
71 & Lu & - & - & $\sim {2.62} \times 10^{-5}$ & ${3.26}^{+0.10}_{-3.26} \times 10^{-3}$ & $\sim {1.37} \times 10^{-4}$ \\
72 & Hf & - & - & $\sim {5.86} \times 10^{-5}$ & ${1.57}^{+0.21}_{-1.57} \times 10^{-2}$ & ${5.95}^{+47.26}_{-5.78} \times 10^{-4}$ \\
73 & Ta & - & - & $\sim {5.08} \times 10^{-6}$ & ${2.70}^{+0.72}_{-2.70} \times 10^{-3}$ & $\sim {9.74} \times 10^{-5}$ \\
74 & W & - & - & $\sim {7.91} \times 10^{-5}$ & ${6.10}^{+1.51}_{-6.10} \times 10^{-2}$ & ${2.17}^{+1.23}_{-0.14} \times 10^{-3}$ \\
75 & Re & - & - & $\sim {2.30} \times 10^{-5}$ & ${2.10}^{+0.53}_{-2.10} \times 10^{-2}$ & ${7.41}^{+35.39}_{-5.94} \times 10^{-4}$ \\
76 & Os & - & - & $\sim {9.10} \times 10^{-5}$ & ${1.77}^{+0.33}_{-1.77} \times 10^{-1}$ & ${6.15}^{+0.50}_{-0.10} \times 10^{-3}$ \\
77 & Ir & - & - & $\sim {1.06} \times 10^{-5}$ & ${3.15}^{+0.54}_{-3.15} \times 10^{-2}$ & ${1.09}^{+10.96}_{-2.17} \times 10^{-3}$ \\
78 & Pt & - & - & $\sim {8.74} \times 10^{-6}$ & ${4.70}^{+2.24}_{-4.70} \times 10^{-2}$ & ${1.62}^{+3.68}_{-0.74} \times 10^{-3}$ \\
79 & Au & - & - & $\sim {1.41} \times 10^{-6}$ & ${9.53}^{+9.46}_{-9.53} \times 10^{-3}$ & $\sim {3.28} \times 10^{-4}$ \\
80 & Hg & - & - & $\sim {2.98} \times 10^{-8}$ & ${1.01}^{+19.08}_{-1.01} \times 10^{-3}$ & $\sim {3.47} \times 10^{-5}$ \\
81 & Tl & - & - & - & ${7.59}^{+18.57}_{-7.59} \times 10^{-5}$ & $\sim {2.60} \times 10^{-6}$ \\
82 & Pb & - & - & - & ${1.54}^{+4.94}_{-1.54} \times 10^{-3}$ & $\sim {5.29} \times 10^{-5}$ \\
83 & Bi & - & - & - & ${5.07}^{+18.73}_{-5.07} \times 10^{-4}$ & $\sim {1.74} \times 10^{-5}$ \\
84 & Po & - & - & - & ${6.61}^{+30.07}_{-6.61} \times 10^{-8}$ & $\sim {2.27} \times 10^{-9}$ \\
86 & Rn & - & - & - & ${7.57} \times 10^{-5}$ & $\sim {2.59} \times 10^{-6}$ \\
87 & Fr & - & - & - & $\sim {7.49} \times 10^{-9}$ & - \\
88 & Ra & - & - & - & ${5.73} \times 10^{-4}$ & ${1.96}^{+19.96}_{-1.74} \times 10^{-5}$ \\
89 & Ac & - & - & - & ${2.05}^{+38.40}_{-5.73} \times 10^{-4}$ & ${7.02} \times 10^{-6}$ \\
90 & Th & - & - & - & ${3.66}^{+13.38}_{-2.05} \times 10^{-4}$ & ${1.25} \times 10^{-5}$ \\
91 & Pa & - & - & - & ${1.55}^{+45.97}_{-3.66} \times 10^{-4}$ & ${5.32}^{+36.68}_{-2.59} \times 10^{-6}$ \\
92 & U & - & - & - & ${2.48}^{+24.18}_{-1.55} \times 10^{-4}$ & ${8.51} \times 10^{-6}$ \\
\enddata
\end{deluxetable*}

\begin{figure*}[!ht]
    \centering
    \includegraphics[width=0.47\textwidth]{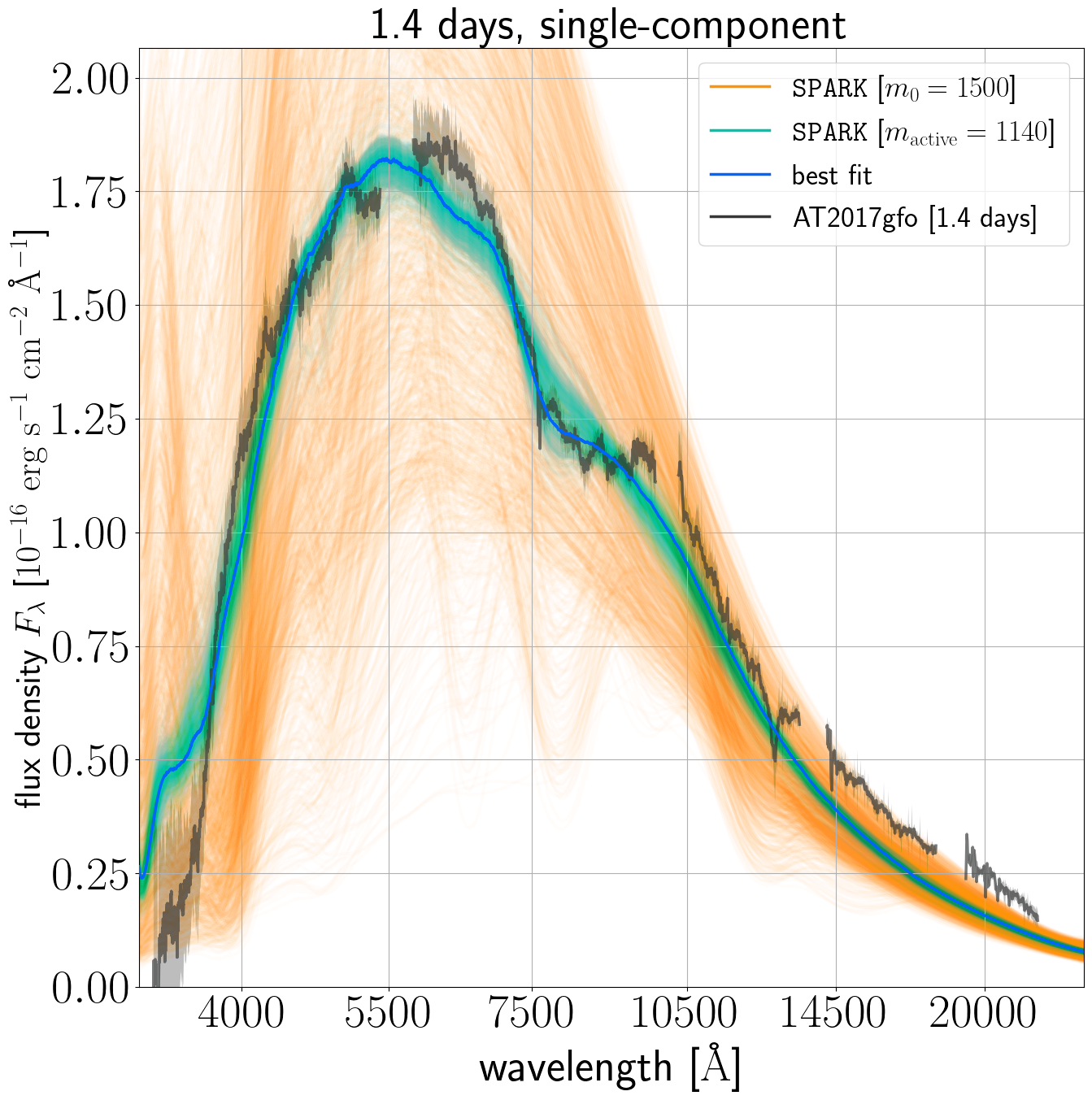}
    \includegraphics[width=0.47\textwidth]{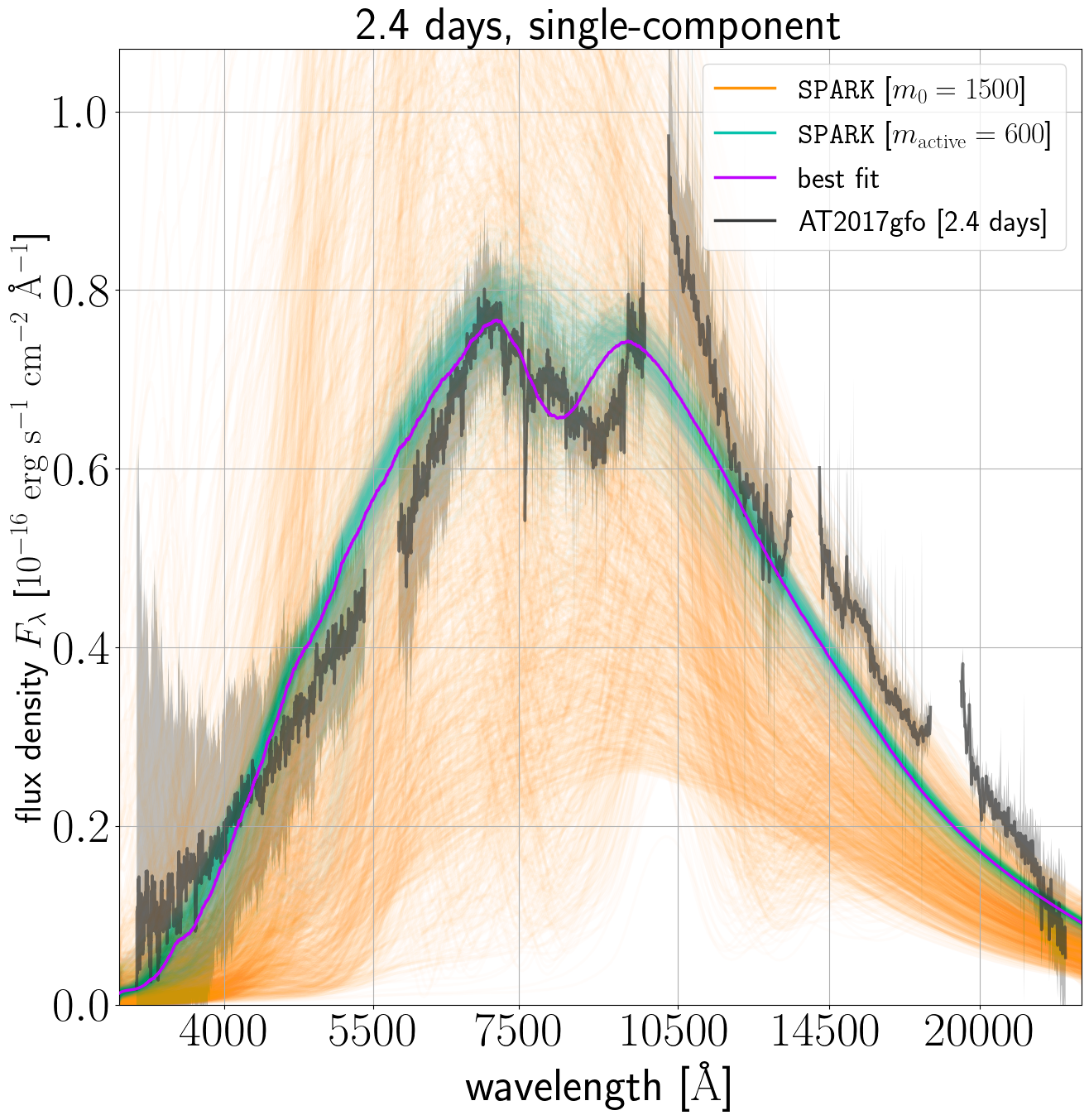}
    \includegraphics[width=0.47\textwidth]{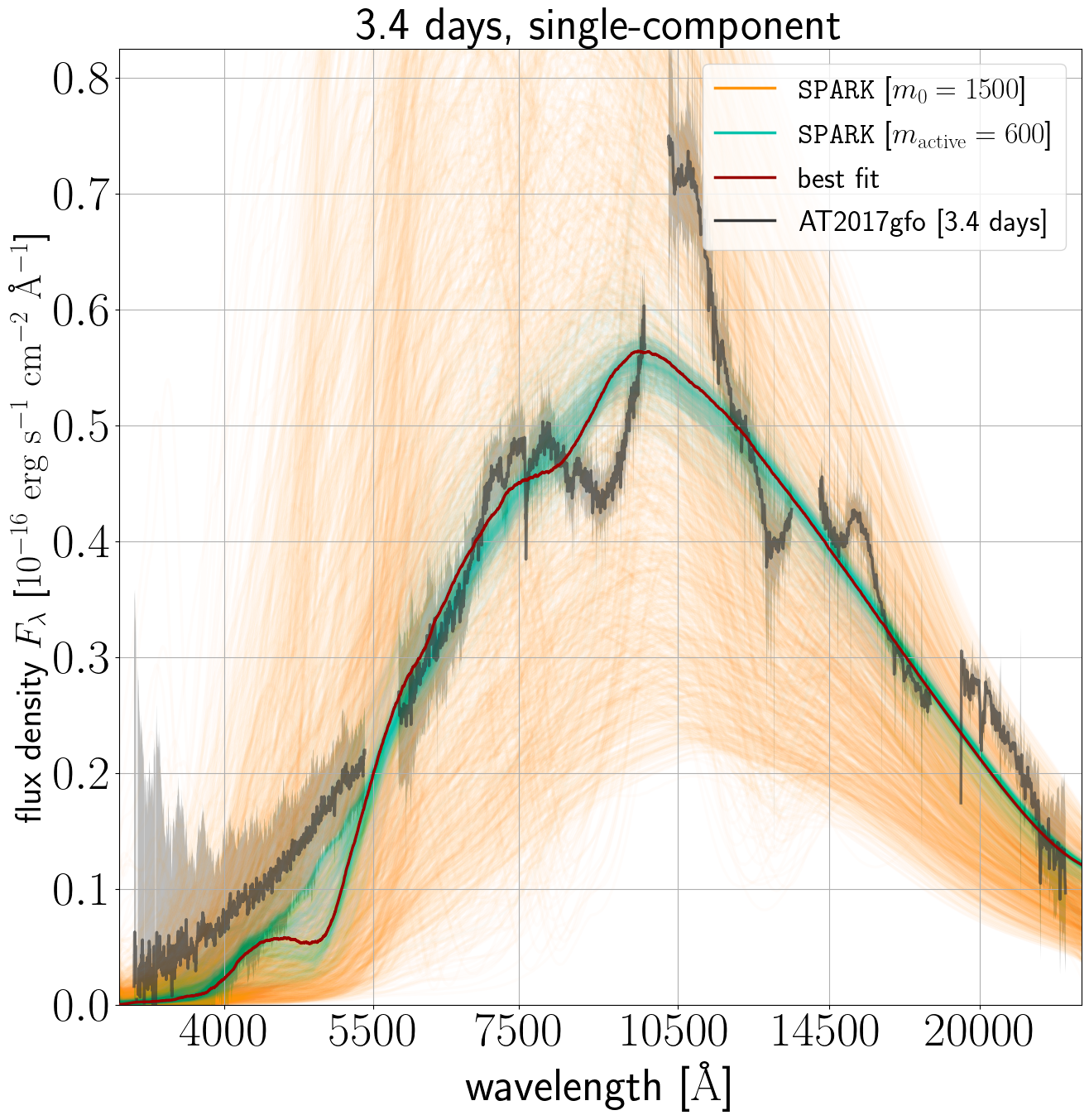}
    \figcaption{\textbf{All spectra in the training set, and the best fits, for the single-component \SPARK~runs at 1.4, 2.4, and 3.4 days.} \textit{Top left:} spectra corresponding to the $m_0 = 1500$ initial Latin hypercube samples to explore parameter space and $m_{\mathrm{active}} = 1140$ additional active learning points selected using BAPE, for the fit to the 1.4-day spectrum of AT2017gfo, taken from \V23. We also show the best fit, as obtained from the median of the posteriors. \textit{Top right:} the same as the top left panel, but for $m_0 + m_{\mathrm{active}} = 1500 + 600$ spectra for the 2.4-day fit. \textit{Bottom center:} the same as the top left panel, but for the 3.4-day fit. The 1.4- and 2.4-day single-component models (shown here) are favored over the multicomponent models; the 3.4-day single-component model is disfavored with respect to the multicomponent model shown in Figure~\ref{fig:all_spec_multi}.}\label{fig:all_spec_single}
\end{figure*}

\begin{figure*}[!ht]
    \centering
    \includegraphics[width=0.47\textwidth]{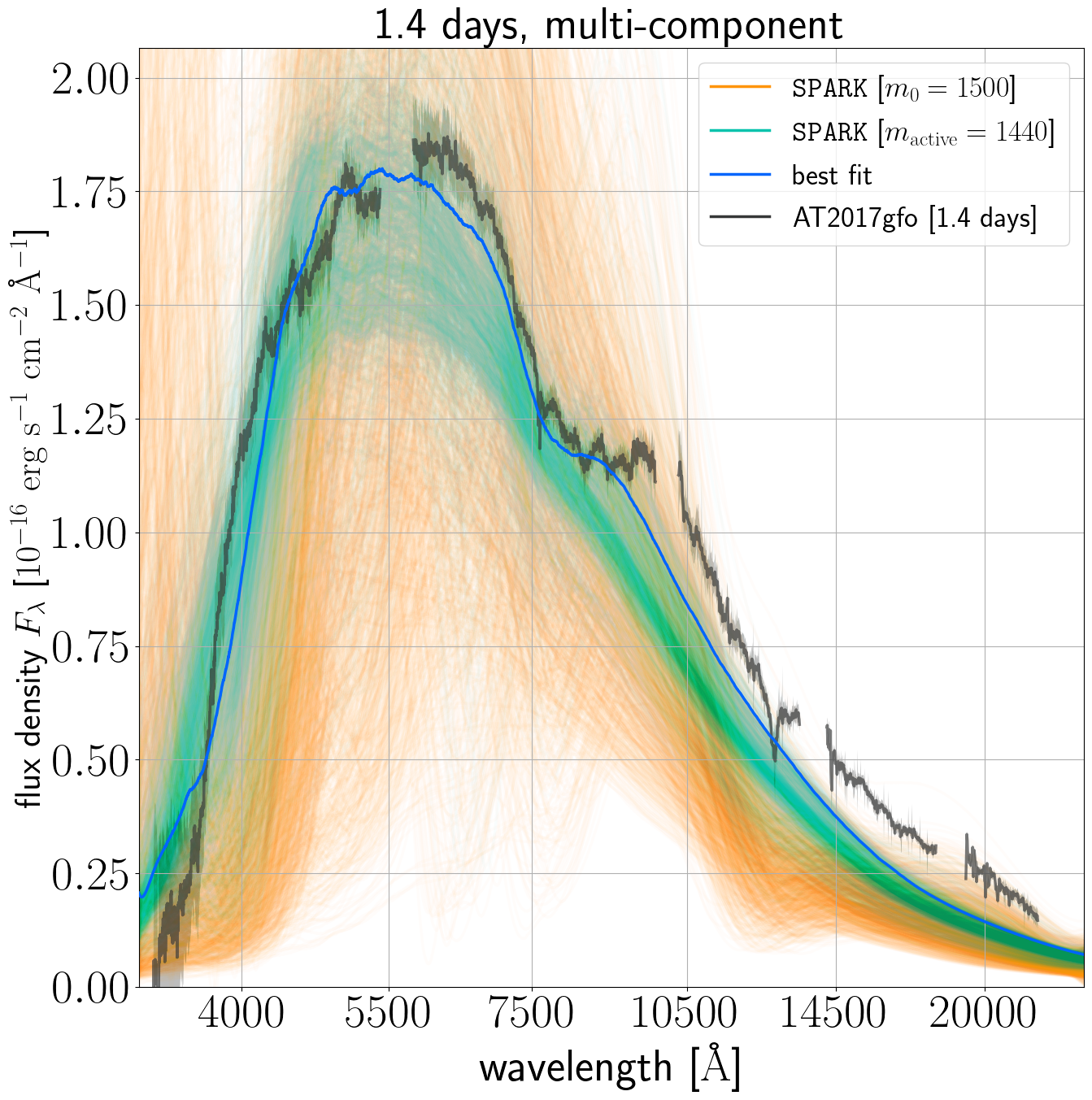}
    \includegraphics[width=0.47\textwidth]{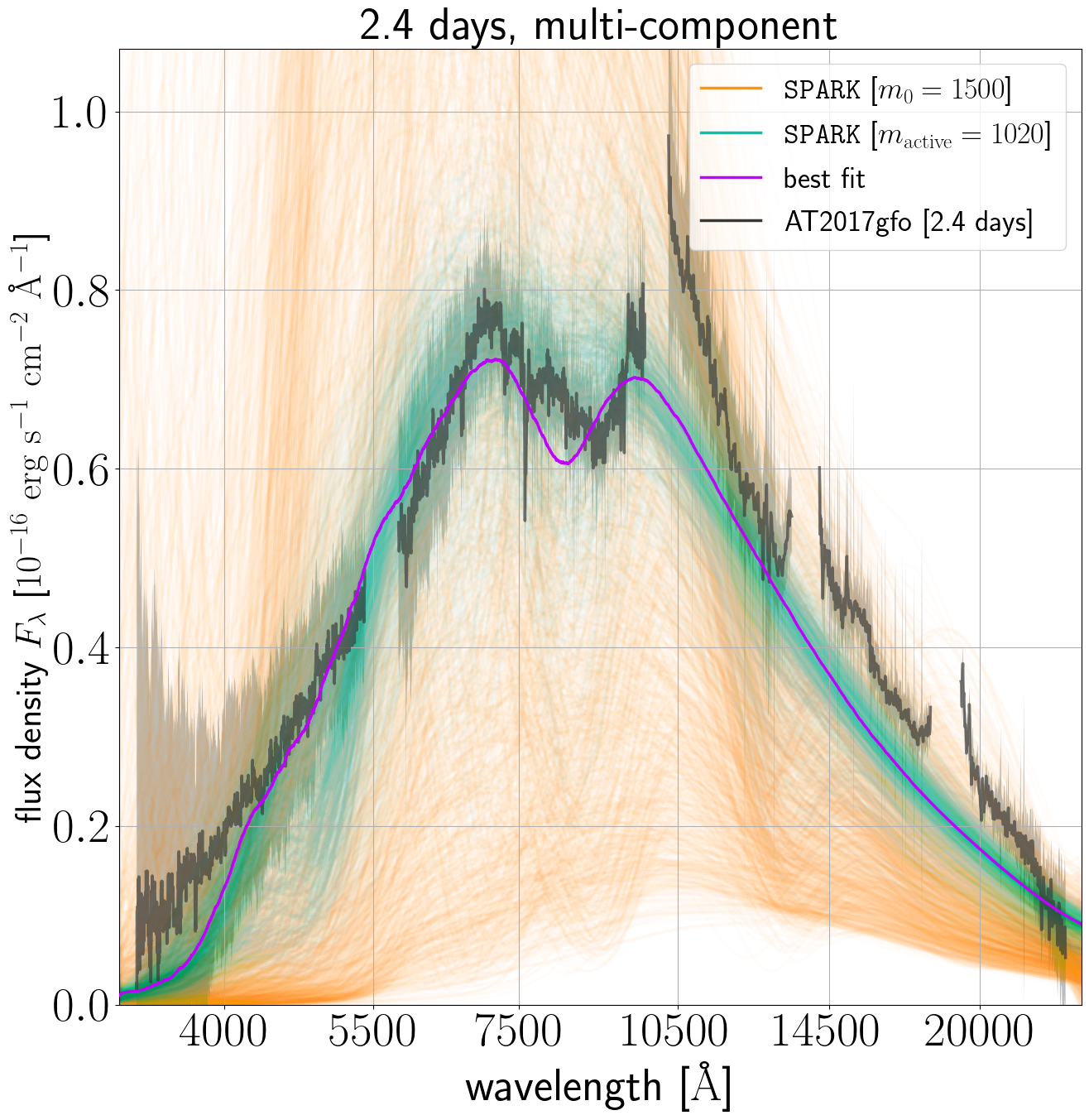}
    \includegraphics[width=0.47\textwidth]{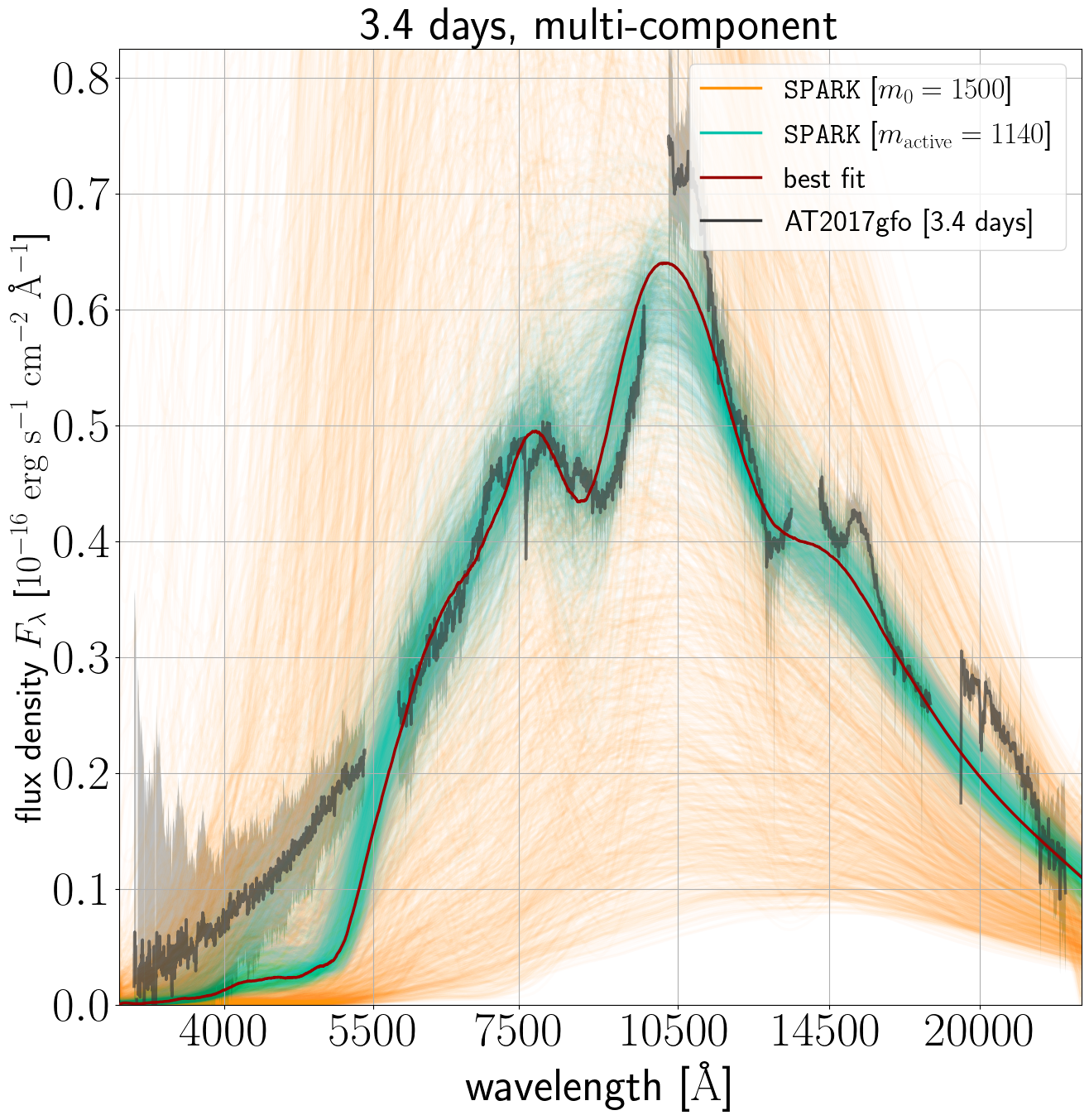}
    \figcaption{\textbf{All spectra in the training set, and the best fits, for the multicomponent \SPARK~runs at 1.4, 2.4, and 3.4 days.} \textit{Top left:} spectra corresponding to the $m_0 = 1500$ initial Latin Hypercube samples to explore parameter space and $m_{\mathrm{active}} = 1440$ additional active learning points selected using BAPE, for the fit to the 1.4-day spectrum of AT2017gfo. We also show the best fit, as obtained from the median of the posteriors. \textit{Top right:} the same as the top left panel, but for the 2.4-day fit, where $m_{\mathrm{active}} = 1020$ active learning samples are required. \textit{Bottom center:} the same as the top left panel, but for the 3.4-day fit, where $m_{\mathrm{active}} = 1140$ active learning samples are required. The 1.4- and 2.4-day multicomponent models shown here are disfavored with respect to the single-component models; the 3.4-day multicomponent model shown here is favored over the single-component model.}\label{fig:all_spec_multi}
\end{figure*}

\begin{figure*}[!ht]
    \includegraphics[width=0.95\textwidth]{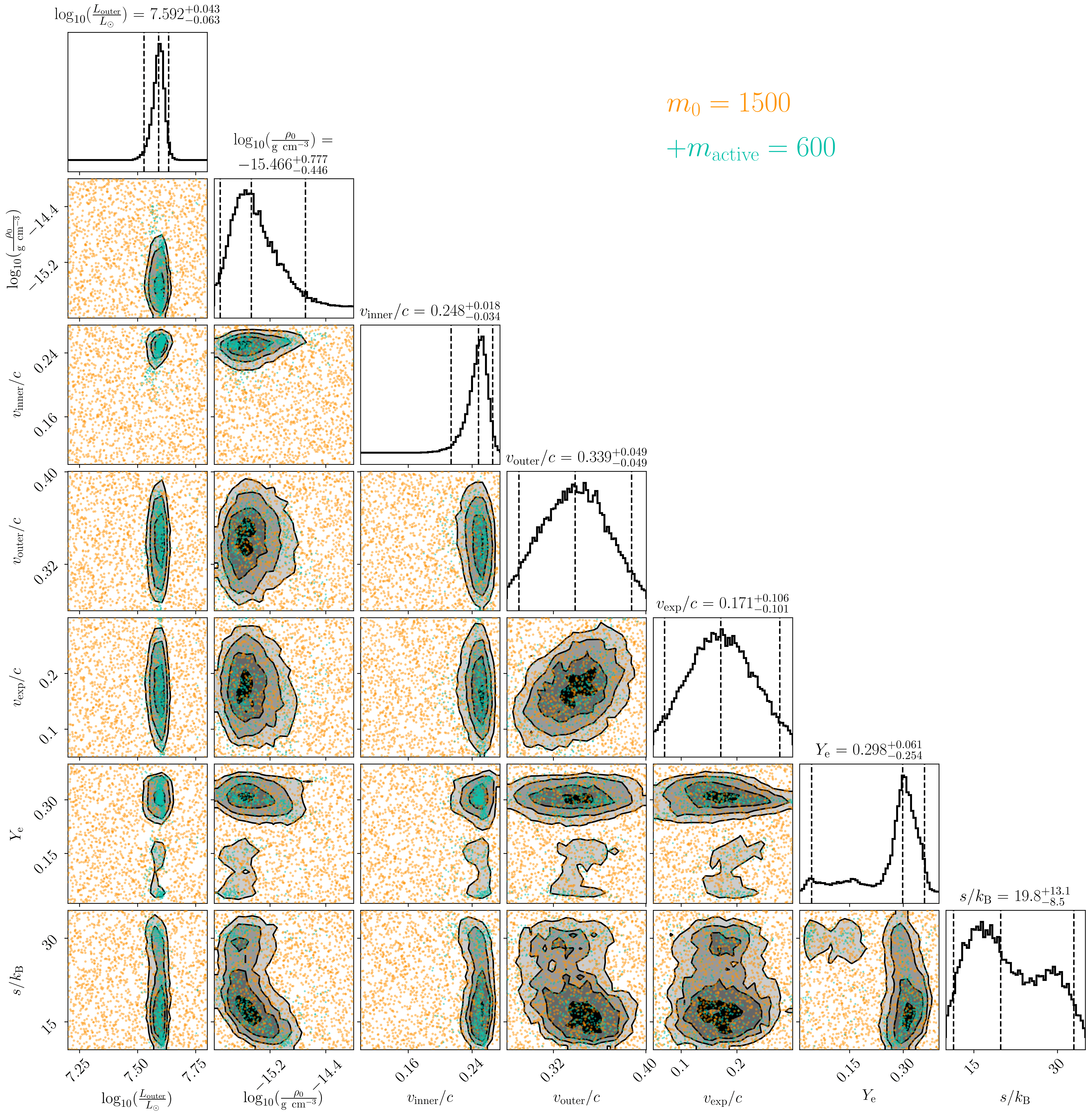}
    \figcaption{\textbf{Posterior from the single-component \SPARK~run at 2.4 days.} All posteriors are obtained through dynamic nested sampling of the surrogate GP using \texttt{dynesty} (\citealt{speagle20}). The posterior presents some bimodality, most evident in the $s/k_{\mathrm{B}}$ dimension. We select samples with $s/k_{\mathrm{B}} \leqslant 25.0$ to select the lower-entropy mode of the posterior, which is a better fit and is our preferred model in Table~\ref{tab:bestfit_single}.}\label{fig:corner_single_2.4}
\end{figure*}

\begin{figure*}[!ht]
    \includegraphics[width=0.95\textwidth]{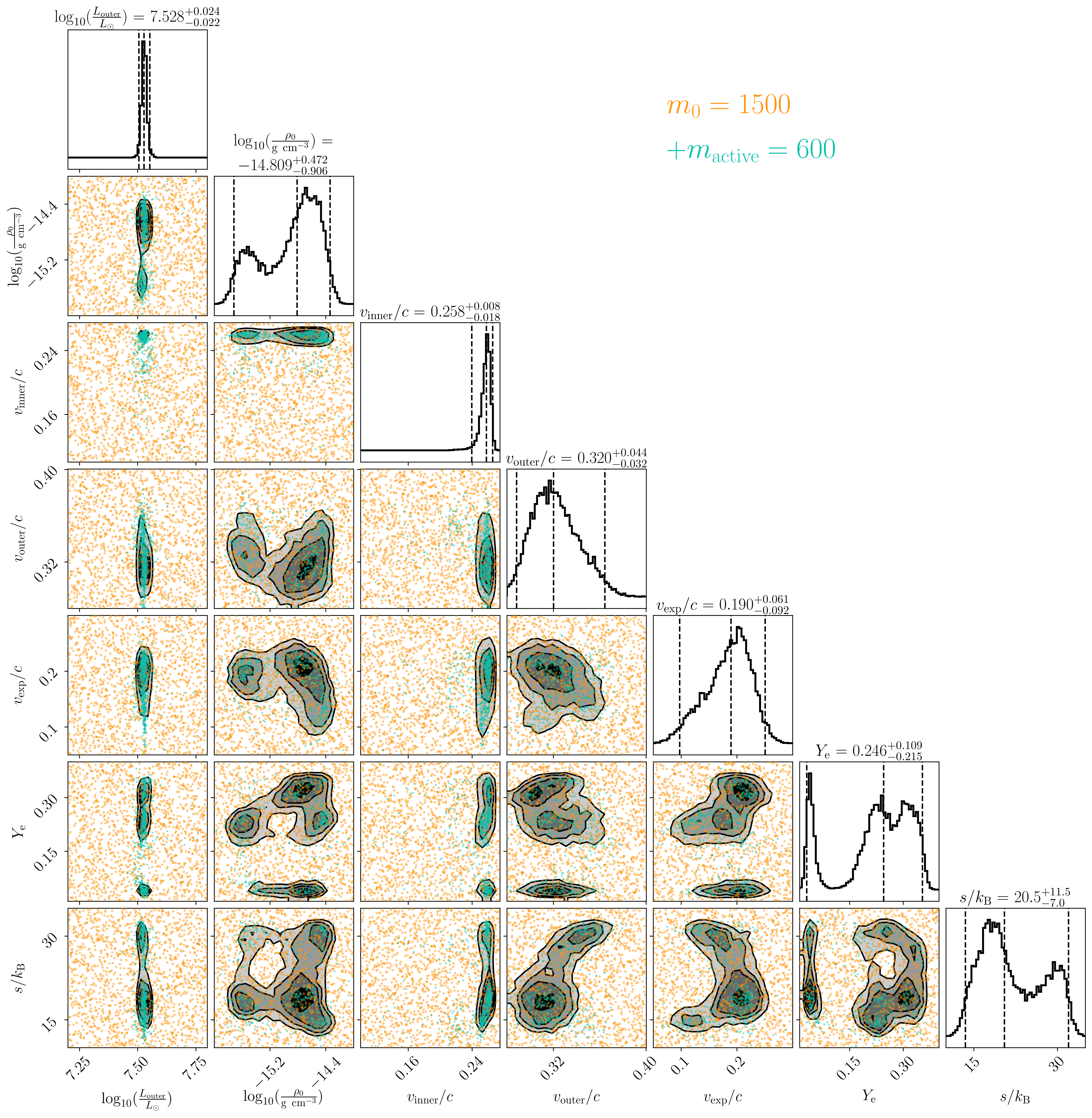}
    \figcaption{\textbf{Posterior from the single-component \SPARK~run at 3.4 days.} The posterior at this epoch is highly multimodal, as most evident in the $\rho_0,~Y_{\mathrm{e}},~\mathrm{and}~s/k_{\mathrm{B}}$ dimensions. Although none of these modes in the posterior correspond to a good fit and the multicomponent model is favored at 3.4 days, the higher-$\rho_0$, mid-$Y_{\mathrm{e}}$, low-$s$ mode produces the best possible fit with single-component ejecta. We select this mode of the posterior by selecting only samples in the range $\log_{10}(\rho_0 / \mathrm{g~cm^{-3}}) \in [-15.0, -14.0]~\cup~Y_{\mathrm{e}} \in [0.13, 0.29]~\cup~s/k_{\mathrm{B}} \in [10.0, 25.0]$. The parameters of this mode are given in Table~\ref{tab:bestfit_single}.}\label{fig:corner_single_3.4}
\end{figure*}

\begin{figure*}[!ht]
    \includegraphics[width=0.95\textwidth]{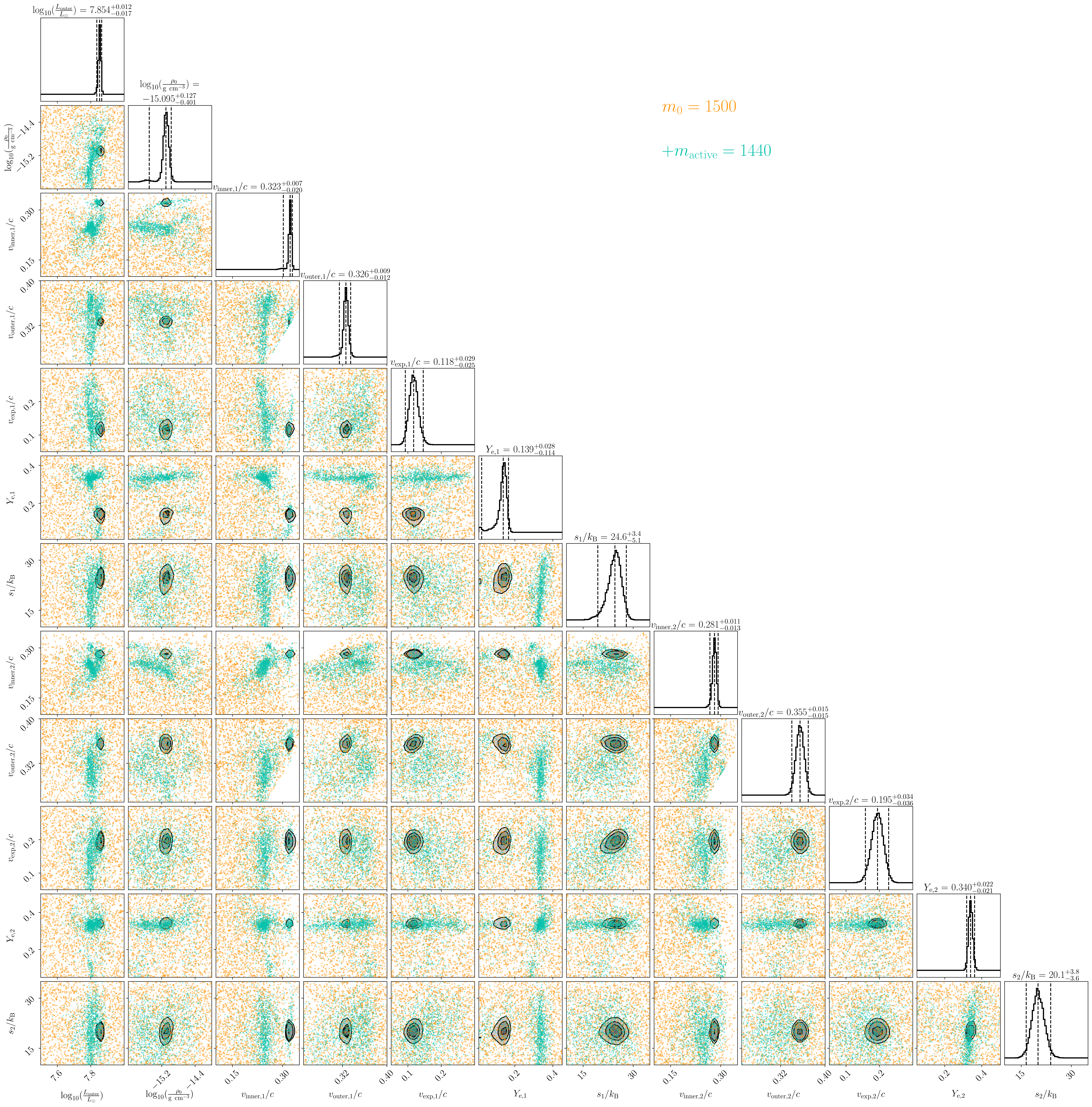}
    \figcaption{\textbf{Posterior from the multicomponent \SPARK~run at 1.4 days.} The region of parameter space sampled using active learning is substantially larger than the higher-probability region of the posterior, which is relatively narrow.}\label{fig:corner_multi_1.4}
\end{figure*}

\begin{figure*}[!ht]
    \includegraphics[width=0.95\textwidth]{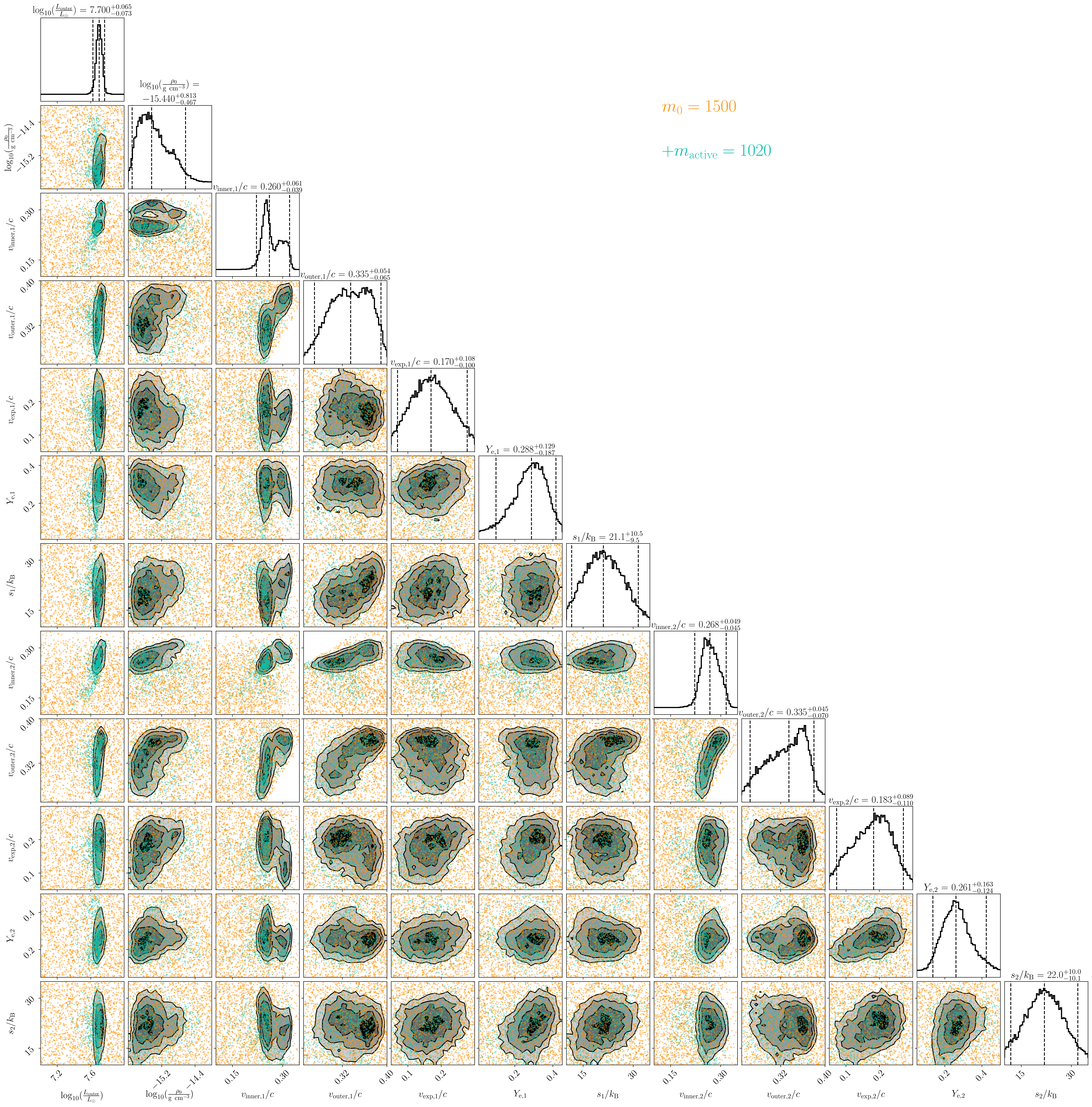}
    \figcaption{\textbf{Posterior from the multicomponent \SPARK~run at 2.4 days.} The posterior contains one lower-probability mode, most evident in the $v_{\mathrm{inner},1}$ dimension. We ignore this mode, as the median of the higher-probability mode yields our best fit, as given in Table~\ref{tab:bestfit_multi}.}\label{fig:corner_multi_2.4}
\end{figure*}

\begin{figure*}[!ht]
    \includegraphics[width=0.95\textwidth]{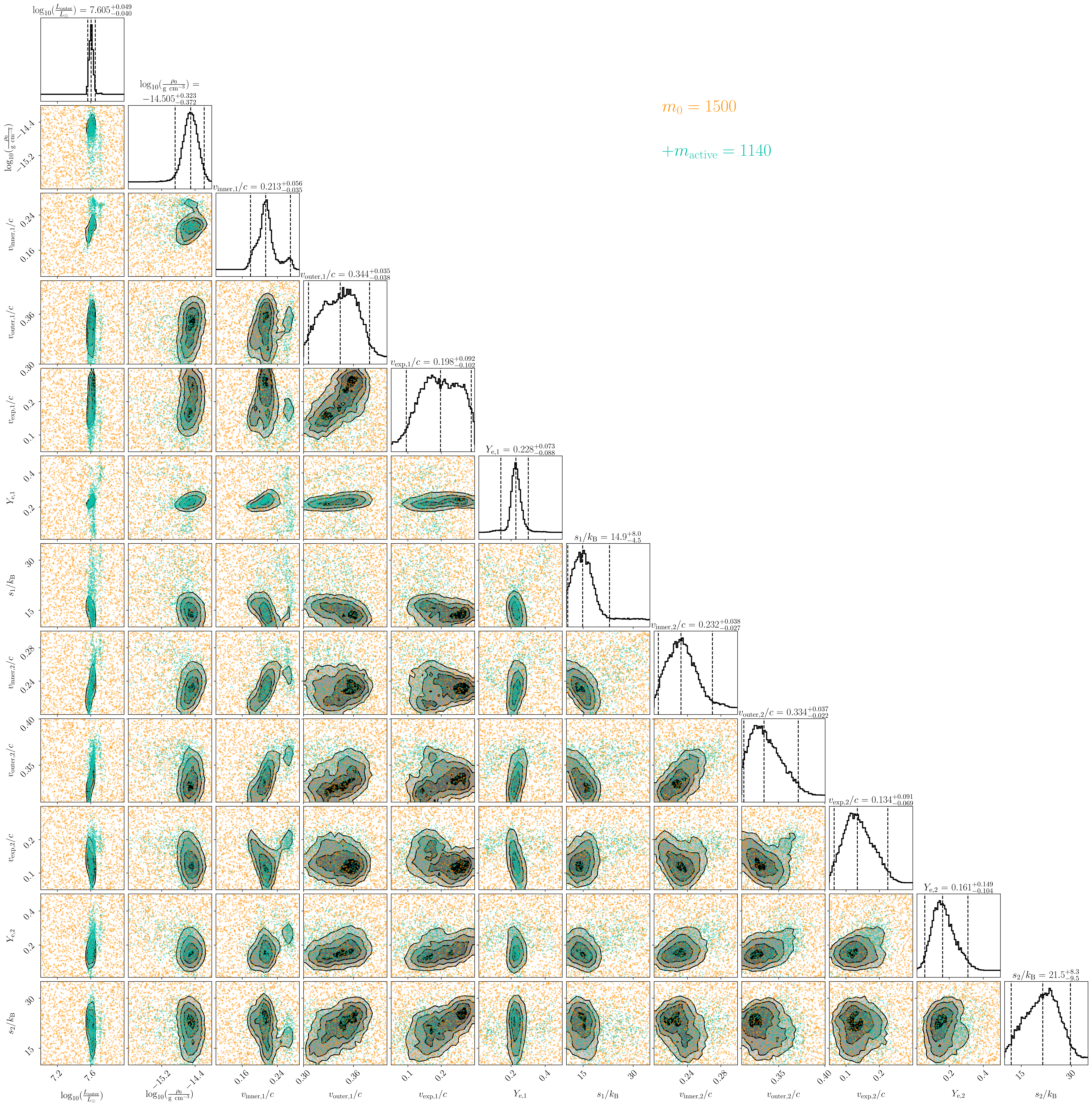}
    \figcaption{\textbf{Posterior from the multicomponent \SPARK~run at 3.4 days.} As at 2.4 days, the posterior contains one lower-probability mode, most evident in the $v_{\mathrm{inner},1}$ dimension. We once again ignore this mode, as the median of the higher-probability mode yields our best fit, as given in Table~\ref{tab:bestfit_multi}.}\label{fig:corner_multi_3.4}
\end{figure*}


\end{document}